\documentclass[10pt]{article}
\usepackage[a4paper,bindingoffset=0.2in,%
            left=1in,right=1in,top=1in,bottom=1in,%
            footskip=.25in]{geometry}
\usepackage{graphicx,xcolor}
\usepackage{amsmath}
\usepackage{amsthm}
\usepackage{gensymb}
\usepackage{amssymb, latexsym, mathrsfs}
\usepackage{resizegather}
\usepackage{enumerate}
\usepackage{subcaption}
\usepackage{algorithm}
\usepackage{indentfirst}
\usepackage{color}
\usepackage{transparent}
\usepackage{url}
\usepackage{cite}

\usepackage[affil-it]{authblk}
\usepackage[colorlinks=true,citecolor=blue]{hyperref}
\begin{document}

     \title{\textbf{Voltage Control of DC Islanded Microgrids: Scalable Decentralised $\mathcal{L}_1$ Adaptive Controllers}}
\author[1]{Daniel O'Keeffe\thanks{Research is supported by the Irish Research Council enterprise partnership scheme (Award No. R16920) in collaboration with University College Cork, Ireland and United Technologies Research Centre Ireland Ltd.}\thanks{Email: \tt\small{danielokeeffe@umail.ucc.ie}; Corresponding author}}
\author[2]{Stefano Riverso\thanks{Email: \tt\small RiversS@utrc.ucc.com}}
\author[2]{Laura Albiol-Tendillo\thanks{Email: \tt\small AlbiolL@utrc.ucc.com}}
\author[1,3]{Gordon Lightbody\thanks{Email: \tt\small g.lightbody@ucc.ie}}
\affil[1]{\textnormal{Control \& Intelligent Systems Group, School of Engineering,
University College of Cork, Ireland}}
\affil[2]{\textnormal{United Technologies Research Centre Ireland Ltd, 4th Floor Penrose Business Centre, Cork, Ireland}}
\affil[3]{\textnormal{MaREI-SFI Research Centre, University College Cork, Ireland}}

     \date{\textbf{Technical Report}\\ January, 2018}

     \maketitle

     \begin{abstract}
       Voltage stability is a critical feature of an efficiently operating power distribution system such as a DC islanded microgrid. Large-scale autonomous power systems can be defined by heterogeneous elements, uncertainty and changing conditions. This paper proposes a novel scalable decentralised control scheme at the primary level of the typical hierarchical control architecture of DC islanded microgrids with arbitrary topology. Local state-feedback $\mathcal{L}_1$ adaptive controllers are retrofitted to existing baseline voltage controllers of DC-DC boost converters, which interface distributed generation units with loads. Furthermore, local controller synthesis is modular as it only requires approximate information about the line parameters that couple neighbouring units. The performance of the proposed architecture is evaluated using a heterogeneous DC islanded-microgrid that consists of 6 DC-DC boost converters configured in a radial and meshed topology. The use of $\mathcal{L}_1$ adaptive controllers achieves fast and robust microgrid voltage stability in the presence of plug-and-play operations, unknown load and voltage reference changes, and unmodelled dynamics. Finally, sufficient conditions for global stability of the overall system are provided.  
     
          \textbf{Keywords:} \emph{Decentralised Control,  Low-Voltage DC Islanded Microgrid, Robust-Adaptive Control, Scalable Design, Voltage Stability}
     \end{abstract}

\newpage

\section{Introduction}
          Over the last decade, considerable efforts have been made to transform the current passive electricity grid into a dynamic, adaptable and resilient Smart Grid (SG) \cite{Adam}. The SG will be the future cornerstone for increased autonomy, reliability and distribution efficiency \cite{Symanski}. To achieve such features, intelligent interoperability between electrical, control and communication systems must be coordinated \cite{Farhangi2010}. 

The paradigm-shift towards distributed generation and storage units (DGUs/DSUs), market liberalisation, bi-directional transmission and demand-side interaction requires a distributed solution to manage future power networks. Islanded microgrids (ImGs) have emerged as a smart-grid initiative to autonomously integrate power-electronic-interfaced DGU/DSUs with loads, and provide ancillary services to the utility grid \cite{Lasseter2002,Guerrero2009,Quintero2009,Guerrero2013,Hogan2014a}.  Research and development of AC mGs has naturally progressed as AC power distribution deeply embedded in society \cite{Guerrero2013}, \cite{Microgrids2013}. Advances in DC-DC power electronics, has led to the promising emergence of DC mGs \cite{DeDoncker2014}. DC power distribution avoids inherent issues associated with AC such as harmonic compensation, reactive power and synchronisation; thus improving power quality, efficiency and reliability. Furthermore, the use of DC can reduce the weight of a power network by 10 tons/MW compared to AC components \cite{DeDoncker2014}; important for application such as the More Electric Aircraft (MEA) and electric vehicles. Recently, DC ImGs have been deployed in low-voltage DC (LVDC) networks such as telecommunication towers, occupied interior spaces, data centres and traction systems \cite{Symanski,Patterson2012,Becker2011,Elsayed2015}. The next wave of DC mG applications are expected in large-scale residential, commercial and industrial (C\&I) buildings, and aerospace \cite{Abdelhafez2009, Wheeler2014}.

 Key control features of large-scale mGs include; (i) Voltage stability and accurate load-sharing, (ii) Scalability: the ability to design controllers independent of the size and topology of the mG, (iii) Plug-and-play (PnP) operations: the ability to reconfigure DGU/DSUs without compromising global stability conditions, and (iv) Robustness to uncertainty within a heterogeneous system \cite{Anuradha2013,Meng2017}.

Voltage stability and accurate load-sharing of the DC-DC power converters that interface DGUs, DSUs with loads is integral to the safe and efficient operation of the ImG. A distributed hierarchical control architecture, utilising classical controllers and low-bandwidth communications, has become the standard within mG research \cite{Guerrero2013,Vasquez2016,Meng2017}. Though feature (i) is achieved using this approach, (ii)-(iv) are limited. Stability conditions are only satisfied for specific radial and bus-connected topologies, while homogeneous subsystems are only considered. As identified in \cite{Tucci2016c} and demonstrated in \cite{OKeeffe2017a}, the approach lacks scalability, PnP capabilities and robustness to uncertainty.

Recent mG research has addressed features (ii) and (iii). PnP control designs, first outlined in \cite{Stoustrup2009}, have successfully been deployed as primary and secondary controllers in the standard hierarchical control structure of AC \cite{Riverso2015,Riverso2017a} and DC ImGs \cite{Tucci2016c,Tucci2017g}. Primary controllers are locally responsible for stable power distribution, while secondary controllers coordinate system voltage levels and improve load-sharing accuracy using low-bandwidth communications (LBC).
PnP controllers maintain operation stability when DGUs and loads are reconfigured without requiring \textit{a priori} knowledge. Global asymptotic stability (GAS) is guaranteed by checking the viability of DGU plug-in/out operations through an off-line optimisation problem using linear matrix inequalities (LMIs). Furthermore, the technique is scalable as local controllers depend only on knowledge of corresponding DGU and line-couplings. Once DGU plug-in/out, neighbouring controllers are required to retune off-line, resulting in limited robustness. Recently, line-independent \cite{Tucci2016e, Han2017a} and robust \cite{Sadabadi2017a} PnP controllers were proposed to overcome this. However, these PnP techniques are computationally extensive, controller gains are required to discontinuously switch after off-line stability checks are performed, and robustness to network uncertainty is limited.
 
Adaptive control strategies
have recently been proposed to accommodate the heterogeneity and privacy requirements of large-scale mGs,
where dynamics, system reconfigurability, coupling and loads
can be uncertain or unknown \cite{Nasirian2014a,Josep2014,Vu2017}. These applications implement adaptive controls based on premeditated conditions or linear controllers to provide small-signal adjustments to droop resistances for dynamic performance when achieving system objectives such as voltage coordination and load-sharing. Here, the adaptive laws are adapting to uncertainty
of the droop parameters, as opposed to uncertainty concerning the system dynamics. Furthermore, these techniques depend on accurate system models, specific mG topologies, and do not address well-documented adaptive control issues, as outlined in \cite{Anderson2005}. These include guaranteed stability in the presence of uncertainty and fast adaptation. To address these issues, robust-adaptive control techniques, such as the $\mathcal{L}_1$ adaptive controller ($\mathcal{L}_1$AC) \cite{Cao2006b,Cao2006,L12010,Yoo2010}, have recently been developed and successfully deployed in various applications \cite{Michini2009,Gregory2010,Svendsen2012,Li2009,Zhao2014}.

This paper proposes a scalable decentralised $\mathcal{L}_1$AC to ensure fast and robust voltage control by augmenting baseline primary voltage controllers. The rationale for implementing an augmentation approach as opposed to a fully
adaptive one is that in real systems it is common to have baseline controllers designed to provide reference tracking and
disturbance rejection during nominal operation. Though the proposed design is not line-independent, due to its adaptive nature, conditions on \textit{a priori} parameter knowledge are relaxed. The paper aims to address features (i)-(iv) in the following context: \begin{itemize}
\item Heterogeneous DC ImG consisting of grid-forming DC-DC boost converters.
\item Parametric uncertainty of system dynamics i.e. network topology, line couplings and loads.
\item Reconfiguration of DGUs and loads through PnP operations
\end{itemize}

This paper is structured as follows. In section 2, the DC ImG state-space model is developed using a arbitrary load-connected topology and Quasi-Stationary Line approximations. State-feedback baseline controllers are also designed. In section 3, an overview of the $\mathcal{L}_1$AC architecture is discussed, and the decentralised augmenting controllers are subsequently designed with local and global stability conditions provided. Section 4 reviews criticisms of  $\mathcal{L}_1$AC theory. Finally, section 5 describes the simulation tests carried out, including PnP operations, robustness to unknown load changes and unmodelled dynamics, and voltage reference tracking.

A version of this work has been accepted to the 2018 European Control Conference.

 \section{DC Islanded Microgrid Model}
\subsection{DGU Electrical Model}
\begin{figure}[!htb]
\centering
\includegraphics[width=4in]{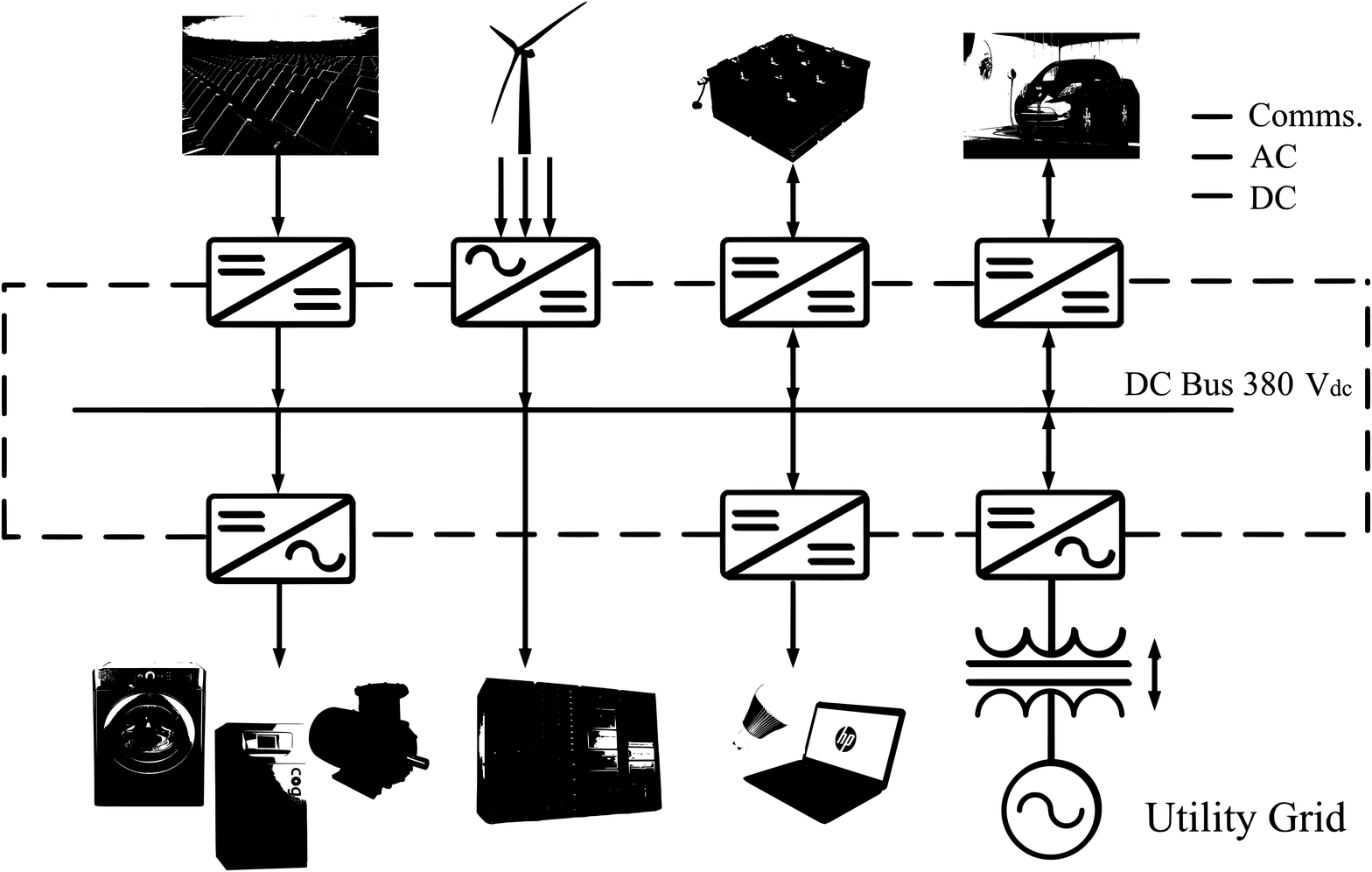}
\caption{Bus-connected LVDC microgrid. Adapted from \cite{Lu2014}.}
\label{fig:MG}
\end{figure}

Microgrids are generally coupled to a stiff utility grid. Transformers couple AC buses to the grid in AC mGs. However in the case of DC mGs, power limited converter units interface the mG with the utility grid, thus effectively islanding them i.e. the inertia of the stiff utility grid is buffered. A fundamental feature of any islanded-mG (ImG) is the ability to convert power between different DGUs, DSUs and loads automatically and efficiently. This is performed using different DC-DC boost and buck converter topologies. Fig. \ref{fig:MG} presents a bus-connected DC mG which can exchange power with the utility grid.

This work considers boost converters, which step-up low voltages to high voltages. Initial DC mG research investigated buck converters \cite{Vasquez2016}, \cite{Guerrero2011}, \cite{Tucci2016c}, \cite{Lu2014}, \cite{Shafiee2014} as they are commonly interfaced with low-power loads, and are easier to control. Boost converter controllers are notoriously difficult to tune in mGs due to their non-minimum phase action and have only received attention recently \cite{OKeeffe2017a}, \cite{Wang2016}.

\begin{figure*}[!htb]    
\centering
\includegraphics[width=\textwidth, height=4cm]{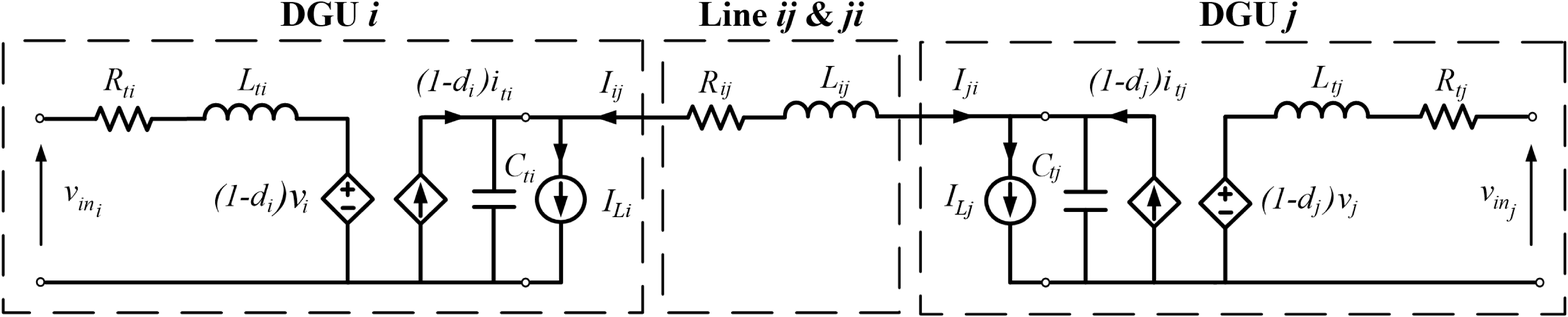}
\caption{Averaged model of DC ImG composed of two radially coupled boost converter DGUs with unknown loads.}
\label{fig:MG2}
\end{figure*}

For simplicity the DC ImG can be modelled as a  two-node network, and subsequently generalised to a network of \textit{N}-nodes. Fig. \ref{fig:MG2} represents the averaged model, which considers dynamics over both on/off switching states, of two boost converters $i$ and $j$ coupled via resistive and inductive power lines, for $k \in \{i, j\}$. 

%The ImG model in \cite{Tucci2016c} considers the average model of buck converters, which mitigates the model's dependency on pulse-width modulated (PWM) duty cycles $d_k$, acting as the control signals. However, due to the indirect energy transfer topology of boost converters, the duty cycles $d_k$ become stability parameters i.e. both multiply the inductor current $i_{t_{k}}$, and capacitor voltage states $v_{k}$. Therefore, the small-signal model must be considered (explain why better?).

The ImG of Fig. \ref{fig:MG2} is arranged in a general load-connected topology where each DGU supplies power to a local load at the point of common coupling (PCC). DGUs can be mapped to load-connections via the Kron Reduction method \cite{D??rfler2013}, \cite{KronBah1965} which preserves the profile of electrical parameters at the PCC. This is a positive feature, as the model of each DGU is not dependent on the load, which could be unknown (e.g. non-linear/linear resistive, interfacing buck converter or variable speed motor drive). Instead, Fig. \ref{fig:MG2} represents the load as a current disturbance, $I_{L_k}$. 

Applying Kirchoff's voltage and current laws to the DC ImG of Fig. \ref{fig:MG2} yields the following set of averaged differential equations:
\begin{subequations}
\begin{equation}
\textrm{DGU $i$:}
\begin{cases}
	     	     \dfrac{d I_{ti}}{dt} = \dfrac{1}{L_{ti}}V_{in_{i}} - \dfrac{(1-d_i)}{L_{ti}} V_{dc_{i}} - \dfrac{R_{ti}}{L_{ti}} I_{ti} \\
	     	     \\
	     \dfrac{d V_{dc_{i}}}{dt} =  \dfrac{(1-d_i)}{C_{ti}} I_{ti} + \dfrac{1}{C_{ti}} I_{ij} - \dfrac{1}{C_{ti}}I_{Li}
	     \end{cases}
	     \label{eq:DGU i}
\end{equation}
\begin{equation}
\textrm{DGU $j$:}
\begin{cases}
	     	     \dfrac{d I_{tj}}{dt} = \dfrac{ 1}{L_{tj}}V_{in_{j}} - \dfrac{(1-d_j)}{L_{tj}} V_{dc_{j}} - \dfrac{R_{tj}}{L_{tj}} I_{tj}, \\
	     	     \\
	     \dfrac{dV_{dc_{j}}}{dt} =  \dfrac{(1-d_j)}{{C_{tj}}} I_{tj} + \dfrac{1}{C_{tj}} I_{ji} - \dfrac{1}{C_{tj}}I_{Lj}
	     \end{cases}
	     \label{eq:DGU j}
\end{equation}
\begin{equation}
\textrm{Line $ij$:}
\begin{cases}
L_{ij}\dfrac{d I_{ij}}{dt} = V_{dc_{j}} - R_{ij} I_{ij} - V_{dc_{i}}.
\end{cases}
\label{eq:LineIJ}
\end{equation}
\begin{equation}
\textrm{Line $ji$:}
\begin{cases}
L_{ji}\dfrac{d I_{ji}}{dt} = V_{dc_{j}} - R_{ji} I_{ji} - V_{dc_{i}}
\end{cases}
\label{eq:LineJI}
\end{equation}
\label{eq:AvMG}
\end{subequations}
\vspace{3mm} 

\textbf{Remark 1:} \textit{As in \cite{Tucci2016c}, lines $ij$ and $ji$ physically couple DGU $i$ to DGU $j$ and vice versa, therefore $R_{ij}$ = $R_{ji}$ and $L_{ij}$ = $L_{ji}$. Hence, in steady-state, $I_{ij}$ = $-I_{ji}$.}
\vspace{3mm} \newline
The system of (\ref{eq:AvMG}) can be represented in state space form as,

\begin{equation}
\begin{aligned}
\dot{x}(t) = Ax(t)+Bu(t)+Ed(t)
\\
y(t) = Cx(t)
\end{aligned}
\label{eq:AvMGSS}
\end{equation}
where, $x(t) =[
I_{ti}, V_{dc_{i}}, I_{tj}, V_{dc_{j}}, I_{ij}, I_{ji}]^T$, is the state vector, $u(t) =
[V_{in_{i}}, V_{in_{j}}
]^T$ is the input, $d(t) = [
I_{L_{i}}, I_{L_{j}}
]^T$ is the load current disturbance, and $y(t) =
[V_{dc_{i}}, V_{dc_{j}}]^T$ is the measurable output. Matrices of (\ref{eq:AvMGSS}) are detailed in section \ref{Model Matrices}.

\subsection{Quasi Stationary Line Model}\label{sec:QSL}
If the time constant of the line transients is very fast, i.e. assuming $L_{ij}$ and $L_{ji}$ are significantly small, then line dynamics can be neglected. This type of model is known as a Quasi-Stationary Line (QSL) approximation. This is usually a good approximation for small-scale mGs where the lines are predominantly resistive. In open-loop, global stability can be inferred by ensuring local DGU stability, as detailed in section \ref{Model Matrices}.
Line equations (\ref{eq:LineIJ}) and (\ref{eq:LineJI}) are represented in steady-state form using QSL approximations, i.e. $\dfrac{d I_{ij}}{dt} = \dfrac{d I_{ji}}{dt}= 0$:

\begin{equation}
I_{ij} = \frac{V_{dc_{j}} - V_{dc_{i}}}{R_{ij}},
\label{eq:i_ij}
\end{equation} 

\begin{equation}
I_{ji} = \frac{V_{dc_{i}} - V_{dc_{j}}}{R_{ji}}.
\label{eq:i_ji}
\end{equation} 

Replacing line current variable $I_{ij}$ of  equation (\ref{eq:DGU i}) 
with equation (\ref{eq:i_ij}) yields the following model for DGU $i$,

\begin{equation}
\textrm{DGU $i$:}
\begin{cases}
	     	     \dfrac{d I_{ti}}{dt} = \dfrac{1}{L_{ti}}V_{in_{i}} - \dfrac{(1-d_i)}{L_{ti}} V_{dc_{i}} - \dfrac{R_{ti}}{L_{ti}} I_{ti} \\
	     	     \\
	     \dfrac{dV_{dc_{i}}}{dt} =  \dfrac{(1-d_i)}{{C_{ti}}} I_{ti} + \dfrac{V_{dc_{j}}}{R_{ij}C_{ti}} - \dfrac{V_{dc_{i}}}{R_{ij}C_{ti}} - \dfrac{1}{C_{ti}}I_{Li}
	     \end{cases}
	     \label{eq:DGU i 2}
\end{equation}

Interchanging indexes $i$ and $j$ yields the model for DGU $j$. Representing (\ref{eq:DGU i 2}) in a general compact state space form, the dynamics of DGU $i$ are,

\begin{equation}
\Sigma_{[i]}^{\textrm{DGU}}:
\begin{cases}
\dot{x}_{[i]}(t) = \left[ \begin{array}{cc}
-\frac{R_{ti}}{L_{ti}} & -\frac{(1-d_i)}{L_{ti}}\\
\frac{(1-d_j)}{C_{tj}} & -\frac{1}{R_{ij}C_{ti}}
\end{array} \right]x_{[i]}(t)+\left[ \begin{array}{c}
\frac{1}{L_{t_{i}}}\\
0
\end{array} \right]V_{in_{i}} + \left[ \begin{array}{c}
0\\
-\frac{1}{C_{ti}}
\end{array} \right]I_{Li} + \left[ \begin{array}{cc}
0 & 0 \\
0 & \frac{1}{R_{ij}C_{ti}}
\end{array} \right]x_{[j]}(t)
\\
y_{[i]}(t) = C_ix_{[i]}(t)
\end{cases}
\label{eq:DGUSS}
\end{equation}
where ${x}_{[i]}(t)= [I_{t_{i}}, V_{dc_{i}}]^T$ , $I_{Li}$ is the exogenous current disturbance. Unlike with the buck converter, where the averaged state space model of (\ref{eq:DGUSS}) is equivalent to the small-signal state space model, the boost converter is different. From the state matrix of above, the duty-cycle control input is a product of the state vector. As a result, the duty-cycle operating point directly influences stability. The averaged model is therefore non-linear and must be linearised about the duty-cycle operating point by forming a small-signal model\footnote{Note: each average quantity can be expressed as the sum of its steady state and small-signal values e.g. $d_k = D_k + \tilde d_k$, $V_{dc_{k}} = \bar{V}_{dc_{k}} + \tilde v_{dc_{k}}$.}.

\begin{equation}
\Sigma_{[i]}^{\textrm{DGU}}:
\begin{cases}
\dot{x}_{[i]}(t) = A_{ii}x_{[i]}(t)+B_{i}u_{[i]}(t) + E_{i}d_{[i]}(t) + \zeta_{[i]}(t) + \gamma_{[i]}(t) \\
y_{[i]}(t) = C_ix_{[i]}(t)
\end{cases}
\label{eq:DGUSS2}
\end{equation}
where ${x}_{[i]}(t)= [\tilde i_{t_{i}},\tilde v_{dc_{i}}]^T$ , is the small-signal state vector, $u_{[i]}(t) = \tilde{d}_i(t)$ is the small-signal PWM control signal, $d_{i}(t) = \tilde{i}_{Li}$ is the small-signal exogenous current disturbance, $\zeta_{[i]}(t) = A_{ij}x_{j}(t)$ represents coupling with DGU $j$ and $\gamma_i(t) = \frac{\tilde v_{in_{i}}}{L_{ti}}$ is the small-signal input voltage disturbance. It is assumed that changes in input voltages $V_{in_{k}}$ are very slow, and thus can be neglected\footnote{As the input voltage to power converters in a mG is usually from renewable power or storage devices. The dynamics of these devices are much slower than the fast switching dynamics of power converters, therefore it is a safe assumption to neglect small-signal changes in input voltage}. Therefore $\gamma_i(t) = 0$.

The matrices of (\ref{eq:DGUSS2}) are,
\begin{equation*}
A_{ii}=
\left[ \begin{array}{cc}
-\frac{R_{ti}}{L_{ti}} & -\frac{(1-D_i)}{L_{ti}}\\
\frac{(1-D_j)}{C_{tj}} & -\frac{1}{R_{ij}C_{ti}}
\end{array} \right]
A_{ij}=\left[ \begin{array}{cc}
0 & 0 \\
0 & \frac{1}{R_{ij}C_{ti}}
\end{array} \right]
B_{i} = \left[ \begin{array}{c}
\frac{\bar{V}_{dc_{i}}}{L_{t_{i}}}\\
\frac{-\bar{I}_{t_{i}}}{C_{ti}}
\end{array} \right]
E_i = 
\left[ \begin{array}{c}
0\\
-\frac{1}{C_{ti}}
\end{array} \right]
C_i =
\left[ \begin{array}{cc}
0 & 1
\end{array} \right]
\end{equation*}

where $\bar{V}_{dc_{i}} = \frac{\bar{V}_{in_{i}}}{(1-D_i)}$ and $\bar{I}_{t_{i}} = \frac{\bar{V}_{in_{i}}}{(1-D_i)^2R_{L_{i}}}$.

\subsection{QSL Model DC Islanded Microgrid Composed of \textit{N} DGUs}
In this section, the two DGU network of Fig. \ref{fig:MG2} is generalised to an ImG composed of \textit{N} converter DGUs. \cite{OKeeffe2017a} demonstrated that converter coupling dynamics predominantly manifest from physical power lines; duty-cycle  coupling is weak. Neighbouring DGUs are thus defined if they are coupled by the $RL$ power line of Fig. \ref{fig:MG2}. Letting $\mathcal{D} = \{1,...,\textit{N} \}$, $\mathcal{N}_i \subset \mathcal{D}$ denotes a neighbour-subset for DGU $i$. As before, assuming QSL approximation of all line dynamics $(i, j)\in \mathcal{D}$, the DC ImG model is represented by (\ref{eq:DGUSS}), with $\zeta_{[i]}(t) = \sum_{j\in\mathcal{N}_i}A_{ij}x_{[j]}(t)$. The only change in (\ref{eq:DGUSS}) is the local state vector matrix $A_{ii}$, becoming:
 \begin{equation}
A_{ii}=
\left[ \begin{array}{cc}
-\frac{R_{ti}}{L_{ti}} & -\frac{(1-D_i)}{L_{ti}}\\
\frac{(1-D_i)}{C_{ti}} & \sum_{j\in\mathcal{N}_i}-\frac{1}{R_{ij}C_{ti}}
\end{array} \right]
\label{eqn:Aii}
 \end{equation}
The overall global model of the $N$ DGU ImG can be given by,
 
 \begin{equation}
 \mathbf{\Sigma}^{DGU}_{[N]}:
 \begin{cases}
 \dot{\textbf{x}}(t)= \textbf{Ax}(t)+\textbf{Bu}(t) + \textbf{Ed}(t) \\
  \textbf{y}(t) = \textbf{Cx}(t)
  \end{cases}
  \label{eqn:LSSMIMO}
 \end{equation}
where $\textbf{x} = (x_{[1]}, x_{[2]},....,x_{[N]}) \in \mathbb{R}^{2n_i}, \textbf{u} = (u_{[1]}, u_{[2]},....,u_{[N]}) \in \mathbb{R}^{n_i}, \textbf{d} = (d_{[1]}, d_{[2]},....,d_{[N]})\in \mathbb{R}^{n_i}, \textbf{y} = (y_{[1]}, y_{[2]},....,y_{[N]})\in \mathbb{R}^{n_i}$. Matrices $\textbf{A}$, $\textbf{B}$, $\textbf{C}$ and $\textbf{E}$ are detailed in section \ref{Model Matrices}.

         \subsection{Decentralised Baseline Voltage Control}
 Power converter designers and manufacturers design cascaded current and voltage loop controllers for nominal operation. Such baseline controllers are intended to track voltage references and asymptotically reject unknown load disturbances when operating without uncertainty \cite{Lavretsky2012}. The idea of this work is to retrofit each DGU, with decentralised $\mathcal{L}_1$ adaptive voltage controllers, in order to enhance the performance of each DGU during operations deviating from the nominal case i.e. parametric uncertainty, PnP operations, unknown load changes. Effectively,  the closed-loop DGU can be treated as a black-box\footnote{Within reason; some \textit{a priori} bound or subset must be known}. This section details the design of two conventional decentralised baseline controllers; a static state-feedback (DeSSf) controller with integral action, and a type III output voltage compensator.

\subsubsection{Decentralised Static State-Feedback Controller}
 
 Baseline controllers are designed for standalone decoupled converters, assuming a connection to a linear resistive load. The state space matrices, of the same form as (\ref{eq:DGUSS2}) but without coupling term $\zeta_{[i]}(t)$, are given as,
 
 \begin{equation*}
 A_{ii}^{nom}=
 \left[ \begin{array}{cc}
 -\frac{R_{t_{i}}}{L_{t_{i}}} & -\frac{(1-D_{i})}{L_{t_{i}}}\\
 \frac{(1-D_{i})}{C_{t_{i}}} & -\frac{1}{R_{L_{i}}C_{t_{i}}}
 \end{array} \right]
 A_{ij}^{nom}=
  \left[ \begin{array}{cc}
 0 & 0\\
  0 & \frac{1}{R_{ij}C_{t_{i}}}
  \end{array} \right]
 B_{i}^{nom} = \left[ \begin{array}{c}
 \frac{\bar{V}_{dc_{i}}}{L_{t_{i}}}\\
 \frac{-\bar{I}_{t_{i}}}{C_{t_{i}}}
 \end{array} \right]
 E_{i}^{nom} = 
 \left[ \begin{array}{c}
 0\\
 -\frac{1}{C_{t_{i}}}
 \end{array} \right]
  \end{equation*}
  
 \textbf{Remark 2:} \textit{Power converter manufacturers design baseline controllers for nominal operation with a priori parametric knowledge. However, due to privacy requirements and changing conditions the subsequently designed augmenting $\mathcal{L}_1$ adaptive controllers does not have a priori parametric knowledge, and therefore must be designed within a known subset.}
\vspace{3mm} \newline
 In order to track constant voltage references in the presence of constant current disturbances, integral state error between the reference voltage and output voltage is added to the local DGU model. The dynamics are defined as,
 
 \begin{equation}
 \xi_{[i]}(t) = \int_{0}^{t}(V_{ref_{[i]}} - y_{[i]}(t)) dt = \int_{0}^{t}(V_{ref_{[i]}} - C_{i}x_{[i]}(t)) dt
 \end{equation} 
 The DeSSf control law with integral action becomes,

\begin{equation}
\mathcal{C}_{[i]} : u_{[i]}^{bl}(t) = -K_{i}^{bl}\bar{x}_{[i]}(t)
\label{eq:SFBL}
\end{equation}

where $K_{i}^{bl} = [K_{i}^{i}, K_{i}^{v}, K_{i}^{\xi}] \in \mathbb{R}^{3}$ is the DeSSf control gain vector. Subsequently, the open-loop model augmented with the integral state $\xi_{[i]}(t)$ becomes third order, hence $\bar{x}_{[i]}(t) = [[x_{[i]}(t)]^T, \xi_{[i]}(t)]^T \in \mathbb{R}^{3}$ is the augmented open-loop state vector. The state-space model of DGU $i$ can now be defined as,

\begin{equation}
 \bar{\Sigma}_{[i]}^{\textrm{DGU}}:
 \begin{cases}
 \dot{\bar{x}}_{[i]}(t) = \bar{A}_{ii}\bar{x}_{[i]}(t)+ \bar{B}_{i}u_{[i]}^{bl}(t) + \bar{E}_{i}\bar{d}_{[i]}(t) + \bar{\zeta}_{[i]}(t) \\
 \bar{y}_{i}(t) = \bar{C}_i\hat{x}_{[i]}(t)
 \end{cases}
 \label{eq:DGUSSCL}
 \end{equation}
 
 where $\bar{d}_{[i]} = [d_{[i]}, V_{ref_{[i]}}]^T \in \mathbb{R}^2$ is the exogenous signal vector, which includes load current disturbance and reference voltage, $\bar{\zeta}_{[i]}(t) = \sum_{j\in\mathcal{N}_i}\bar{A}_{ij}\bar{x}_{[j]}(t)$, and $\bar{y}_{i}(t)$ is the measurable output. The matrices of (\ref{eq:DGUSSCL}) are defined as,
\begin{equation*}
\bar{A}_{ii}
 =
\left[ \begin{array}{cc}
A_{ii} & 0 \\
-C_{i} & 0 
\end{array} \right]
\bar{B}_{i}
 =
\left[ \begin{array}{cc}
B_{i} \\
0 
\end{array} \right]
\bar{E}_{i} = 
\left[ \begin{array}{cc}
E_{i} & 0 \\
0 & 1
\end{array} \right] 
\bar{A}_{ij} =
\left[ \begin{array}{cc}
 A_{ij} & 0 \\
 0 & 0
 \end{array} \right]
 \bar{C}_i = 
 \left[ \begin{array}{ccc}
  0 & 1 & 0\\
  \end{array} \right]
\end{equation*} 
where $\bar{A}_{ii} \in \mathbb{R}^{3\times 3}$, $\bar{B}_{i}\in \mathbb{R}^{3\times 1}$, $\bar{E}_i \in \mathbb{R}^{3\times 3}$,  $\bar{A}_{ij} \in \mathbb{R}^{3\times 3}$ and $\bar{C_i} \in \mathbb{R}^{3}$.

The DeSSf controllers can be tuned via pole placement or using linear quadratic integral (LQI) control. The LQI control technique selects optimal controller gains $[K_{i}^{i}, K_{i}^{v}, K_{i}^{\xi}]$ by weighting the cost of state deviation and control effort using steady-state energy values \cite{Kurucs2015}. In steady-state, the energy stored by the inductor and capacitor equate to,

\begin{equation}
E_{i}^L = \frac{1}{2}L_{t_{i}}I_{t_{i}}^2 \hspace{2mm} ; \hspace{2mm} E_{i}^C = \frac{1}{2}C_{t_{i}}V_{dc_{i}}^2
\end{equation}
 respectively. As the $Q_{i}^{lqi}$ matrix is related to the energy/state deviation cost, the weights were set as,
 \begin{equation}
 Q_{i}^{lqi} = \frac{1}{C_{t_{i}}V_{dc_{i}}^2}\left[\begin{array}{ccc}
 L_{t_{i}}I_{t_{i}}^2 & 0 & 0 \\ 0 & 1 & 0 \\
 0 & 0 & C_{t_{i}}G_{i}
 \end{array} \right]
 \end{equation}
where $G_{i}$ is selected through iterative design.
 \subsubsection{Type III Compensator}
 In \cite{OKeeffe2017a}, each DC-DC boost converter is represented by an ideal duty-cycle to output capacitor voltage transfer function, given as:
 
 \begin{equation}
 \frac{\tilde{v_{i}}(s)}{\tilde{d}_i(s)} = \frac{-V_{in_i}(s-\frac{(1-D_i)R_{Li}}{L_{ti}})}{L_{ti}C_{ti}(s^2+\frac{1} {R_{Li}C_{ti}}s+\frac{(1-D_i)^2}{L_{ti}C_{ti}})}.
 \label{eq:d_vout1}
 \end{equation}
 This is ideal in the sense that it does not include parasitic inductor or capacitor resistance. Each DGU of Fig. \ref{fig:MG2} is modelled with a parasitic inductor resistor $R_{t_{i}}$. The duty cycle to output capacitor voltage transfer function is now represented by,
 
 \begin{equation}
  \frac{\tilde{v_{i}}(s)}{\tilde{d}_i(s)} =
  \frac{\frac{-V_{in_{i}}}{(1-D_i)^2}(s+(R_{t_{i}}-(1-D_i)^2R_{L_{i}}))}{s^2 +(\frac{1}{R_{L_{i}}C_{t_{i}}}+\frac{R_{t_{i}}}{L_{t_{i}}})s+(\frac{(1-D_i)^2}{L_{t_{i}}C_{t_{i}}}+\frac{R_{t_{i}}}{R_{L_{i}}L_{t_{i}}C_{t_{i}}})}
  \label{eq:d_vout2}
  \end{equation}
  
 \textbf{Remark 3}: \textit{Note that these transfer functions are dependent on knowledge of the load, and therefore the load influences the operating-point and hence stability. In this case, the load is represented as a linear resistance.}
 \vspace{3mm} \newline
 Equations (\ref{eq:d_vout1}) and (\ref{eq:d_vout2}) highlight the non-minimum phase property of boost converters, which manifests as a right-half plane (RHP) zero in both transfer functions. As a result, due to the discontinuous energy transfer between inductor and capacitor during switching, the output voltage initially undershoots subsequent to any disturbance or reference change. If the bandwidth of the controllers is very fast, then the RHP zero can attract stable poles into the RHP and destabilise the DGU. Type II and III compensators are commonly used in power converter control where phase injection is required to compensate the phase lag introduced by resonant poles and RHP zeros. As the parasitic inductor resistance introduces some damping, type II compensators can generally be used, though type III compensators will inject more phase margin. The transfer function of a type III compensator is,
 
 \begin{equation}
 \mathcal{C}_{[i]} : C_{v_{i}}(s) = \frac{k_{c_i}}{s}\frac{(s+\omega_{z_i})^2}{(s+\omega_{p_i})^2}.
 \label{eq: TypeIII}
 \end{equation}
 
 \textbf{Remark 4:} \textit{Though the closed-loop state space model of DGU $i$ in (\ref{eq:DGUSSCL}) includes the coupling term $\hat{\zeta}_{[i]}(t)$, decentralised controllers are designed without accounting for this term. Equally, in the case of the classical compensators, the transfer functions of (\ref{eq:d_vout1}) and (\ref{eq:d_vout2}) do not include coupling parameters. The affect on stability by not accounting for DGU interaction is explored in section \ref{Instability}.}
 \vspace{3mm} \newline
 \section{Decentralised $\mathcal{L}_1$ Adaptive Control Augmentation}\label{DeL1AC}
\begin{figure}[!htb]    
 \centering
 \includegraphics[width=8.5cm]{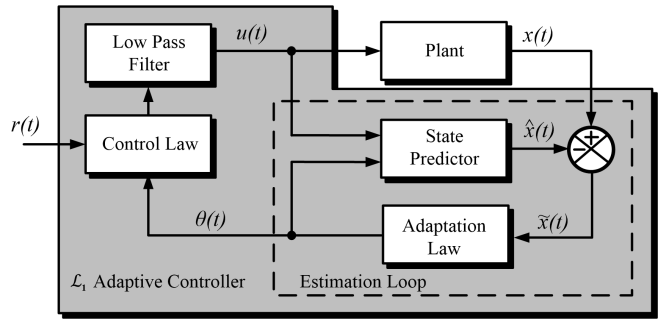}
 \caption{General Architecture of $\mathcal{L}_1$ Adaptive Controller. Adapted from \cite{L12010}}.
 \label{fig:L1Arch}
 \end{figure}

 Conventional MRAC architectures frequently suffer from a trade-off between estimation and robustness \cite{Anderson2005}. Fast estimation/adaptation requires large adaptive gains which can destabilise control-loops. The $\mathcal{L}_1$AC, a modification of the indirect MRAC architecture, decouples this trade-off by inserting a low-pass filter (LPF) at the input to both the plant and state-predictor, as seen in Fig. \ref{fig:L1Arch}. Consequently, robustness instead depends on the choice of filter-bandwidth, thus enabling fast adaptation \cite{L12010}.
 
 Application of the $\mathcal{L}_1$AC has been successful in various safety-critical applications; notably sub-scale NASA aircraft auto-pilots \cite{Gregory2010}, and unmanned water/aerial vehicles \cite{Svendsen2012,Michini2009}. These applications use centralised $\mathcal{L}_1$AC approaches. Recently, a decentralised $\mathcal{L}_1$AC approach has been used to augment aircraft baseline controllers\cite{Kumaresan2016}. $\mathcal{L}_1$AC has also attracted the interest of StatOil Norway for application in managed pressure drilling and rotary steerable systems \cite{Mahdianfar2016}.

 Ultimately, the $\mathcal{L}_1$AC architecture has potential to improve mG voltage control, with uniform performance across an entire operating range which experiences large uncertainties being the key feature.\subsection{$\mathcal{L}_1$ Adaptive Control Architecture}\label{L1ACArch}
 From Fig. \ref{fig:L1Arch} a state-predictor replaces the reference model of the indirect MRAC, and a LPF limits the control signal bandwidth. The state-error dynamics, $\tilde x(t)$, between the plant and state-predictor drives the projection-based adaptation law. This adjusts the control parameters in order to drive $\tilde x(t) \rightarrow 0$.
 
 \subsubsection{Plant structure}
 
 The plant has a known structure, but with unknown parameter values. A matched uncertainty term is introduced to represent parametric uncertainty in the dynamics of $\bar{\Sigma}_{[i]}^{\textrm{DGU}}$, hence (\ref{eq:DGUSSCL}) can be represented as,
 \begin{equation}
  \hat{\Sigma}_{[i]}^{\textrm{DGU}}:
  \begin{cases}
   \dot{\bar{x}}_{[i]}(t) = \hat{A}_{m}\bar{x}_{[i]}(t)+ \bar{B}_{i}(u_{[i]}(t) +\bar{\theta}_{[i]}^T(t)\bar{x}_{[i]}(t))+ F\bar{E}_i\bar{d}_{[i]}(t) \\ 
  \bar{y}_{[i]}(t) = \bar{C}_i\bar{x}_{[i]}(t)
  \end{cases}
  \label{eq:L1SS}
  \end{equation}
 
 where $\bar{x}_{[i]}(t) \in \mathbb{R}^{3}$, is the system measurable state vector; $\hat{A}_{m} \in \mathbb{R}^{3\times 3}$ is the Hurwitz design matrix that specifies the desired closed-loop dynamics; $u(t) \in  \mathbb{R}$ is the control signal; $F = [0, 0, 1]$, and $\bar{\theta}(t)$ is the unknown matched parametric uncertainty vector. This belongs to a known compact convex set of uniform boundedness $\theta \in \Theta \subset \mathbb{R}^3$.
 
 \subsubsection{Control Law}
 The small-signal control input $u(t)$ for $\hat{\Sigma}_{[i]}^{\textrm{DGU}}$ consists of the summation between the baseline and $\mathcal{L}_1$AC control signals, 
 \begin{equation}
 \mathcal{C}_{[i]}^{\mathcal{L}_1} : u_{[i]}(t) = \tilde{d}_{[i]}(t) = u_{[i]}^{bl}(t) + u_{[i]}^{\mathcal{L}_1}(t)
 \end{equation}
 The augmenting $\mathcal{L}_1$AC law, fitted with a first-order LPF, is 
 \begin{equation}
 u_{[i]}^{\mathcal{L}_1}(t) = -C(s)[\hat{\theta}_{[i]}^T\bar{x}_{[i]}](t)
 \label{eqn:C_L1}
 \end{equation}
 where $C(s)=\frac{\omega_c}{s+\omega_c}$, and $\hat{\theta}_{[i]}^T(t)$ is the parametric estimation vector, as defined in section \ref{Adaptive}. The robustness of the $\mathcal{L}_1$AC is dependent on the LPF bandwidth $\omega_c$, as subsequently designed.
 
 \subsubsection{State-predictor}
 The state-predictor generates an estimate of the system states. From the perspective of the $\mathcal{L}_1$AC, the baseline dynamics are combined with the open-loop DGU dynamics to form an augmented closed-loop system.

   \textbf{Assumption 2:} \textit{The design of the decentralised $\mathcal{L}_1$ adaptive voltage controllers can neglect the exogenous disturbance term $\bar{d}_{[i]}$ and coupling term $\bar{\zeta}_{[i]}$.}
   
    Without loss of generality, the state-predictor formulation is proposed for all DGUs as, 
 \begin{equation}
  \mathcal{E}_{[i]}:
  \begin{cases}
  \dot{\hat{x}}_{[i]}(t) = \hat{A}_{m}\hat{x}_{[i]}(t)+\hat{B}(u_{[i]}^{\mathcal{L}_1}(t)+\hat{\theta}_{[i]}^T(t)\bar{x}_{[i]}(t))+ F\hat{E}_i\hat{d}_{[i]}(t) \\
  \hat{y}_{[i]}(t) = \hat{C}_{[i]}\hat{x}_{[i]}(t)
  \end{cases}
  \label{eq:L1SPSS2}
  \end{equation}
 where $\hat{x}_{[i]}(t) \in \mathbb{R}^3$ is the predicted state vector and $\hat{\theta} \in \mathbb{R}^{3}$ is the parametric estimation vector.
 \vspace{3mm} \newline
 \textbf{Remark 5.}\textit{
 $\mathcal{L}_1$AC theory of \cite{L12010} assumes that the input matrix $\hat{B}_i$ is known. However, the $\hat{B}_i$ matrix of (\ref{eq:L1SPSS2}) consists of unknown parameters which cannot be compensated by the adaptive control law. To overcome this, the state-predictor can be transformed into its control canonical form so that a known $\hat{B}$ is attained. As a result, these unknown parameters are transferred to the output matrix $\hat{C}$, which is not required for state-feedback control.}
 \vspace{1mm}\newline
Transforming to control-canonical form, the closed-loop transfer function  from control-input to voltage output is,
 \begin{equation}
 \frac{Y(s)}{U(s)} = \hat{C}(s\mathbb{I} - \hat{A}_m)^{-1}\hat{B}_i
 \end{equation}
 where,
 \begin{equation}
  \hat{A}_{m}=
  \left[\begin{array}{cc}
  A_{m}-\hat{B}_{i}K_{i}^x & \hat{B}_{i}K_{i}^\xi \\
  -\hat{C}_i & 0 
  \end{array} \right]
  \end{equation}
  and $K_{i}^x = [K_{i}^i, K_{i}^v]$. 
  \begin{equation}
   \frac{Y(s)}{U(s)} = \frac{f_{r}s^{r}+f_{r-1}s^{r-1}+...+f_{0}}{s^N + e_{N-1}s^{N-1}+...+e_0}
   \end{equation}
   where,
   \begin{equation}
   \hat{A}_{m}^{CC} = \left[\begin{array}{ccc}
     0 & 1 & 0 \\
     0 & 0 & 1 \\
     -e_0 & -e_1 & -e_2 \\
     \end{array} \right]
     b = \left[\begin{array}{ccc}
          0 \\
          0 \\
          1 \\
          \end{array} \right]
          C_{cc} =  \left[\begin{array}{ccc}
                    f_0 & f_1 & f_2 \\
                    \end{array} \right]
   \end{equation}
   and,
   \begin{equation}
   \begin{aligned}
   e_2 = \frac{1}{{C_{ti}}}(\sum_{j\in\mathcal{N}_i}\frac{1}{R_{ij}} - I_{ti}K_i^v) + \frac{1}{L_{ti}}(R_{ti}+V_{dc_{i}}K_i^i) \\
   e_1 = \frac{1}{L_{ti}C_{ti}}((R_{ti} + V_{dc_{i}}K_i^i)(\sum_{j\in\mathcal{N}_i}\frac{1}{R_{ij}} - I_{ti}K_i^v)+((1-D_i)+V_{dc_{i}}K_i^v)((1-D_i)+I_{ti}(K_i^i - K_i^\xi)))\\
   e_0 = \frac{1}{L_{ti}C_{ti}}(I_{ti}K_i^\xi(R_{ti} + V_{dc_{i}}K_i^i)-V_{dc_{i}}K_i^\xi((1-D_i)+I_{ti}K_i^i))
   \end{aligned}
   \end{equation}
    With this, the closed-loop state-predictor in control canonical form is given as,
      \begin{equation}
      \mathcal{E}_{[i]}^{CC}:
        \begin{cases}
      \dot{\hat{z}}_{[i]}(t) = \hat{A}_m^{CC}\hat{z}_{[i]}(t)+b( u_{[i]}^{\mathcal{L}_1}(t)+ \hat{\theta}_{[t]}^T(t)\bar{z}_{[i]}(t))\\
      \hat{y}_{cc_{[i]}}(t) = C_{cc}\hat{z}_{[i]}(t) 
      \end{cases}
      \label{SPCC}
      \end{equation}
 As (\ref{SPCC}) is dependent on the control canonical form of the plant $\bar{z}_{[i]}(t)$, a transformation from the measured state vector $\bar{x}_{[i]}(t)$ to the new state vector is required. Therefore,
 \begin{equation}
 \bar{z}_{[i]}(t) = T_{i}\bar{x}_{[i]}(t)
 \end{equation}
From state transformation theory, and since controllability has already been assumed, the transformation matrix is computed as,
\begin{equation}
T_{i} = C_{\hat{z}}(C_{\bar{x}})^{-1}
 \label{eqn:Tran}
\end{equation}
where, $C_{\hat{z}} = [b, \hat{A}_m^{CC}b, \hat{A}_m^{CC^2}b]$ and $C_{\bar{x}} = [\bar{B}_i, \hat{A}_m\bar{B}_i , \hat{A}_m^2\bar{B}_i]$ are the controllability matrices associated with the state-predictor in control canonical form and plant. 
\vspace{3mm} \newline
\textbf{Assumption 3.} \textit{The computation of (\ref{eqn:Tran}) requires knowledge of the plant input matrix $\bar{B}_i$, which from Remark 5, is uncertain. However, it is assumed that the adaptation will account for this. Therefore, approximate/nominal parametric values are chosen for $\bar{B}_i$ - see section 3.1.6.}

\subsubsection{Adaptive Law}\label{Adaptive} 
 From hereafter, control-canonical form notation is used. The adaptive law generates an estimate of the plant uncertainties. Defining the state-error and parametric estimation error vectors as, $\tilde{{z}}_{[i]}(t) = \bar{z}_{[i]}(t) - \hat{z}_{[i]}(t)$ and $\tilde{{\theta}}_{[i]}(t) = \theta_{[i]}(t) - \hat{\theta}_{[i]}(t)$, the state-error dynamics, used to drive the adaptive law, can be defined as,
 \begin{equation}
 \dot{\tilde{z}}_{[i]}(t) = \hat{A}_{m}\tilde{z}_{[i]}(t)+b\tilde{\theta}_{[i]}(t)\hat{z}_{[i]}(t)
 \label{eqn:ErrorDyn}
 \end{equation}
 The adaptive law is determined from Lyapunov's second stability method. A quadratic Lyapunov candidate is defined as a function in terms of $\tilde{{z}}_{[i]}(t)$ and $\tilde{{\theta}}_{[i]}(t)$.
 \begin{equation}
 \mathcal{V}_{[i]}(\tilde{z}_{[i]}(t), \tilde{\theta}_{[i]}(t)) = \tilde{z}_{[i]}(t)^TP_{i}\tilde{z}_{[i]}(t) + \tilde{\theta}_{[i]}(t)^T\Gamma_{i}^{-1}\tilde{\theta}_{[i]}(t)
 \label{eqn:LyapVi}
 \end{equation}
 where, $P_{i} \in \mathbb{R}^{3\times 3}$ is a symmetric matrix, such that $P_{i} = P_{i}^T  > 0$ is the solution to the algebraic Lyapunov linear inequality $\hat{A}_m^TP_{i} + P_{i}\hat{A}_m \leq -Q_{i}$, for arbitrary $Q_{i} = Q_{i}^T > 0$, and $\Gamma_{i} \in \mathbb{R}^+$ is the adaptive gain. From \cite{Slotine1991}, if the time-derivative of (\ref{eqn:LyapVi}) is at least negative semi-definite, then each subsystem, in this case each DGU, is locally stable since the energy along the trajectories of state and estimation errors is decreasing. The time-derivative of (\ref{eqn:LyapVi}) is, 
 
 \begin{equation}
 \mathcal{\dot{V}}_{[i]}(\tilde{z}_{[i]}(t), \tilde{\theta}_{[i]}(t))  = \frac{d\mathcal{V}_{[i]}(t)}{d\tilde{x}_{[i]}(t)}\tilde{z}_{[i]}(t) + \frac{d\mathcal{V}_{[i]}(t)}{d\theta_{[i]}(t)}\dot{\theta}_{[i]}(t) =  \dot{\tilde{z}}_{[i]}^T(t)P_{i}\tilde{z}_{[i]}(t)+\tilde{z}_{[i]}^T(t)P_{i}\dot{\tilde{z}}_{[i]}(t) +2\tilde{\theta}_{[i]}^T(t)\Gamma_{i}^{-1}\dot{\tilde{\theta}}_{[i]}(t)
 \label{eqn:dotVi}
 \end{equation}
  Using (\ref{eqn:ErrorDyn}),(\ref{eqn:dotVi}) can be written as,
  
  \begin{equation}
  \mathcal{\dot{V}}_{[i]}(\tilde{z}_{[i]}(t), \tilde{\theta}_{[i]}(t))  = 2(\hat{A}_{m}\tilde{z}_{[i]}(t)+b\tilde{\theta}_{[i]}(t)\bar{z}_{[i]}(t))P_{[i]}\tilde{z}_{[i]} +  2\tilde{\theta}_{[i]}^T\Gamma_{i}^{-1}\dot{\tilde{\theta}}_{[i]}
  \end{equation}
  From the algebraic Lyapunov linear inequality equation, $\hat{A}_{m}^TP_{i} + P_{i}\hat{A}_{m} = 2\hat{A}_{m}P_{i} = - Q_{i}$. Also, since $\dot{\tilde{\theta}}_{[i]} = \dot{\theta}_{[i]} - \dot{\hat{\theta}}_{[i]}$, and $\dot{\theta}_{[i]} = 0$,
  
  \begin{equation}
  \mathcal{\dot{V}}_{[i]}(\tilde{z}_{[i]}(t), \tilde{\theta}_{[i]}(t))  = -\tilde{z}_{[i]}(t)Q_{i}\tilde{z}_{[i]}(t)+ 2\tilde{\theta}_{[i]}(t)(\tilde{z}_{[i]}(t)^TP_{i}b\bar{z}_{[i]}(t)+\dot{\hat{\theta}}_{[i]}(t)\Gamma_{i}^{-1})
  \label{eqn:dotVi2}
  \end{equation}
To ensure (\ref{eqn:dotVi2}) is  at least negative semi-definite, i.e. $2\tilde{\theta}_{[i]}(\tilde{z}_{[i]}^T(t)P_{i}b\bar{z}_{[i]}(t+\dot{\hat{\theta}}_{[i]}(t)\Gamma_{i}^{-1}) = 0$,
  The adaptive law is given by,
  \begin{equation}
  \dot{\hat{\theta}}_{[i]} = -\Gamma_{i}\tilde{x}_{[i]}(t)^TP_{i}b\bar{x}_{[i]}(t)
  \label{eqn:adaptiveLaw1}
  \end{equation}
  
  To prevent parameter drift, the parametric uncertainty estimate is bounded using the projection operator, as detailed in \cite{Yoo2010,L12010,Lavretsky2011}. Therefore, (\ref{eqn:adaptiveLaw1}) becomes, 
  
  \begin{equation}
  \dot{\hat{\theta}}_{[i]}(t) = \Gamma_{i}Proj(\hat{\theta}_{[i]}(t), -\tilde{z}_{[i]}^T(t)P_{i}b\bar{z}_{[i]}(t))
  \label{eqn:adaptiveLaw2}
  \end{equation}
   Finally,
  
  \begin{equation}
  \mathcal{\dot{V}}_{[i]}(\tilde{z}_{[i]}(t), \tilde{\theta}_{[i]}(t))  = -\tilde{z}_{[i]}(t)Q_{i}\tilde{z}_{[i]}(t) \leq 0
  \label{eqn:dotVi3}
  \end{equation}
  
  Hence, the equilibrium of the state error dynamics of (\ref{eqn:ErrorDyn})  and adaptive law of (\ref{eqn:adaptiveLaw2}) is locally stable i.e. $\tilde{z}_{[i]}(t)$ and $\tilde{\theta}_{[i]}(t)$ are bounded. Since $\bar{z}_{[i]}(t) = \hat{z}_{[i]}(t) -  \tilde{z}_{[i]}(t)$, and as the state estimate vector $\hat{z}_{[i]}(t)$ results from a stable design, the plant states $\bar{z}_{[i]}(t)$ are also bounded. Ideally, the plant dynamics are driven to equal the desired predictor dynamics. This warrants convergence of $\tilde{z}_{[i]}(t) \rightarrow 0$.  However (\ref{eqn:dotVi3}) does not prove local asymptotic stability. To show that the state-prediction error converges asymptotically to zero, the second-derivative of $\mathcal{V}_{[i]}(t)$ is computed,
  
  \begin{equation}
  \mathcal{\ddot{V}}_{[i]}(\tilde{z}_{[i]}(t), \tilde{\theta}_{[i]}(t))  = -2\tilde{z}_{[i]}(t)^TQ_{i}\dot{\tilde{z}}_{[i]}(t)
  \label{eqn:Vddot}
  \end{equation}
  From (\ref{eqn:ErrorDyn}), $\dot{\tilde{z}}_{[i]}(t)$ is uniformly bounded by design, and thus (\ref{eqn:Vddot}) is bounded. A bounded second-derivative implies a smooth first-derivative, resulting in a uniformly continuous $\mathcal{\dot{V}}_{[i]}$. By invoking Barbalat's lemma in section A.6.1 of \cite{L12010}, it follows that $\lim\limits_{t \rightarrow \infty} \tilde{z}_{[i]}(t) = 0$. Subsequently local asymptotic stability can be guaranteed. 

 \subsubsection{Filter Design}\label{LPF Design}
\begin{figure}[!htb]    
 \centering
 \includegraphics[width=14cm]{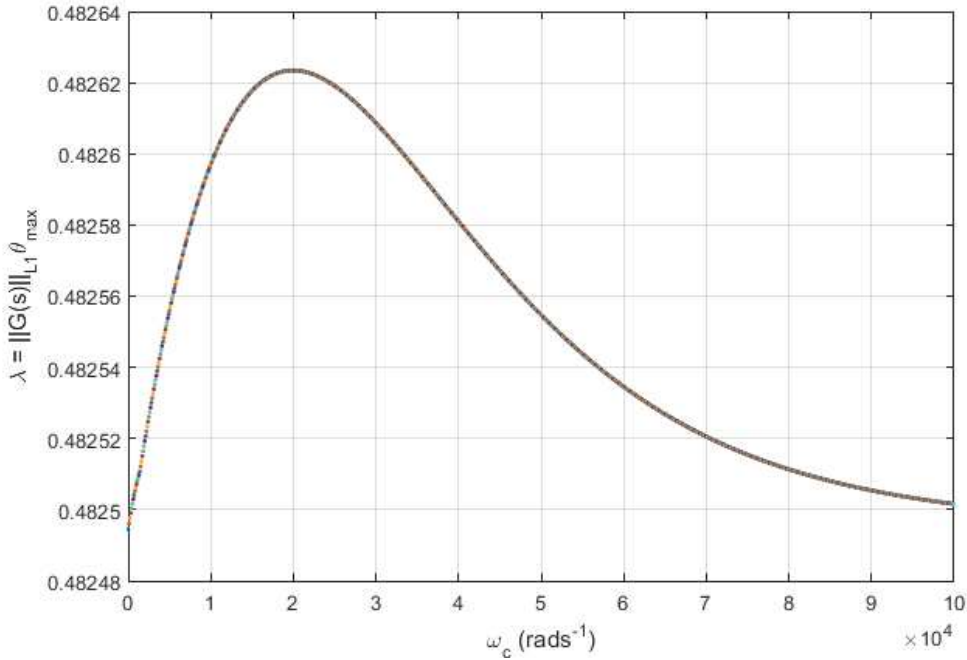}
 \caption{Product of the uncertainty bound and $\mathcal{L}_1$ norm of the high-pass filtered closed-loop system $G(s).$}
 \label{fig:L1norm}
 \end{figure}

 The key feature of the $\mathcal{L}_1$AC is the synthesis of a LPF structure which decouples robustness from adaptation. At this point, local asymptotic stability has been guaranteed during nominal operation i.e. baseline controller design, and adaptation. Here, stability is further guaranteed when the LPF is inserted to filter the control signal. The LPF bandwidth is tuned using the $\mathcal{L}_1$ norm condition. From the perspective of the $\mathcal{L}_1$AC, the baseline controller dynamics are combined with the open-loop DGU dynamics. Therefore, the desired closed-loop reference system in the Laplace domain, where $\hat{A}_{m}$ includes the dynamics of the baseline controller, can be given as, 
 \begin{equation}
 \begin{aligned}
 \bar{z}_{ref_{[i]}}(s) =  (s\mathbb{I}-\hat{A}_{m})^{-1}b(u_{[i]}^{\mathcal{L}_1}(s)+\bar{\theta}_{[i]}^T\bar{z}_{ref_{[i]}}(s))+\bar{z}_{ic_{[i]}}(s)
 \label{eqn:clref1}
 \end{aligned}
 \end{equation}
 where, $\bar{z}_{ref_{[i]}} \in \mathbb{R}^3$, is the reference state vector, $\bar{z}_{ic{[i]}} \in \mathbb{R}^3$ is the initial state vector, $\mathbb{I} \in \mathbb{R}^{3\times 3} $, is the identity matrix. $\hat{A}_{m}$ is defined in section \ref{DesignConsiderations}. Convergence is assumed, i.e. $\hat{\theta}_{[i]}(t) \rightarrow \bar{\theta}_{[i]}$. 
 The desired closed-loop behaviour is represented by the transfer function,
 \begin{equation}
 H(s) = (s\mathbb{I}-\hat{A}_{m})^{-1}
 \end{equation}  
  Combining 
 (\ref{eqn:C_L1}) with (\ref{eqn:clref1}) yields,
 \begin{equation}
 \begin{aligned}
 \bar{z}_{ref_{[i]}}(s)
 = G(s)\bar{\theta}_{[i]}^Tz_{ref_{[i]}}(s)+\bar{z}_{0}(s)
 \label{eqn:clref3}
 \end{aligned}
 \end{equation}
 where, $G(s) = H(s)(1-C(s))$, and $\bar{z}_{0}(s) = (s\mathbb{I}-\hat{A}_{m})^{-1}\bar{z}_{ic_{[i]}}(s)$ 
 As shown in \cite{L12010}, the $\mathcal{L}_1$-norm is now taken on both sides of (\ref{eqn:clref1}),
 \begin{equation}
 ||z_{ref_{[i]}}||_{\mathcal{L}_1} = \frac{||z_0||_{\mathcal{L}_\infty}}{1-||G(s)\theta^T||_{\mathcal{L}_1}} 
 \label{eqn:L1norm}
 \end{equation}
 For the reference states to be bounded, the denominator must be larger than zero. The 1-norm is chosen as the maximum value of $\theta$,
 \begin{equation}
 \theta_{max} = 4\max_{\theta \in \Theta}||\theta||_1
 \label{eqn:ThetaMax}
 \end{equation} 
 $\theta_{max}$ represents the boundary of projection for estimating the parameters when using the adaptation law (\ref{eqn:adaptiveLaw2}). Finally, for the reference states to remain bounded, the following $\mathcal{L}_1$-norm condition must be satisfied,
 \begin{equation}
 \lambda = ||G(s)||_{\mathcal{L}_1}\theta_{max} < 1
 \label{eqn:L1normCond}
 \end{equation} 
 where the degree-of-freedom is $\omega_c$. Inserting the LPF attenuates any HF content in the control channel resulting from large adaptive gains, and compensates LF uncertainty. Fig. \ref{fig:L1norm} shows the result of (\ref{eqn:L1normCond}) when using the parameters in Table  \ref{table:parameters}.
 
 From this, an arbitrary value for $\omega_c$ can be chosen. Choosing an overly large $\omega_c$ may result in high gain feedback which can reduce robustness margins. However, part of the reason why an arbitrary bandwidth can be used is due to large closed-loop gain and phase margins i.e. transient and steady-state gain of closed-loop system are very small (which also allows for the handling of large uncertainty).
 
 The overall decentralised voltage primary control scheme is shown below,
 \begin{figure}[!htb]    
 \centering
 \includegraphics[width=12.5cm]{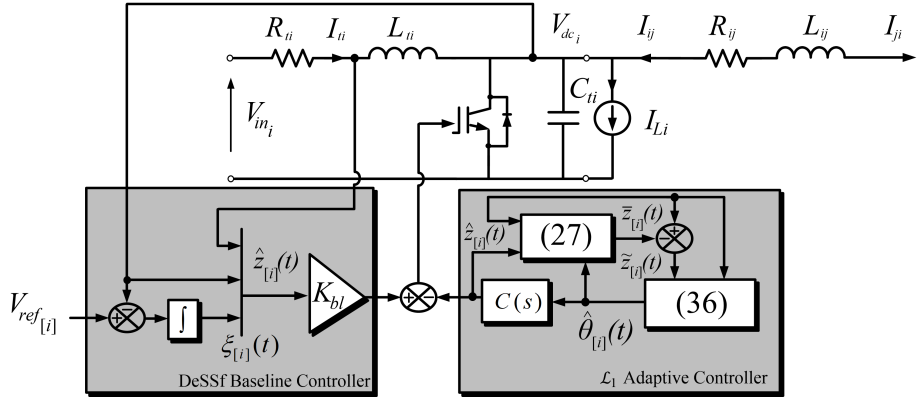}
 \caption{Overall Control Architecture of $ \hat{\Sigma}_{[i]}^{\textrm{DGU}}$.}
 \label{fig:OverallControlArch}
 \end{figure}
 \subsubsection{Design Considerations}\label{DesignConsiderations}
 As opposed to the baseline controller, which is nominally designed for decoupled operation, the desired closed-loop dynamics of the state-predictor are designed for nominal operation of the DGUs when coupled to neighbouring DGUs, and is applied to all DGUs. \textit{A priori} knowledge of the real-time number of neighbouring DGUs is not known, therefore the maximum possible number of couplings, within the set $\mathcal{D}$, is chosen, as in (\ref{eqn:Aii}). The desired closed-loop dynamics in, are selected as (in normal form, i.e. not control-canonical form),
 \begin{equation}
 \begin{aligned}
 \bar{A}_{m}=
 \left[\begin{array}{cc}
 \hat{A}_m-bK_{iv}^{nom} & BK_{\xi}^{nom} \\
 -C_{nom} & 0 
 \end{array} \right]
 \bar{B} = \left[ \begin{array}{c}
 B\\
 0
 \end{array} \right]
 \hat{A}_m =
 \left[ \begin{array}{cc}
 -\frac{R_{t}^{nom}}{L_{t}^{nom}} & -\frac{(1-D_{i}^{nom})}{L_{t}^{nom}}\\
 \frac{(1-D_{i}^{nom})}{C_{t}^{nom}} & -\sum_{j\in\mathcal{M}_i}\frac{1}{R_{ij}^{nom}C_{t}^{nom}}
 \end{array} \right]\\
 B = \left[ \begin{array}{c}
 \frac{\bar{V}_{dc}^{nom}}{L_{t}^{nom}}\\
 \frac{-\bar{I}_{t}^{nom}}{C_{t}^{nom}}
 \end{array} \right]
 \hspace{3mm} ;
 \hspace{3mm}
 K_{iv}^{nom^T} = \left[ \begin{array}{c}
 K_{i}^{nom}\\
 K_{v}^{nom}
 \end{array} \right] 
 \label{eqn:AdesBdes1}
 \end{aligned}
 \end{equation}
 The nominal parameters represent an estimate of where the uncertain dynamics lie within the polytope. While baseline DeSSf controllers are designed using the pole-placement method (baseline controllers are designed to place poles for fast closed-loop performance), within the predictor the estimate of $K_{iv}^{nom}$ is designed using the LQI method. It is worth noting that, by considering the load as an exogenous disturbance, the eigenvalues are dependent only on the QSLs. As the number of couplings increases, the eigenvalues of $\hat{A}_m$ become faster, increasing the closed-loop bandwidth.
 The parameters of $\bar{B}$ are chosen based on an approximate expectation of the steady-state output voltage $\bar{V}_{dc}^{nom}$ and inductor current $\bar{I}_{t}^{nom}$. This expectation is suitable for $\bar{V}_{dc}^{nom}$ as output voltage tracking is the primary control objective. However, from (\ref{eq:DGUSS2}), $\bar{I}_{ti}$ is load dependent. As the real-time load and effective QSL resistance are unknown, $\bar{I}_{t}^{nom}$ is conservatively designed for the smallest expected load power and maximum number of couplings. 
 
 The parameter bounds, or the maximal deviation from the desired dynamics is calculated as,
 \begin{equation}
 \begin{aligned}
 \bar{A}_{m} = A-B_i\theta^T 
 \hspace{3mm} ;
 \hspace{3mm}
 A = \left[ \begin{array}{cc}
 A_{ii}-B_{i}K_{i}^{iv} & B_{i}K_{i}^\xi \\ -C_{i} & 0 
 \end{array} \right]
 \end{aligned}
 \end{equation}
 where the DeSSf control gain vector $[K_{i}^{iv}, K_{i}^{\xi}] \in \mathbb{R}^3$ is calculated for a wide range of different parameters in $A_{ii}$ and $B_i$. The parameters that bring the closed-loop dynamics $A$ to the edge of the boundary of uncertainty represents the maximal deviation from the desired dynamics. Consequently, the parameter bound $\theta_{max}$ is calculated according to (\ref{eqn:ThetaMax}).
 
 Ultimately, $K_x^{nom}$ is calculated using the LQR method, such as in \cite{Kurucs2015}.
 \subsubsection{Conditions for Global Asymptotic Stability}
 Despite decentralised controllers guaranteeing local asymptotic stability, as shown in section \ref{Instability}, global asymptotic stability can be violated due to the presence of unaccounted DGU coupling. Conventional decentralised control theory \cite{Lunze, Bakule1988a} treats coupling terms as disturbances and suggests that controller design should be robust to neighbouring states in order to achieve global asymptotic stability. We demonstrated this in \cite{OKeeffe2017a}, where type III compensators were detuned to provide a wider performance range when coupled to neighbouring DGUs. However, global knowledge of coupling parameters, load dynamics, and real-time information about the number of neighbours are required \textit{a priori}, i.e. not scalable.
 
 This section aims to provide offline conditions for guaranteeing global asymptotic stability using decentralised controllers $\mathcal{C}_{[i]}^{\mathcal{L}_1}$. Most decentralised controllers are based on the idea of small-couplings or weakly coupled subsystems, where coupling terms are small and within some known bounded subset. This introduces degrees of conservativity. 
 
 The conventional connective stability method, described in \cite{Lunze, Bakule1988a}, can be used to construct sufficient global stability conditions using aggregated interconnection models and exploiting Lyapunov functions. However, we showed in \cite{OKeeffe2018a} that conditions are only satisfied when the small-gain theorem is satisfied, and demonstrated that interconnections typically have large-gains in DC ImGs. This explains why the decentralised PnP controllers of \cite{Tucci2016c,Sadabadi2017a} require $P_i$ in the form, 
 \begin{equation}
 P_i = 
 \left[
 \begin{array}{c|cc}
   \eta_i & 0 & 0\\
   \hline
   0 & \ast & \ast  \\
    0 & \ast & \ast
 \end{array}
 \right]
 \label{eqn:PiLMI}
 \end{equation}
 where $\eta_i$ is a local design parameter and $\ast$ denotes arbitrary values, in order to neutralise such interactions between DGUs i.e. term $(b)$ in {\ref{eq:OverallVdot2}} equals zero. Furthermore, we described a distributed architecture, using robust-adaptive controllers that provides sufficient conditions for global asymptotic stability in the presence of large-gain interconnections. However, $\mathcal{C}_{[i]}^{\mathcal{L}_1}$ requires information about $\mathcal{C}_{[j]}^{\mathcal{L}_1}$. 
 
 From the perspective of decentralised, communication-less control, global asymptotic stability must instead be determined by deriving global stability conditions offline and incorporating these into the design. The overall Lyapunov function candidate that describes the global system can be written as,
\begin{equation}
 \mathcal{V}(t) = \sum\limits_{i=0}^{N} \left(\tilde{z}_{[i]}^T(t)P_{i}\tilde{z}_{[i]}(t) + \tilde{\theta}_{[i]}^T(t)\Gamma_{i}^{-1}\tilde{\theta}_{[i]}(t)\right)
 \label{eqn:OverallLyap}
 \end{equation}
 \textbf{Assumption 4:} \textit{We assume local controllers exploit (\ref{eqn:adaptiveLaw2}), and  plant dynamics have converged to desired dynamics.}
 \vspace{3mm} \newline
The derivative of (\ref{eqn:OverallLyap}) is,
 \begin{equation}
  \dot{\mathcal{V}} = -\sum\limits_{i=0}^{N}\tilde{z}_{[i]}^T(t)Q_{i}\tilde{z}_{[i]}(t)
  \label{eq:OverallVdot}
 \end{equation}
 if and only if matrix $\mathbf{P}$ satisfies the Lyapunov inequality equation,
 inequality equation,
 \begin{equation} \underbrace{\hat{\textbf{A}}_{\textbf{m}}^T\textbf{P}+\textbf{P}\hat{\textbf{A}}_{\textbf{m}}}_{(\textrm{a})}+\underbrace{\hat{\textbf{A}}_{\textbf{C}}^T\textbf{P}+\textbf{P}\hat{\textbf{A}}_{\textbf{C}}}_{(\textrm{b})} < 0
  \label{eq:OverallVdot2}
 \end{equation} 
 where $\textbf{P} = $ diag$({P}_{i}) \in \mathbb{R}^{3M\times 3M}$; $\hat{\textbf{A}}_{\textbf{m}} =$ diag$(\hat{A}_{m}) \in \mathbb{R}^{3M\times3M}$ represents the overall desired dynamics; $\hat{\textbf{A}}_{\textbf{C}} = \hat{\textbf{A}} - \hat{\textbf{A}}_{\textbf{m}} \in \mathbb{R}^{3M\times3M}$ represents the coupling dynamics only. As each DGU is designed to be locally asymptotically stable, the matrices of (a) are negative definite. Therefore, for global asymptotic stability, the matrices of (b) need to be negative definite. The use of LMIs in \cite{Tucci2016c,Sadabadi2017a} systematically ensures (b) $<$ 0 through design of $\textbf{K}$. However, here the design of $\textbf{K}$ is performed iteratively offline to ensure $||(a)||$ $>$ $||(b)||$. This typically results in detuned controller gains, as expected due to the conservativeness requirements of decentralised systems. Furthermore, this method can suffer when system size expands as the retuning of $\textbf{K}$ becomes more difficult, despite the advantage of designing the desired dynamics as the same for all DGUs. 
 
 \section{Review of $\mathcal{L}_1$ Adaptive Control Criticisms}
 This section reviews published criticisms of $\mathcal{L}_1$AC theory for the purpose of informing the reviewers of our paper submitted to the European Control Conference. The information provided is based on original insights, and section A.4 of \cite{Altn2016}, which also attempted to review these criticisms. 
 
Recently, there has been an on-going debate concerning the relationship between standard MRAC and $\mathcal{L}_1$AC architectures; in particular the robustness margins associated with inserting the LPF. In 2012, the authors of \cite{Ioannou2014} submitted a version of this journal paper to IEEE Transactions on Automatic Control with 'unsubstantiated and wrong claims' concerning $\mathcal{L}_1$AC theory. The inventors of the theory, Professor's Naira Hovakimyan and Chengyu Cao of University of Illinois and University of Connecticut respectively, were invited to review \cite{Ioannou2014} but instead published their own technical report \cite{Hovakimyan2006} rebuking the claims of \cite{Ioannou2014}. In this report, the main properties of $\mathcal{L}_1$AC and the key differences between conventional MRAC and $\mathcal{L}_1$AC architectures, originally described in \cite{L12010}, are again highlighted. Furthermore, the report demonstrates a correct implementation of the example used in \cite{Ioannou2014}. Eventually, \cite{Ioannou2014} was published in 2014 after three revisions.

Some of the misinterpretations made by \cite{Ioannou2014} include;
\begin{itemize}
  \item Mistaking the $\mathcal{L}_1$AC architecture as an input-filtered direct MRAC.
  \item Derivation of unbounded performance bounds.
  \item Incorrectly suggesting that the introduction of the LPF reduces robustness to unmodelled dynamics.
  \item Suggestion that stiff differential equations due to large adaptive gains cannot be remedied.
\end{itemize}
Each point is addressed in the following four paragraphs.

A typical control objective of any system is to asymptotically track reference signals with very small errors during transients. Theoretically, this can be obtained in MRAC systems using large adaptive gains. As mentioned before and also noted in \cite{Ioannou2014}, large adaptive gains lead to high-gain feedback. As a result, high frequency oscillations and instability can occur in the control channel. \cite{Ioannou2014} correctly understands that the purpose of the LPF is to attenuate such oscillations. However, \cite{Ioannou2014} misinterprets the insertion point of the LPF as only occurring at the input to the reference model of the adaptive controller i.e. an input-filtered direct MRAC. Taking this interpretation, subsequent analysis shows that relative degree matching between plant and MRAC is violated, and thus reference input and output tracking cannot be satisfied. In reality, the $\mathcal{L}_1$AC problem formulation deliberately inserts the LPF at the input to the respective inputs of both the predictor and plant precisely to avoid relative degree mismatch. As a result, the $\mathcal{L}_1$AC is a modified indirect MRAC architecture. Therefore, the results in \cite{Ioannou2014} do not represent an application of $\mathcal{L}_1$AC theory.

In addition to the LPF design, a key feature of the $\mathcal{L}_1$AC architecture is the state-predictor, whose prescribed dynamics converge to a reference model if tracking and estimation errors are small. The performance bounds derived from the architecture show these errors can be made arbitrarily small by increasing the adaptive gains the. Unlike standard MRAC systems, which cannot provide any transient performance guarantees, the error between reference and actual control inputs is also bounded and made arbitrarily small by increasing the adaptive gain. Effectively, the $\mathcal{L}_1$ adaptive non-linear closed-loop system follows an LTI reference system where classical frequency domain analysis can be performed conveniently. In \cite{Ioannou2014}, the authors claim that performance bounds, that are similarly inversely proportional to the adaptive gain, can be derived when the LPF is removed. However, this can be proved to be untrue. In fact both tracking and control input errors between the real and reference system become unbounded when the the LPF is removed (due to a non-strictly proper transfer function). 

Section III of \cite{Ioannou2014}, suggests that robustness is reduced when the LPF is inserted in $\mathcal{L}_1$AC architectures. Admittedly, phase-lag is introduced by the LPF which reduces phase and time-delay margins. However, from equation (15) in \cite{Ioannou2014}, as there is no upper limit on the LPF bandwidth, increasing it\footnote{While also ensuring not to increase the bandwidth too much such that the closed-loop system would become a high-gain feedback system and become susceptible to measurement noise.} would reduce the effect unmodelled dynamics have on the system. \cite{Ioannou2014} also claims that the design of the LPF requires knowledge of the unmodelled dynamics, indicated by equation (2.130) in \cite{L12010}. However, (2.130) is only used for analysing theoretical bounds; equation (2.139) defines the implemented controller with a LPF designed independent of the unmodelled dynamics. Finally, section 2.3 in \cite{L12010} demonstrates that the $\mathcal{L}_1$AC outperforms the MRAC in the presence of unmodelled dynamics using the classical Rohr's example.

\cite{Ioannou2014} correctly points out that the use of large adaptive gains can also lead to stiff differential equations of the predictor and adaptive laws. As a result, the differential solvers of the control software can fail or generate erroneous/high-frequency content. Nevertheless, the LPF is able to attenuate any resulting oscillations in the control channel. To counteract this phenomenon, it is recommended to bound the rate at which the parameter estimate can change - see assumption 2.2.2 in \cite{L12010}. For example, this can be done by placing a rate limiter block at the output of the projection operator in Simulink.

Though \cite{Ioannou2014} is the most widely known critique of $\mathcal{L}_1$AC theory, there are other notable papers, such as \cite{Bo2013,Ortega2016,Ortega2014b}, that have commented on the theory without fully understanding the design and correct implementation of $\mathcal{L}_1$ACs different architectures. 

In \cite{Bo2013}, a comparison between state-feedback direct MRAC, indirect MRAC and $\mathcal{L}_1$AC is shown using a simple first-order plant with matched uncertainties and zero disturbances. The paper demonstrates that the '$\mathcal{L}_1$-controller is significantly outperformed by the indirect MRAC algorithm' in the presence of time-varying references and with or without time-delay. The performance indices used to make this statement are, I believe, inappropriate. For example, the index for prediction-errors is given as $J_e = \int_{0}^{T}|x(t)-x_m(t)|dt$. As the state-prediction error exponentially decays then $J_e = \int_{0}^{T}Ae^{-\alpha t}dt = -(A/\alpha)e^{-\alpha t}+C$. This is not an indication of how the three controllers perform, as regardless of adaptation to uncertainty and guaranteed robustness of transient and stead-state performance the index converges to the initial state-error of the system, which can be large. In fact, $J_e = \infty$ when the $\mathcal{L}_1$AC is tracking a constant reference with a time-delay of 0.6 s. As it is proven in \cite{L12010} that steady-state errors converge to zero for the $\mathcal{L}_1$ adaptive closed-loop system in the presence of constant references, this puzzling index must come from either the inappropriately chosen performance index or the time-delay. In reality, it could be attributed to both, as the time-delay examples in this paper should not be considered as representative of any $\mathcal{L}_1$AC architecture since the design procedure developed in \cite{Cao2007,Cao2010a} for ensuring transient and steady-state robustness in the presence of time-delays, whereby the predictor is designed to include an additional matched disturbance, has not been followed. On a final note, as the $\mathcal{L}_1$AC is a modification of an indirect MRAC, the $\mathcal{L}_1$AC is not expected to perform better than the indirect MRAC anyway when fast adaptation is not required. The example does not require fast adaptation, indicated by the indirect MRAC using an adaptive gain of 30 (meanwhile, the $\mathcal{L}_1$AC uses an adaptive gain of 10,000). A fair comparison of the architectures would be if the example indeed required fast adaptation when uncertainty levels result in an unstable system. Furthermore, none of the architectures can guarantee asymptotic tracking of time-varying references. Though the LPF of $\mathcal{L}_1$AC architectures introduce additional phase-lag, it is shown in \cite{L12010} that the architecture has a uniform 90$\degree$ phase-margin as the adaptation gain increases, while \cite{Cao2010a,Kharisov2014} show that a uniform time-delay margin, which is lower-bounded by a value proportional to uncertainty, is ensured by satisfying a lower bound on the adaptive gain. \cite{Kharisov2014} also shows that the time-delay margin exponentially increases as sampling-rates increase.

In \cite{Ortega2014,Ortega2016} it was suggested that certain scalar LTI systems and reference models cannot satisfy the $\mathcal{L}_1$ stability condition in (\ref{eqn:L1normCond}). This hypothesis, which required the filter bandwidth to be less than the bandwidth of the reference model, was refuted in \cite{W.Fichter2013}. \cite{W.Fichter2013} argues that ``the filter acts as an additional actuator'', and ``if its dynamics are slower than the plant, this will limit both the performance and the robustness of the closed-loop system''. As a result (\ref{eqn:L1normCond}) can be satisfied in such cases when the filter bandwidth is set sufficiently large enough.

The same authors have claimed in \cite{Ortega2014a,Ortega2014c} that ``adaptation is unnecessary'' in a simple class of $\mathcal{L}_1$AC architectures\footnote{The class of $\mathcal{L}_1$ adaptive controller in question is one which compensates matched time-varying uncertainty.} since it approximates a model-following PI controller after adaptation has converged. This statement should not be misinterpreted as claiming that the $\mathcal{L}_1$AC is not adaptive. In fact, the standard MRAC also approximates a model-following PI controller when equipped to system (1) in \cite{Ioannou2014}. Effectively, the large adaptive gains of the adaptation loop ensures that the non-linear closed-loop system converges to a linear reference model, also known as 'limiting' behaviour, provided operation during conditions such as inactive projection bounds and actuator saturation. Indeed the works of \cite{Ortega2014b,Pettersson2012} are not criticisms of $\mathcal{L}_1$AC, rather they demonstrate that this 'limiting' behaviour resembles well-known architectures such as internal-model controllers and disturbance observers. This mapping of well-understood characteristics to certain classes of $\mathcal{L}_1$AC can benefit the integration of $\mathcal{L}_1$AC, particularly when commissioning controls for safety-critical applications i.e. aerospace \cite{Pettersson2012}. Finally, it is important to note that the works which 'criticise' $\mathcal{L}_1$AC are limited to a certain class of architectures (see footnote) and therefore the analysis cannot be representative of other $\mathcal{L}_1$AC architectures, e.g. output-feedback, unknown input-gain, which remain inherently non-linear \cite{Hovakimyan2006}.
 
Ultimately, $\mathcal{L}_1$AC theory was developed to confront several open problems of conventional adaptive control that were outlined in \cite{Anderson2005}, namely the selection and tuning of adaptive gains, guaranteed transient performance of both states and controls, and guaranteed robustness margins. At no point has $\mathcal{L}_1$AC theory been championed over standard MRAC architectures when fast and robust adaptation is not necessary i.e. a system where the identification time-scale is inherently faster than the plant variation time-scale. This work ultimately chose to explore $\mathcal{L}_1$AC theory as reconfiguration and PnP operations have an instantaneous influence on stability at a control level within the system that requires high-bandwidths, which as a result made fast adaptation and transient performance guarantees desirable. 

     \section{Results}
 Below, a meshed and radial mG topology, similar to that of \cite{Tucci2016c}, is considered. This topology is known to destabilise when $ \hat{\Sigma}_{6}^{\textrm{DGU}}$ is plugged-in using only baseline controllers - section \ref{Instability}. Hence, this set-up can adequately evaluate the performance of the proposed decentralised $\mathcal{L}_1$AC augmentation. Each DGU is equipped with controllers $\mathcal{C}_{[i]}^{\mathcal{L}_1}, i = 1,...,6$.
 
 \begin{figure}[!htb]    
 \centering
 \includegraphics[width=7.5cm]{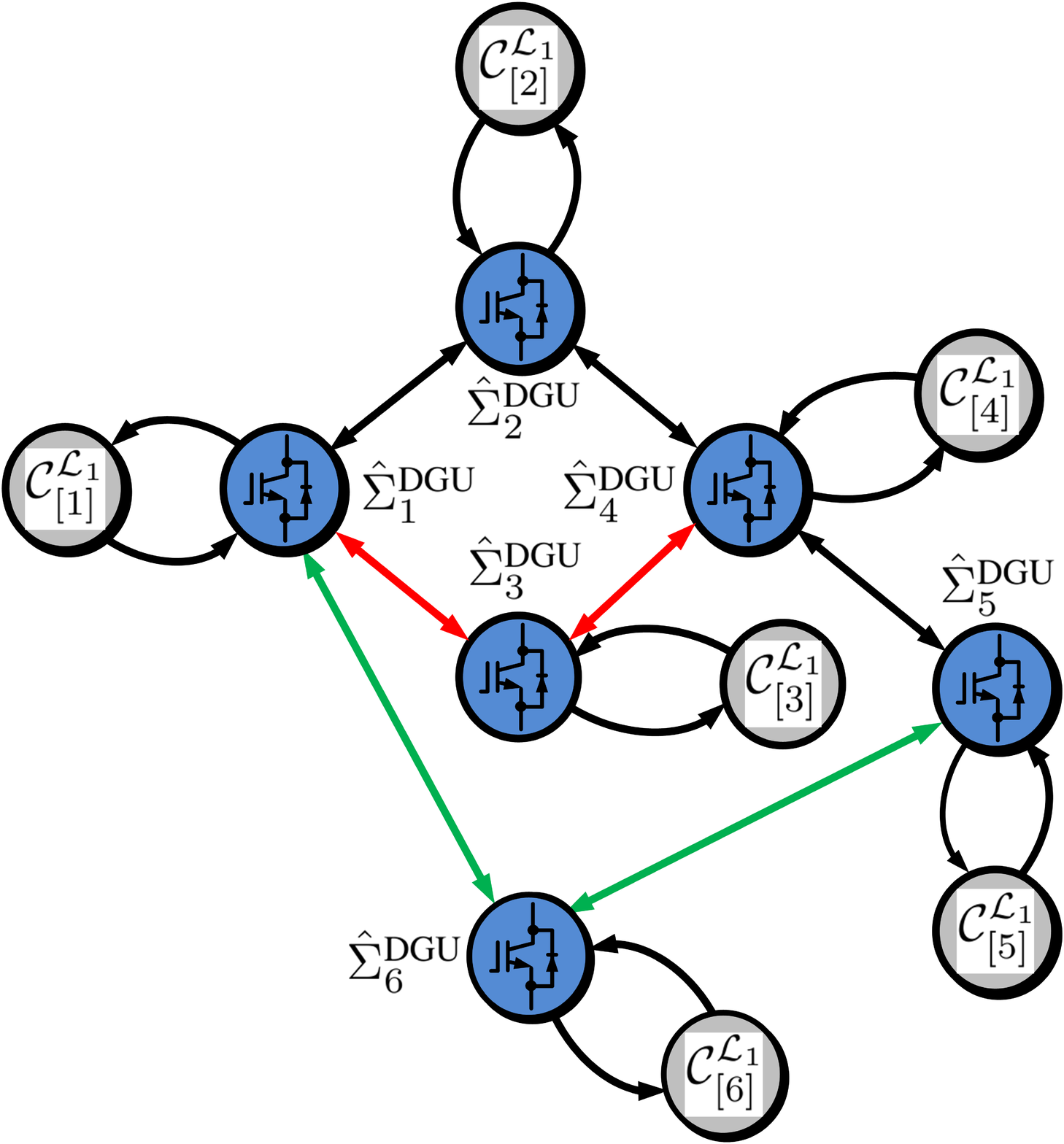}
 \caption{Meshed and radial microgrid configuration - $\hat{\Sigma}_{6}^{\textrm{DGU}}$ plug-in (green) and $\hat{\Sigma}_{3}^{\textrm{DGU}}$ plug-out (red).}
 \label{fig:DGU6DGU3PnPL1}
 \end{figure}
 
 Controllers and simulations were developed in Matlab/Simulink software. For greater accuracy this work uses non-linear PWM driven boost converters, as previously designed in \cite{OKeeffe2017a}, using the simpowersystems toolbox. System parameters are detailed in Table \ref{table:parameters}.

 \begin{table}[!htb]
 \fontsize{8.5}{9.5}\selectfont
 \centering
 % used for centering table
 \caption{System Parameters} 
 \begin{tabular}{c c c c c c c c} % centered columns (4 columns)
 \hline %\hline\hline inserts double horizontal lines
  
 Description & Parameter &  $\hat{\Sigma}_{1}^{\textrm{DGU}}$ & $\hat{\Sigma}_{2}^{\textrm{DGU}}$ & $\hat{\Sigma}_{3}^{\textrm{DGU}}$ & $\hat{\Sigma}_{4}^{\textrm{DGU}}$ & $\hat{\Sigma}_{5}^{\textrm{DGU}}$ & $\hat{\Sigma}_{6}^{\textrm{DGU}}$\\ [0.5ex] % inserts table
 %heading
 \hline\hline
 
 DGU rated power (kW) & $P_{[i]}$ & 5 & 5 & 5 & 5 & 5 & 5 \\
 Local load demand (kW) & $P_{R_{[i]}}$ & 2.5 & 2 & 1.8 & 2.5 & 3 & 2.5 \\
 Input voltage (V)& $V_{in_{[i]}}$  & 95  & 100 & 90 & 105 & 92 & 90 \\
 Reference voltage (V)& $V_{ref_{[i]}}$ & 381 & 380.5 & 380.2 & 379 & 379.5 & 380.7\\
 Switching frequency (kHz)& $f_s$ & 25 & 25 & 25 & 25 & 25 & 25\\
 Duty cycle & $ D_i $ & 0.7507 & 0.7372 & 0.7633 & 0.723 & 0.7576 & 0.7636 \\
 Inductance ($\mu $H)& $L_{ti}$ & 28.47 & 89.62 & 192.5 & 70 & 35 & 93.34 \\
 Capacitance ($\mu $F) & $C_{ti}$ &  37.632 & 51.67 & 40.73 & 37 & 31 & 24.66\\
 Parasitic resistance ($\Omega$)& $R_{ti}$ & 0.02 & 0.04 & 0.02 & 0.2 & 0.4 & 0.5 \\
 Line resistance ($\Omega$)& $R_{ij}$ & 0.5-2-10 & 0.5-4 & 2-4 & 2-4-15 & 15-4 & 10-4 \\
 Line inductance ($\mu $H)& $L_{ij}$ & 10-70-800 & 40-70 & 70-70 & 70-70-25 & 25-90 & 800-90\\
 Nominal duty cycle & $ D_i $ & 0.7368
  & 0.7368
   & 0.7368
    & 0.723 & 0.7368
     & 0.7368
      \\
 Nominal inductance ($\mu $H)& $L_{t^{nom}}$ & 2.794 & 2.794 & 2.794 & 2.794 & 2.794 & 2.794 \\
 Nominal capacitance ($\mu $F) & $C_{t^{nom}}$ &  60.6 & 60.6 & 60.6 & 60.6 & 60.6 & 60.6\\
 Nominal parasitic resistance ($\Omega$)& $R_{t^{nom}}$ & 0.1 & 0.1 & 0.1 & 0.1 & 0.1 & 0.1 \\
 Nominal line resistance ($\Omega$)& $R_{ij^{nom}}$ & 1 & 1 & 1 & 1 & 1 & 1 \\
 Nominal line inductance ($\mu$H)& $L_{ij^{nom}}$ & 10 & 10 & 10 & 10 & 10 & 10 \\
 \hline
 \end{tabular}
 \label{table:parameters} % is used to refer this table in the text
 \end{table}
 
  It should be noted that the dynamics of each DGU are different i.e. the electrical parameters and controller bandwidths are non-identical. Therefore, the system can be defined as heterogeneous. At $t=0 $, $\hat{\Sigma}_{1}^{\textrm{DGU}}, \hat{\Sigma}_{2}^{\textrm{DGU}}, \hat{\Sigma}_{3}^{\textrm{DGU}}$ and $\hat{\Sigma}_{4}^{\textrm{DGU}}$ are connected together through $RL$ power lines in a ring configuration. $ \hat{\Sigma}_{5}^{\textrm{DGU}}$ is connected to $ \hat{\Sigma}_{4}^{\textrm{DGU}}$, while $ \hat{\Sigma}_{6}^{\textrm{DGU}}$ powers a local load exclusively. Tests include PnP operations, robustness to load changes/unmodelled dynamics, and voltage tracking.
 \subsection{Plug-and-Play Operations}
 \subsubsection{Plug-in of DGU} 
\begin{figure}[!htb]
                 \centering
                 \begin{subfigure}[!htb]{0.37\textwidth}
                   \centering
  \includegraphics[width=\textwidth]{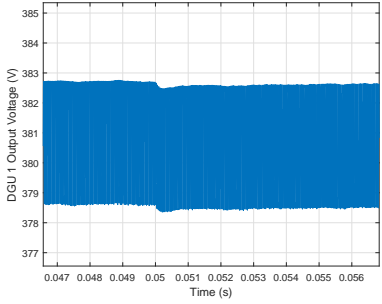}
 \caption{$\hat{\Sigma}_{1}^{\textrm{DGU}}$ output voltage.}
                   \label{fig:DGU1PnPDGU6}
                 \end{subfigure}\hspace*{\fill}
                 \begin{subfigure}[!htb]{0.37\textwidth}
                   \centering
 \includegraphics[width=\textwidth]{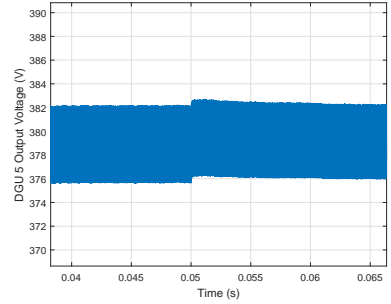}
          \caption{$\hat{\Sigma}_{5}^{\textrm{DGU}}$ output voltage.}
   \label{fig:DGU5PnPDGU6}
                 \end{subfigure}
                 \begin{subfigure}[!htb]{0.37\textwidth}
                   \centering
 \includegraphics[width=\textwidth]{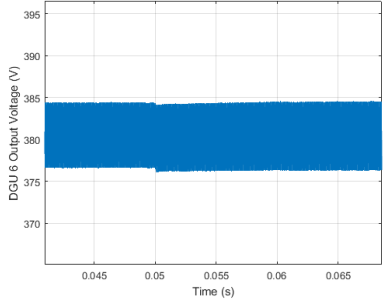}
    \caption{$\hat{\Sigma}_{6}^{\textrm{DGU}}$ output voltage}
 \label{fig:DGU6PnPDGU6}
                 \end{subfigure}
                 \caption{DGU output voltage responses to $\hat{\Sigma}_{6}^{\textrm{DGU}}$ plugging-in.}
                 \label{fig:DGUPnPDGU6}
               \end{figure}

In this section, the PnP capability of the proposed controllers $\mathcal{C}_{[i]}^{\mathcal{L}_1}$ is evaluated. At $t$ = 0.05 s, $\hat{\Sigma}_{6}^{\textrm{DGU}}$ is plugged-in, connecting to $\hat{\Sigma}_{1}^{\textrm{DGU}}$ and $\hat{\Sigma}_{5}^{\textrm{DGU}}$. Fig. \ref{fig:DGUPnPDGU6} plots the responses of the DGUs most directly affected i.e. $\hat{\Sigma}_{1}^{\textrm{DGU}}$, $\hat{\Sigma}_{5}$ and $\hat{\Sigma}_{6}^{\textrm{DGU}}$ \footnote{Non-neighbouring DGU responses are shown in section \ref{SimLargeV}}.

               Fig. \ref{fig:DGUPnPDGU6} shows very good performance when each controller is equipped with $\mathcal{C}_{[i]}^{\mathcal{L}_1}$, with hardly any overshoot and a very fast settling time $\le$ 10 ms.
In the subsequent subsection, $\hat{\Sigma}_{3}^{\textrm{DGU}}$ is unplugged from the ImG. However at $t$ = 0.6 s, $\hat{\Sigma}_{3}^{\textrm{DGU}}$ is subsequently plugged-in, connecting to $\hat{\Sigma}_{5}^{\textrm{DGU}}$ and $\hat{\Sigma}_{6}^{\textrm{DGU}}$.
\begin{figure}[!htb] % "[t!]" placement specifier just for this example
% {Images/Small_Voltage_Ref_Diff/} }
\begin{subfigure}{0.37\textwidth}
\includegraphics[width=\linewidth]{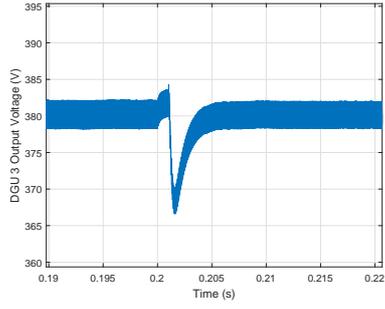}
\caption{$\hat{\Sigma}_{3}^{\textrm{DGU}}$ output voltage} \label{fig:DGU3PnP3_2}
\end{subfigure}\hspace*{\fill}
\begin{subfigure}{0.37\textwidth}
\includegraphics[width=\linewidth]{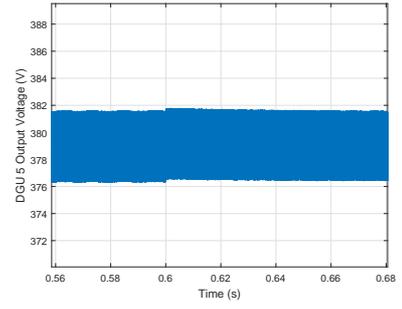}
\caption{$\hat{\Sigma}_{5}^{\textrm{DGU}}$ output voltage} \label{fig:fig:DGU5PnP3_2}
\end{subfigure}
\medskip
\centering
\begin{subfigure}{0.37\textwidth}
\includegraphics[width=\linewidth]{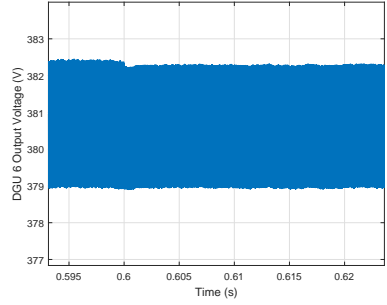}
\caption{$\hat{\Sigma}_{6}^{\textrm{DGU}}$ output voltage} \label{fig:DGU6PnP3_2}
\end{subfigure}

\caption{DGU output voltage responses to $\hat{\Sigma}_{3}^{\textrm{DGU}}$ plugging-in.} \label{fig:DGUPnPDGU6_2}
\end{figure}
 \subsubsection{Unplugging of DGU \label{sec:PnPunplug}}
  At $t$ = 0.2 s, $\hat{\Sigma}_{3}^{\textrm{DGU}}$ is disconnected from $\hat{\Sigma}_{1}^{\textrm{DGU}}$ and $\hat{\Sigma}_{4}^{\textrm{DGU}}$. 
 
 \begin{figure}[!htb] % "[t!]" placement specifier just for this example
 % {Images/Small_Voltage_Ref_Diff/} }
 \begin{subfigure}{0.37\textwidth}
 \includegraphics[width=\linewidth]{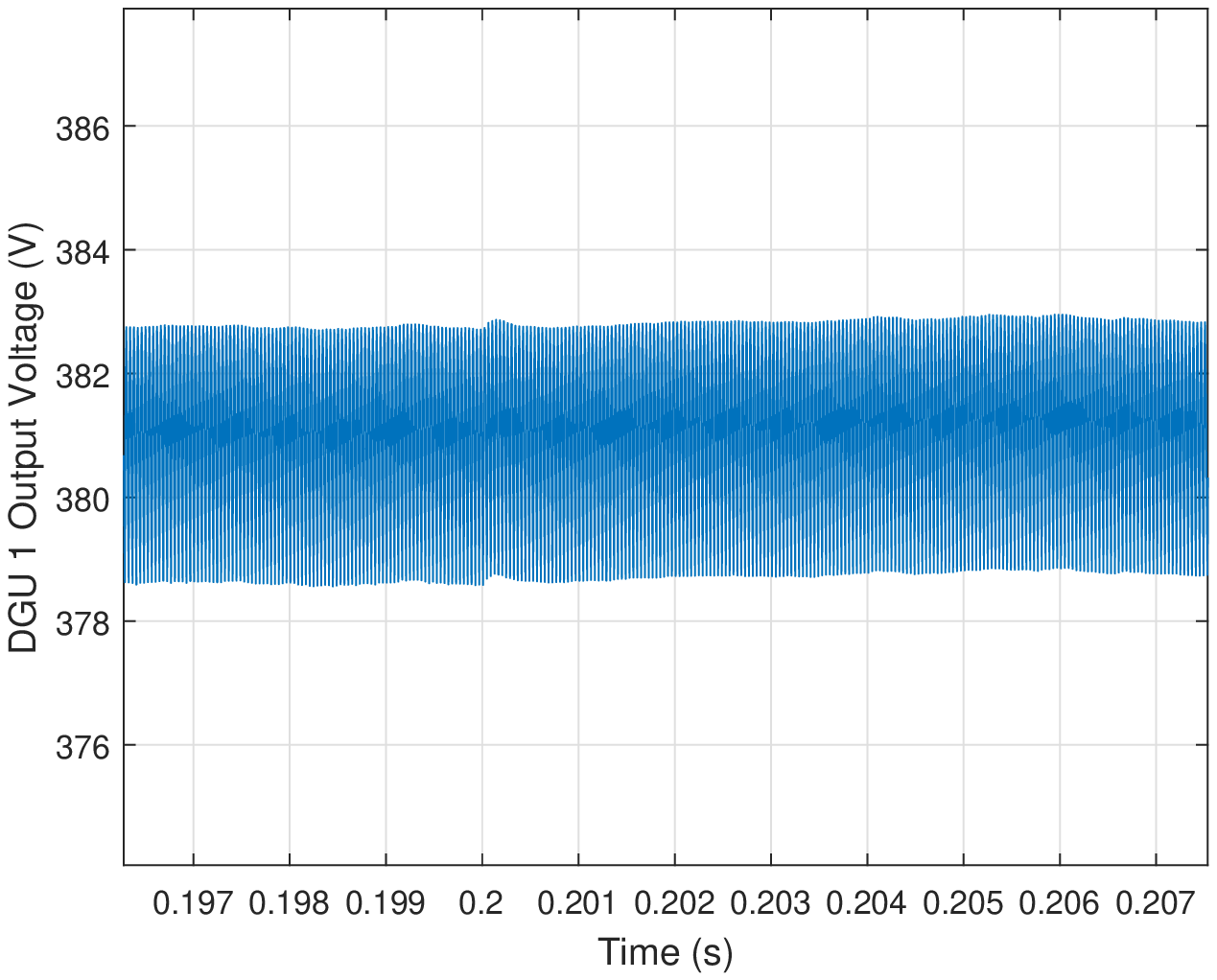}
 \caption{$\hat{\Sigma}_{1}^{\textrm{DGU}}$ output voltage} \label{fig:DGU1PnP3}
 \end{subfigure}\hspace*{\fill}
 \begin{subfigure}{0.37\textwidth}
 \includegraphics[width=\linewidth]{DGU3PnPDGU3i.eps}
 \caption{$\hat{\Sigma}_{3}^{\textrm{DGU}}$ output voltage} \label{fig:DGU3PnP3}
 \end{subfigure}
 \medskip
 \centering
 \begin{subfigure}{0.37\textwidth}
 \includegraphics[width=\linewidth]{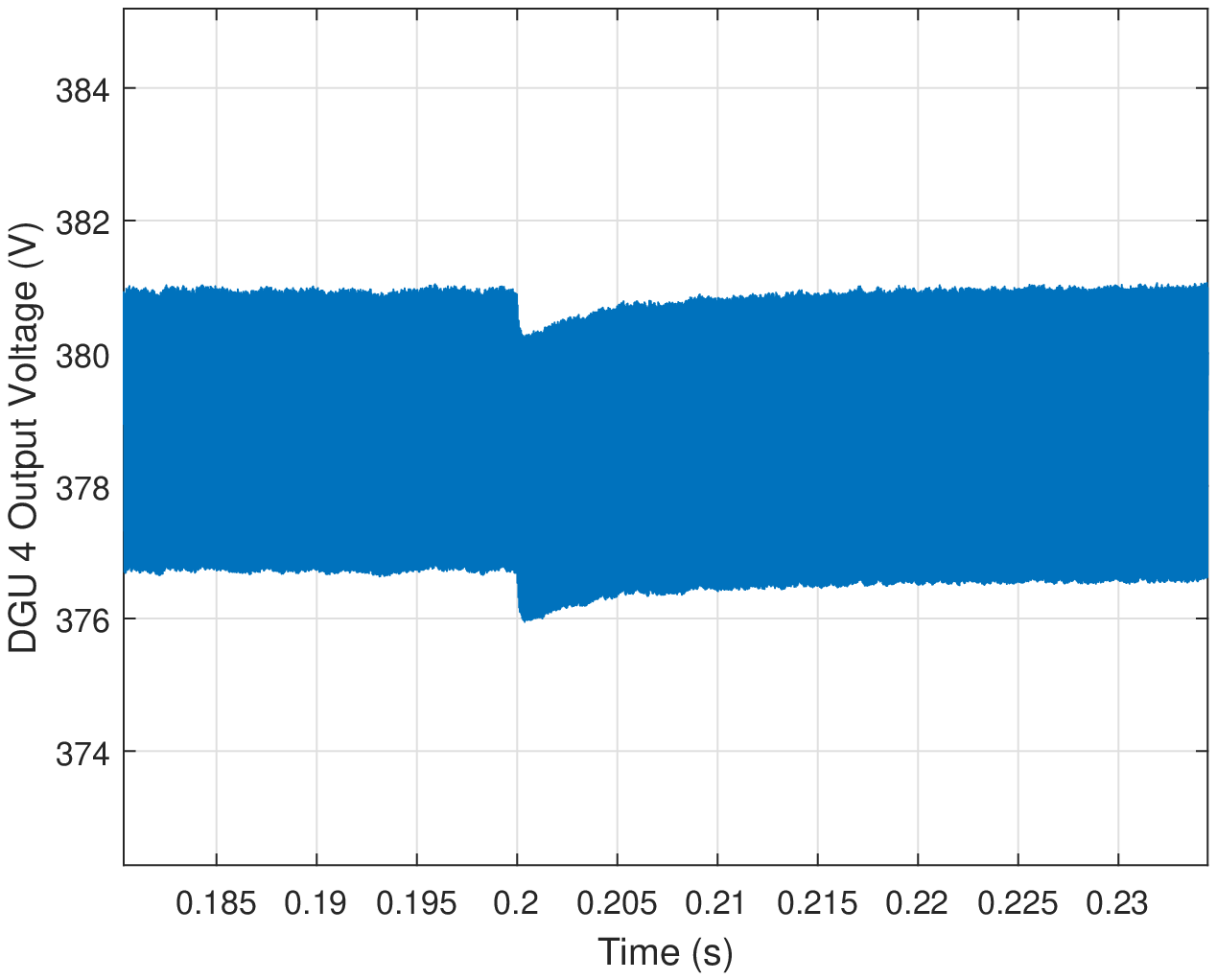}
 \caption{$\hat{\Sigma}_{4}^{\textrm{DGU}}$ output voltage} \label{fig:DGU4PnP3}
 \end{subfigure}
 
 \caption{DGU output voltage responses to $\hat{\Sigma}_{3}^{\textrm{DGU}}$ plugging-out.} \label{fig:DGUPnPDGU3}
 \end{figure}
 
 Fig. \ref{fig:DGUPnPDGU3} highlights good performance during the plug-out operation. The settling times of $\hat{\Sigma}_{1}^{\textrm{DGU}}$ and $\hat{\Sigma}_{4}^{\textrm{DGU}}$ are 1 ms and 20 ms respectively, which again are fast for primary voltage control. As shown in section \ref{SimLargeV}, when $\hat{\Sigma}_{3}^{\textrm{DGU}}$ is plugged-out while equipped with $\mathcal{C}_{[3]}^{\mathcal{L}_{1}}$, large oscillations are induced for 100 ms before settling. Ultimately, the baseline controller can handle the dynamics of being plugged-out to control its own load. Therefore the adaptation loop is turned-off at $t$ = 0.201 s, as shown in Fig. \ref{fig:DGU3PnP3}.
 
 \textbf{Remark 6:} \textit{It should be noted that, the power line resistances used in this test range from $0.5 - 15 \Omega$. Such power lines are applicable in large-scale systems where cabling lengths can be up to 1000 ft or 300 m. For example, a households average cable length is 30 m (section 2.3.1.3 of \cite{Webb2013}), which for 12 AWG cabling has a resistance of 0.16 $\Omega$.} 
 \vspace{3mm} \newline
 Therefore, large line resistances naturally impede current disturbances from neighbouring DGUs. Nonetheless, as seen in section \ref{sec:VrefTrack}, the change in line-currents upon output voltage reference changes are relatively significant. Global-asymptotic stability is maintained when the closed-loop dynamics of each DGU is changed during PnP operations.
 
 \subsection{Robustness to Unknown Load Change}
 
 In order to examine the robustness of DC-ImG to unknown load changes, the load at $\hat{\Sigma}_{6}^{\textrm{DGU}}$ is stepped from 2.5 kW to 800 W at $t$ = 0.3 s. The responses of each DGU are plotted below.
 \begin{figure}[!htb] % "[t!]" placement specifier just for this example
 % {Images/Small_Voltage_Ref_Diff/} }
 \begin{subfigure}{0.5\textwidth}
 \includegraphics[width=\linewidth]{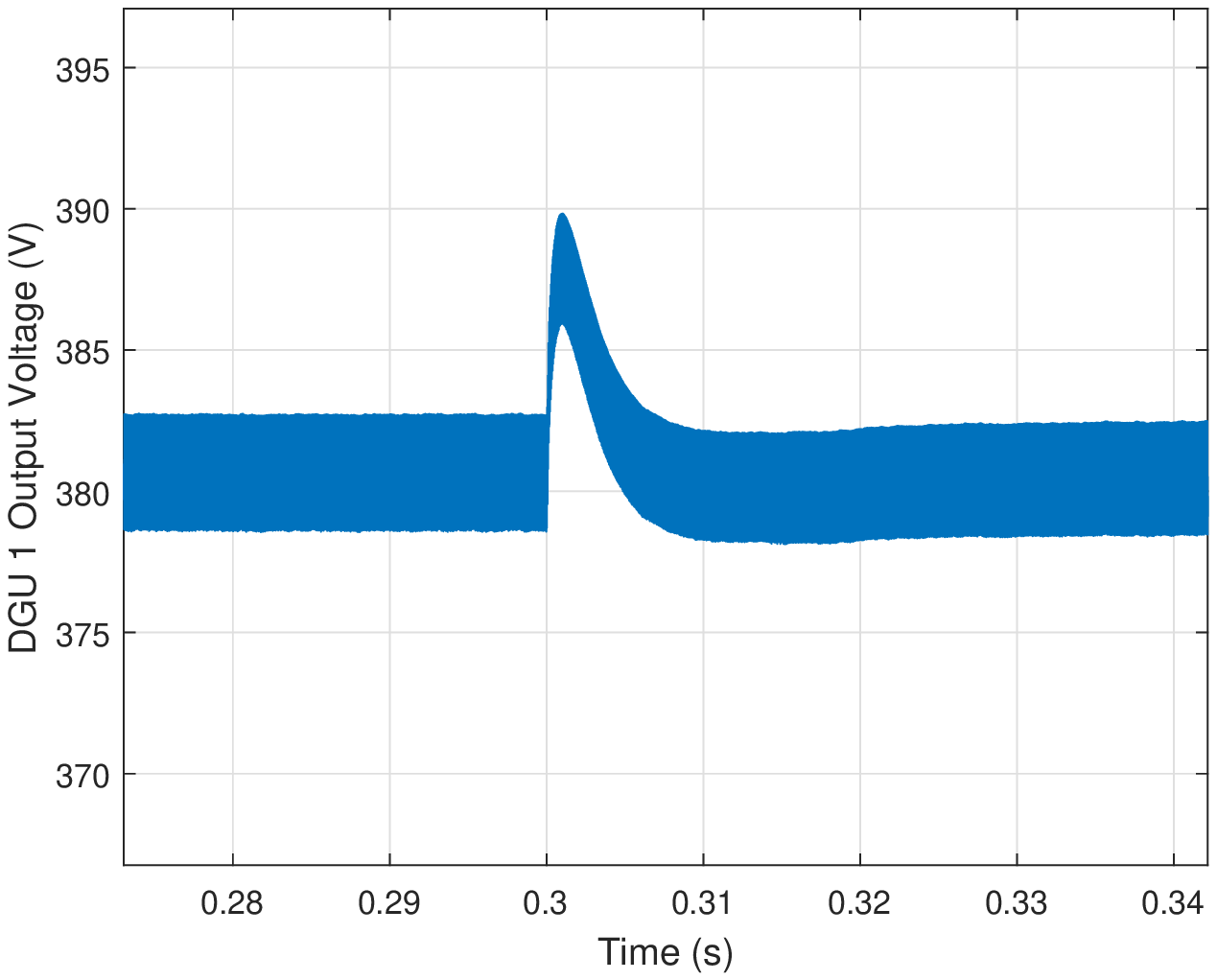}
 \caption{$\hat{\Sigma}_{1}^{\textrm{DGU}}$ output voltage} \label{fig:DGU1LoadChange1}
 \end{subfigure}\hspace*{\fill}
 \begin{subfigure}{0.5\textwidth}
 \includegraphics[width=\linewidth]{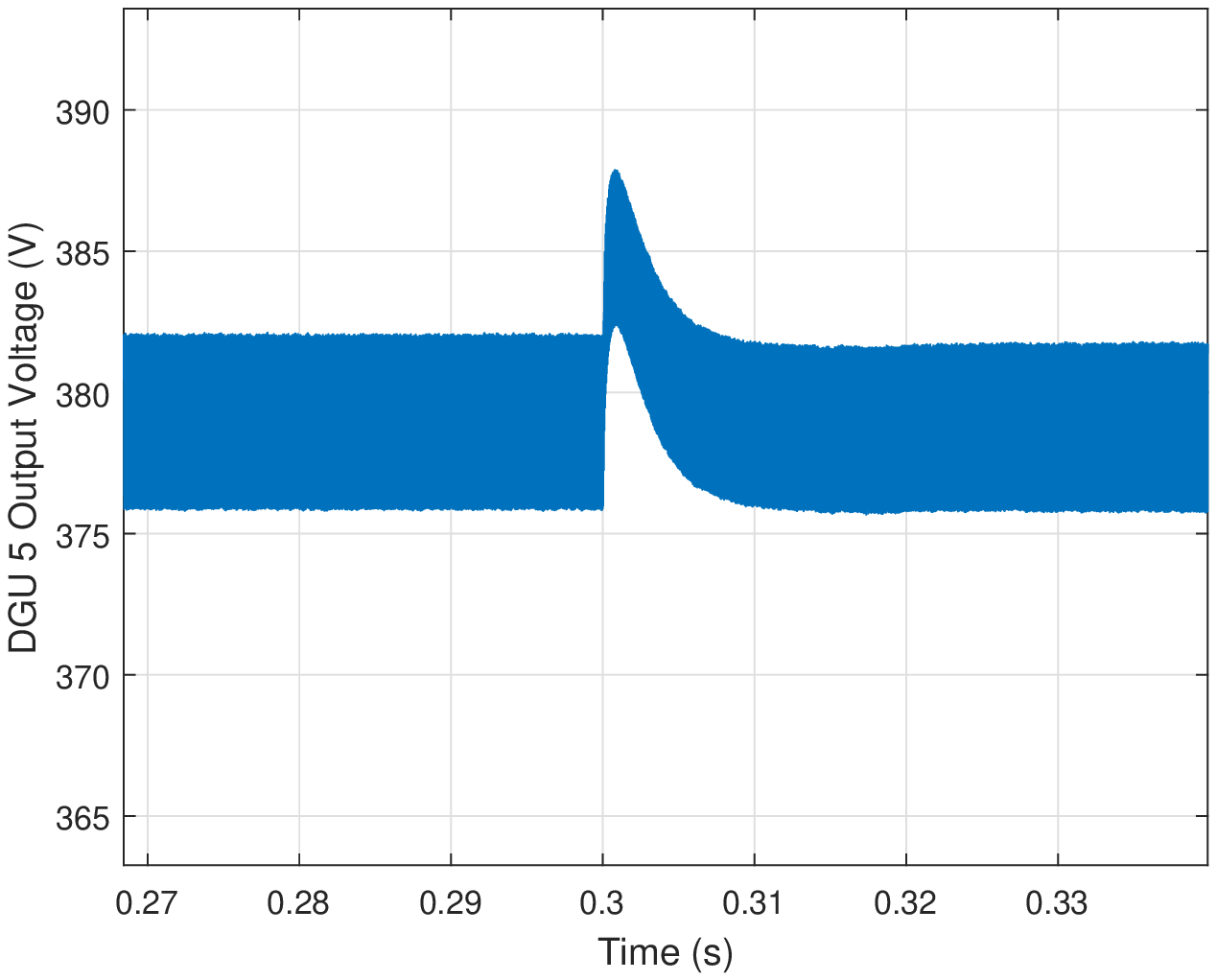}
 \caption{$\hat{\Sigma}_{5}^{\textrm{DGU}}$ output voltage} \label{fig:DGU5LoadChange1}
 \end{subfigure}
 \end{figure}
 
 \begin{figure}[!htb]\ContinuedFloat
 % {Images/Small_Voltage_Ref_Diff/} }
 \centering
 \begin{subfigure}{0.5\textwidth}
 \includegraphics[width=\linewidth]{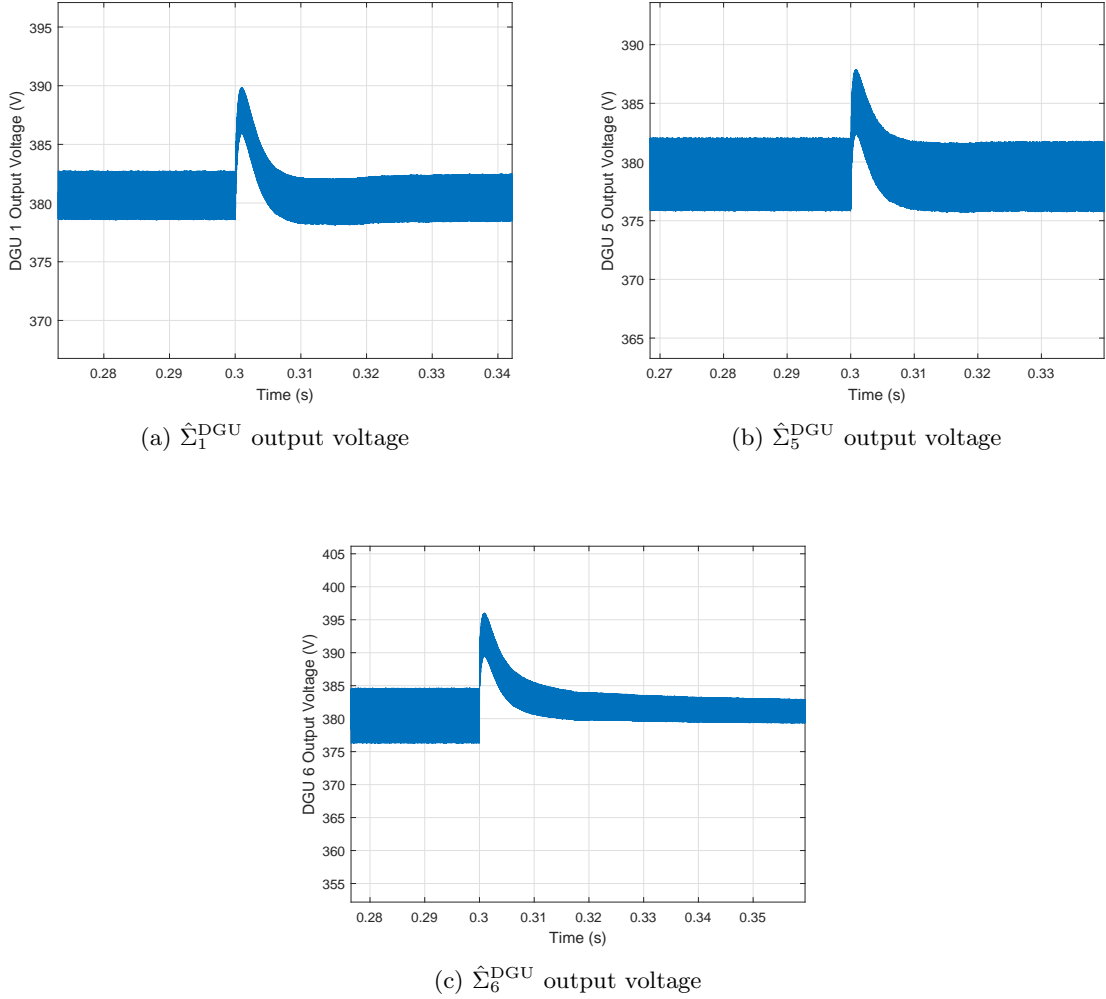}
 \caption{$\hat{\Sigma}_{6}^{\textrm{DGU}}$ output voltage} \label{fig:DGU6LoadChange1}
 \end{subfigure}
 \caption{DGU output voltage responses to $\hat{\Sigma}_{3}^{\textrm{DGU}}$ plugging-out.} \label{fig:DGUPLoadChange1}
 \end{figure}
 
 \vspace{20mm}
 The responses of neighbouring DGUs $\hat{\Sigma}_{1}^{\textrm{DGU}}$ and $\hat{\Sigma}_{5}^{\textrm{DGU}}$ show very good robustness to unknown load changes within the ImG. Settling times are fast, within 30 ms, while overshoot is limited to less than 3.8 $\%$. The response of $\hat{\Sigma}_{6}^{\textrm{DGU}}$ is similarly favourable, with overshoot limited to less than 4 $\%$ and settling time within 30 ms. The load dependent voltage ripple has also reduced.
 
 \subsection{Voltage reference tracking}\label{sec:VrefTrack}
 
 The hierarchical structure of ImG control architectures requires primary voltage reference changes, using commands from secondary controllers, in order to control the power flows amongst DGUs within the ImG, as well as regulate the state-of-charge of batteries. Therefore, a key metric of the proposed system is the performance of the system in response to voltage reference changes. This is evaluated by stepping the voltage reference of $\hat{\Sigma}_{1}^{\textrm{DGU}}$ from 381 V to 375 V at $t$ = 0.8 s. Since the impedance of the $RL$ lines is small the voltage decrease is enough to pull an appreciable amount of power from neighbouring DGUs,
 causing current disturbances to cascade throughout the ImG. Though the $RL$ line parameters that are used correlate to long cabling lengths within a large-scale power system, the resistances are still small enough to cause considerable line currents to flow. For example, when $t\le$ 0.8 s, the steady-state line current $I_{21} = \frac{V_1-V_2}{R_{21}}=\frac{381-380.5}{0.5} = 1 A$. After $t = 0.8 s$  $I_{21} = \frac{375-380.5}{0.5} = -11 A$.
 
  Therefore, it is important that neighbouring DGUs are robust to this unknown disturbance. The responses of $\hat{\Sigma}_{1}^{\textrm{DGU}}$ and its neighbours $\hat{\Sigma}_{2}^{\textrm{DGU}}$ and $\hat{\Sigma}_{6}^{\textrm{DGU}}$ are plotted below. 
 
 \begin{figure}[!htb] % "[t!]" placement specifier just for this example
 % {Images/Small_Voltage_Ref_Diff/} }
 \begin{subfigure}{0.525\textwidth}
 \includegraphics[width=\linewidth]{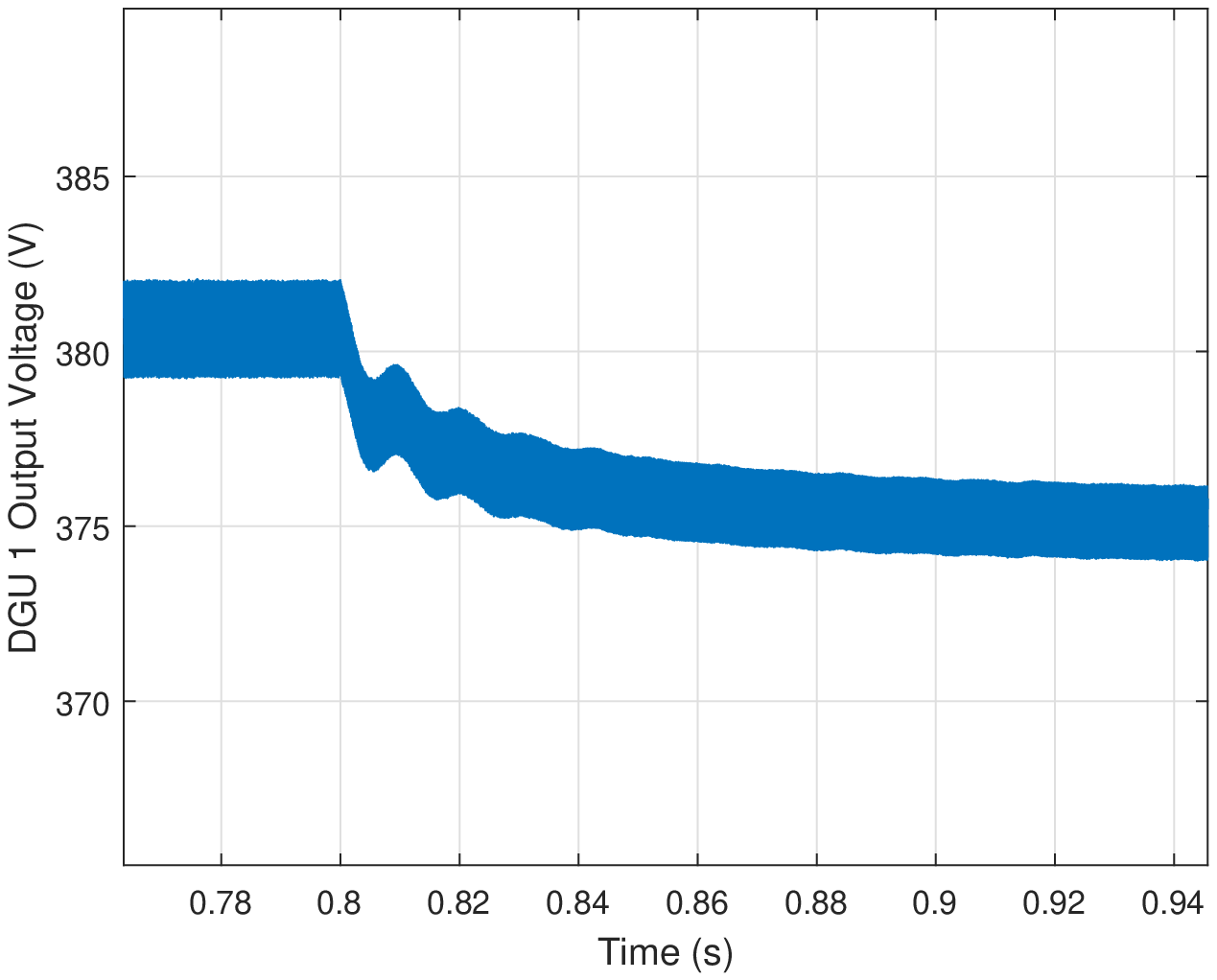}
 \caption{$\hat{\Sigma}_{1}^{\textrm{DGU}}$ output voltage} \label{fig:DGU1Vchange}
 \end{subfigure}\hspace*{\fill}
 \begin{subfigure}{0.525\textwidth}
 \includegraphics[width=\linewidth]{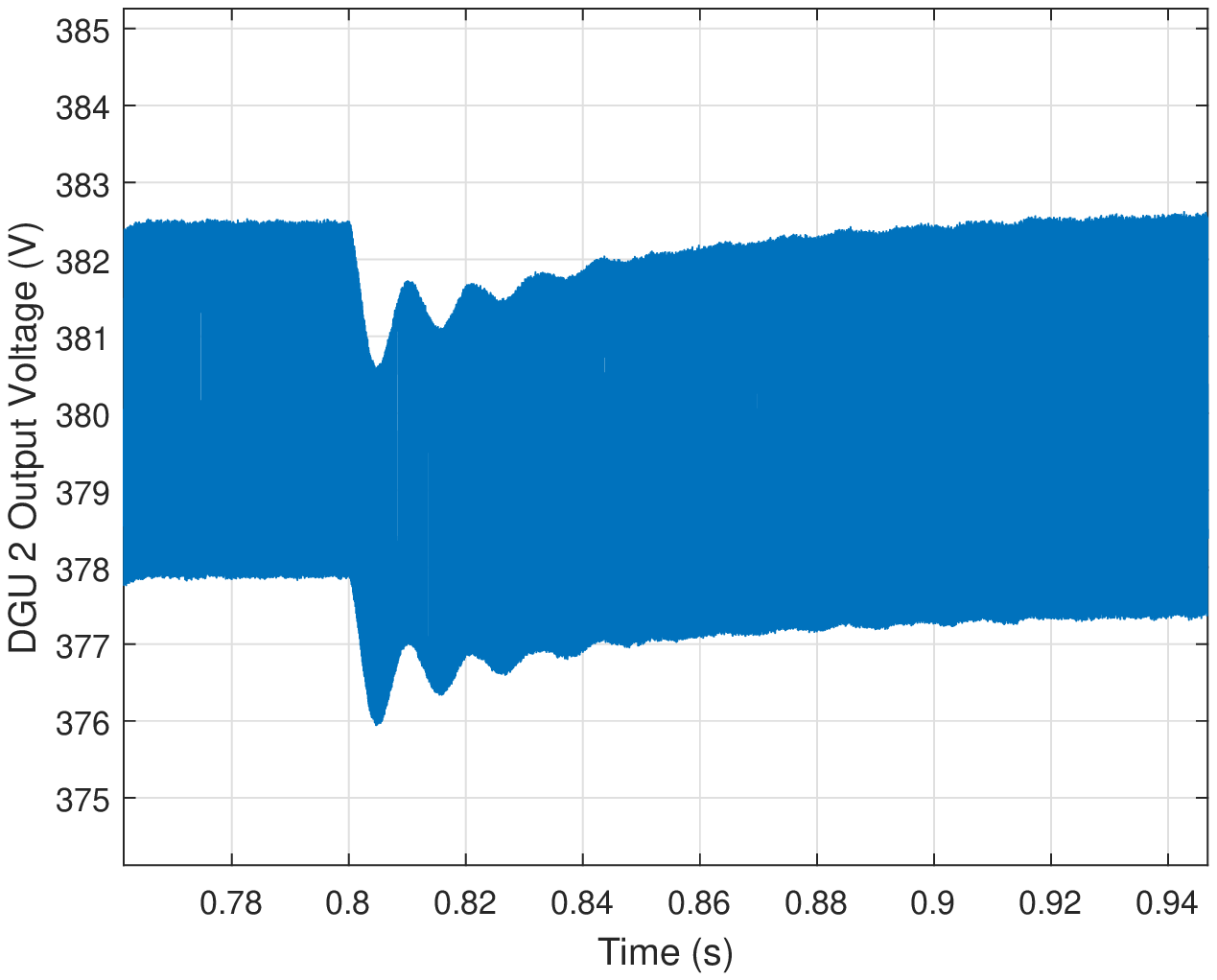}
 \caption{$\hat{\Sigma}_{2}^{\textrm{DGU}}$ output voltage} \label{fig:DGU2VChangeDGU1}
 \end{subfigure}
 \end{figure}
 
 \begin{figure}[!htb] \ContinuedFloat%
 % {Images/Small_Voltage_Ref_Diff/} }
 \centering
 \begin{subfigure}{0.525\textwidth}
 \includegraphics[width=\linewidth]{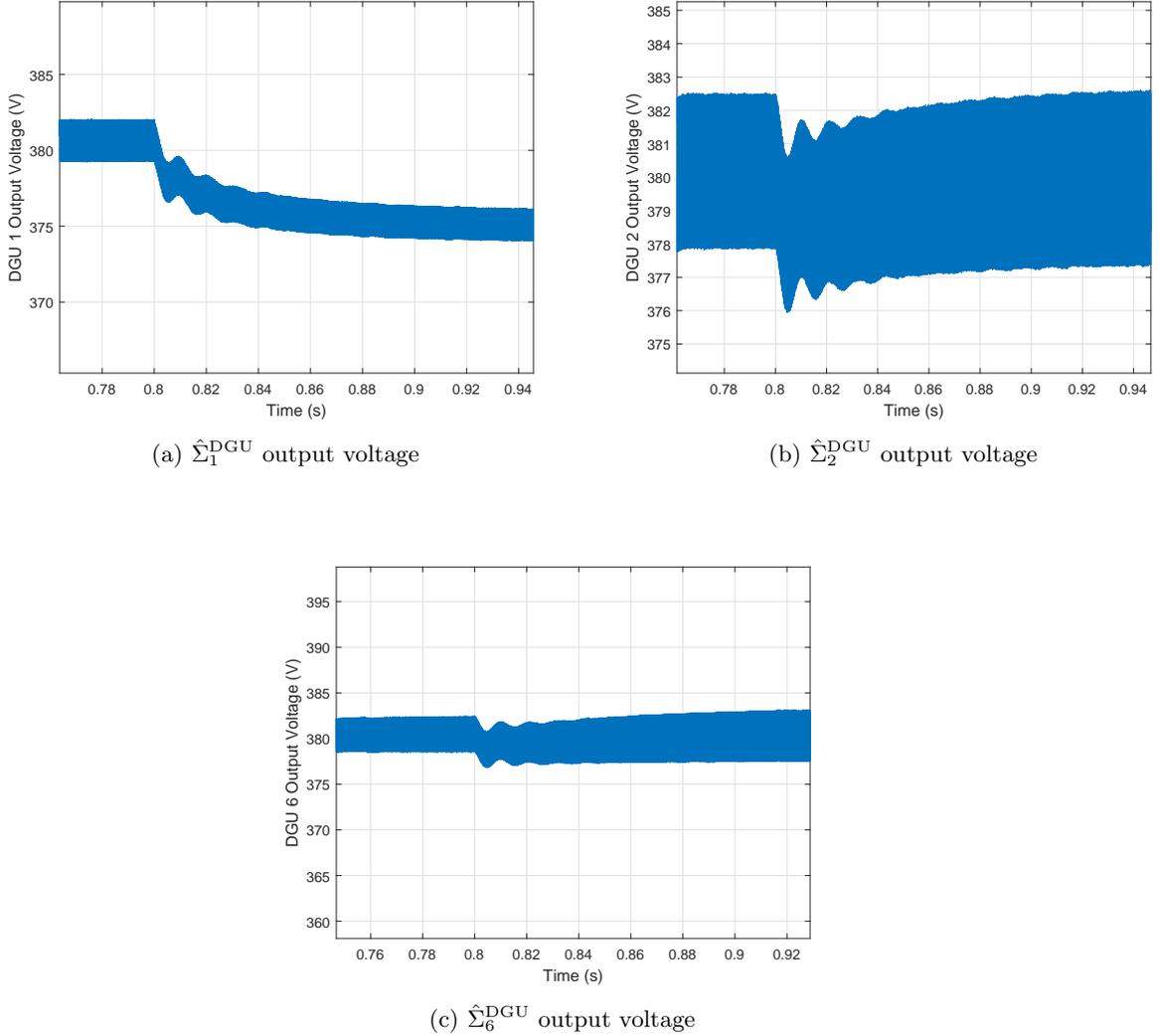}
 \caption{$\hat{\Sigma}_{6}^{\textrm{DGU}}$ output voltage} \label{fig:DGU6VChangeDGU1}
  \end{subfigure}
 \caption{DGU output voltage responses to $\hat{\Sigma}_{1}^{\textrm{DGU}}$ voltage reference step from 381 V to 375 V.}
 \end{figure} 
 Fig. \ref{fig:DGU1Vchange} demonstrates the fast voltage reference tracking capabilities of the system. After small transient oscillations, the settling time is reached within 100 ms. Figs. \ref{fig:DGU2VChangeDGU1} and \ref{fig:DGU6VChangeDGU1} show the interactions between coupled DGUs during the step test are minimal with each DGU showing good robustness to current disturbances.
 
 \subsection{Comparison with average model}
 Previous results were attained  using non-linear PWM switching converter models built using the simpowersystems toolbox of Matlab/Simulink. These models typically lead to very long simulation times i.e. one second can take between 36-48 hours. This is associated with the use of large PWM switching frequencies, non-linear projection operator, and large number of adaptively detected zero-crossings.
 
 To speed up simulation times, the average model of each DGU can be used by constructing the system using the differential equations of (\ref{eq:AvMG}). The following results correlate with the tests performed previously.
 
 \begin{figure}[!htb] % "[t!]" placement specifier just for this example
 % {Images/Average_Model_Results/} }
 \begin{subfigure}{0.37\textwidth}
 \includegraphics[width=\linewidth]{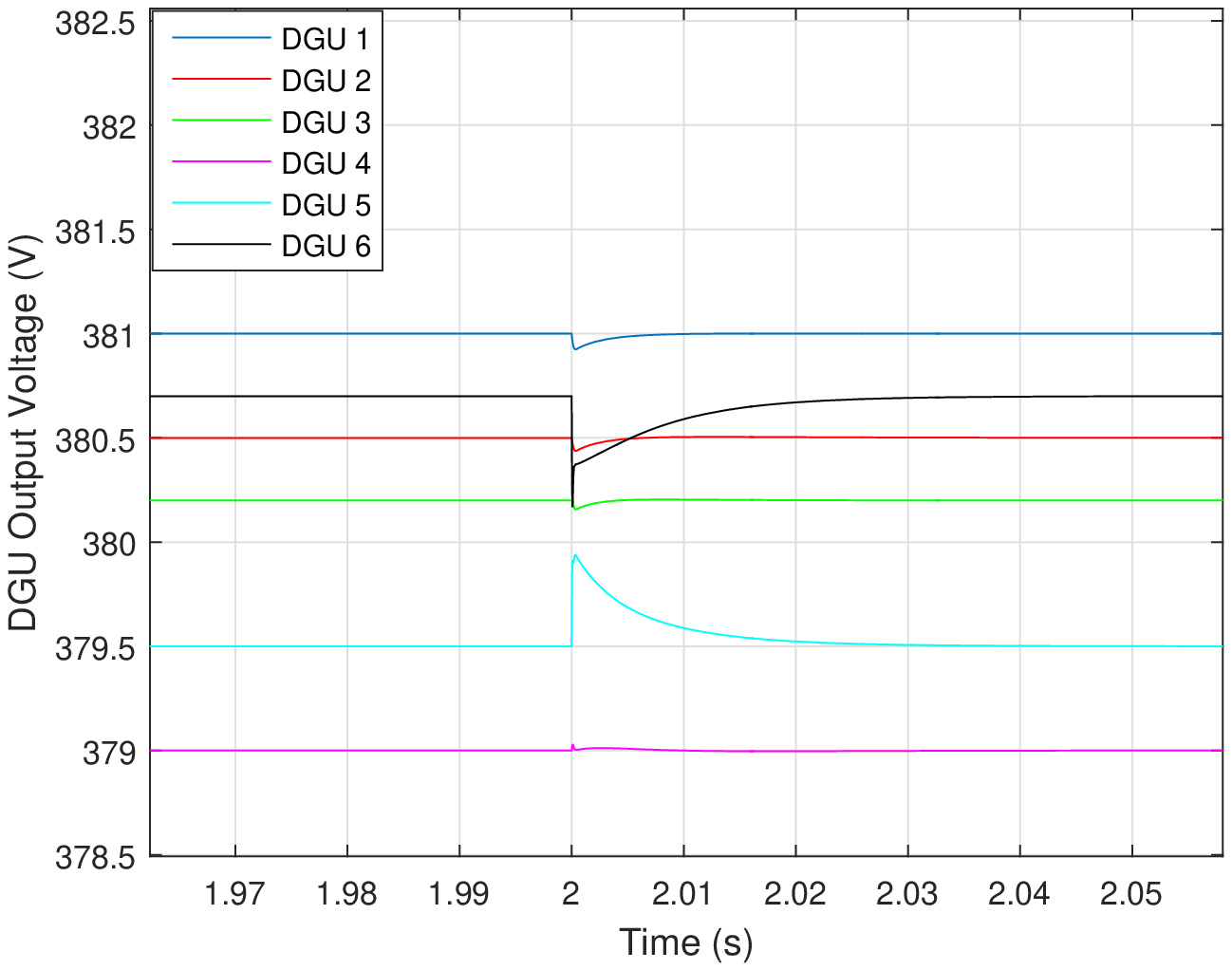}
 \caption{DGU output voltage responses to $\hat{\Sigma}_{6}^{\textrm{DGU}}$ plug-in.} \label{fig:DGUPnPDGU6AV}
 \end{subfigure}\hspace*{\fill}
 \begin{subfigure}{0.37\textwidth}
 \includegraphics[width=\linewidth]{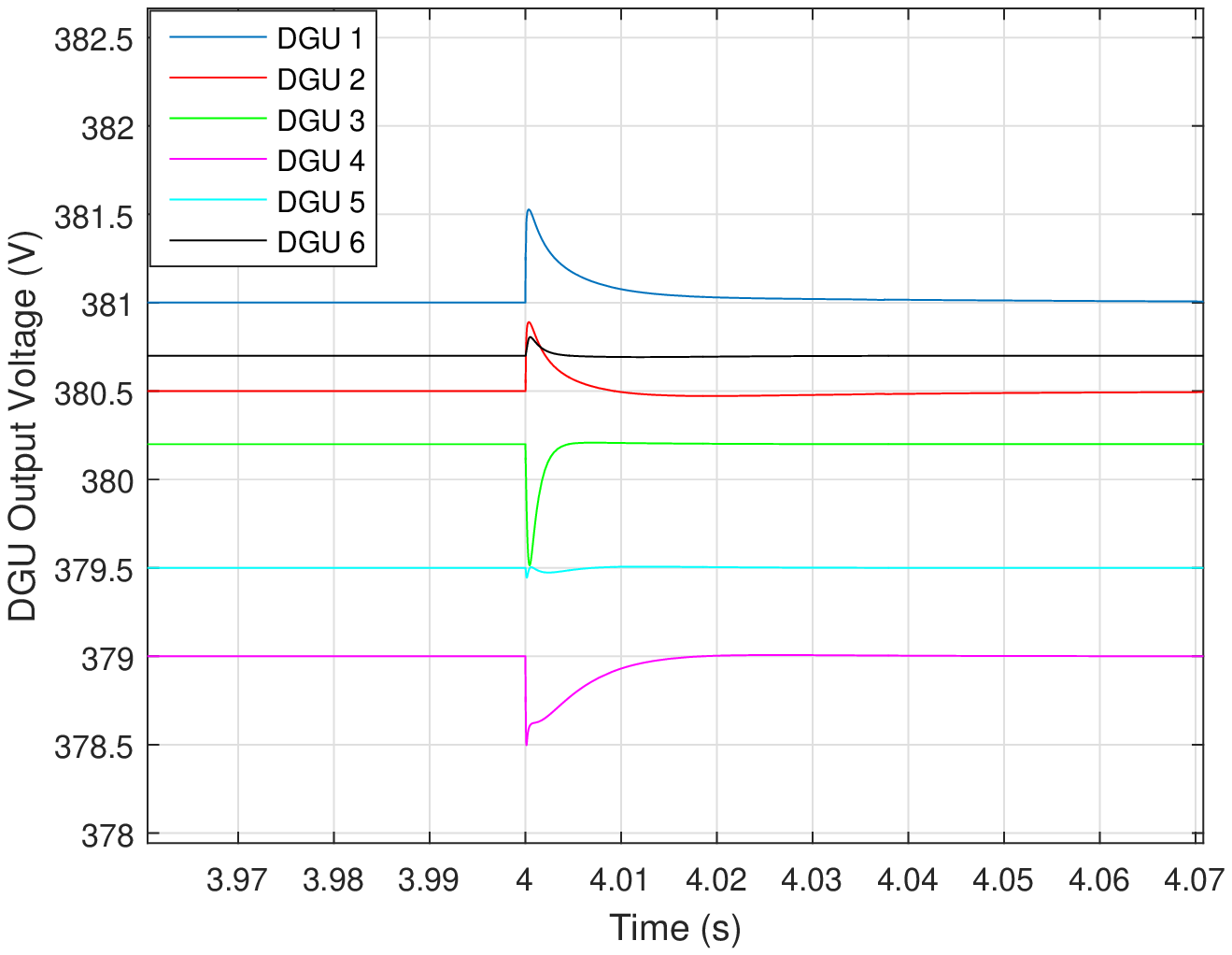}
 \caption{DGU output voltage responses to $\hat{\Sigma}_{3}^{\textrm{DGU}}$ plug-out.} \label{fig:DGUPnPDGU3AV}
 \end{subfigure}
 \begin{subfigure}{0.37\textwidth}
 \includegraphics[width=\linewidth]{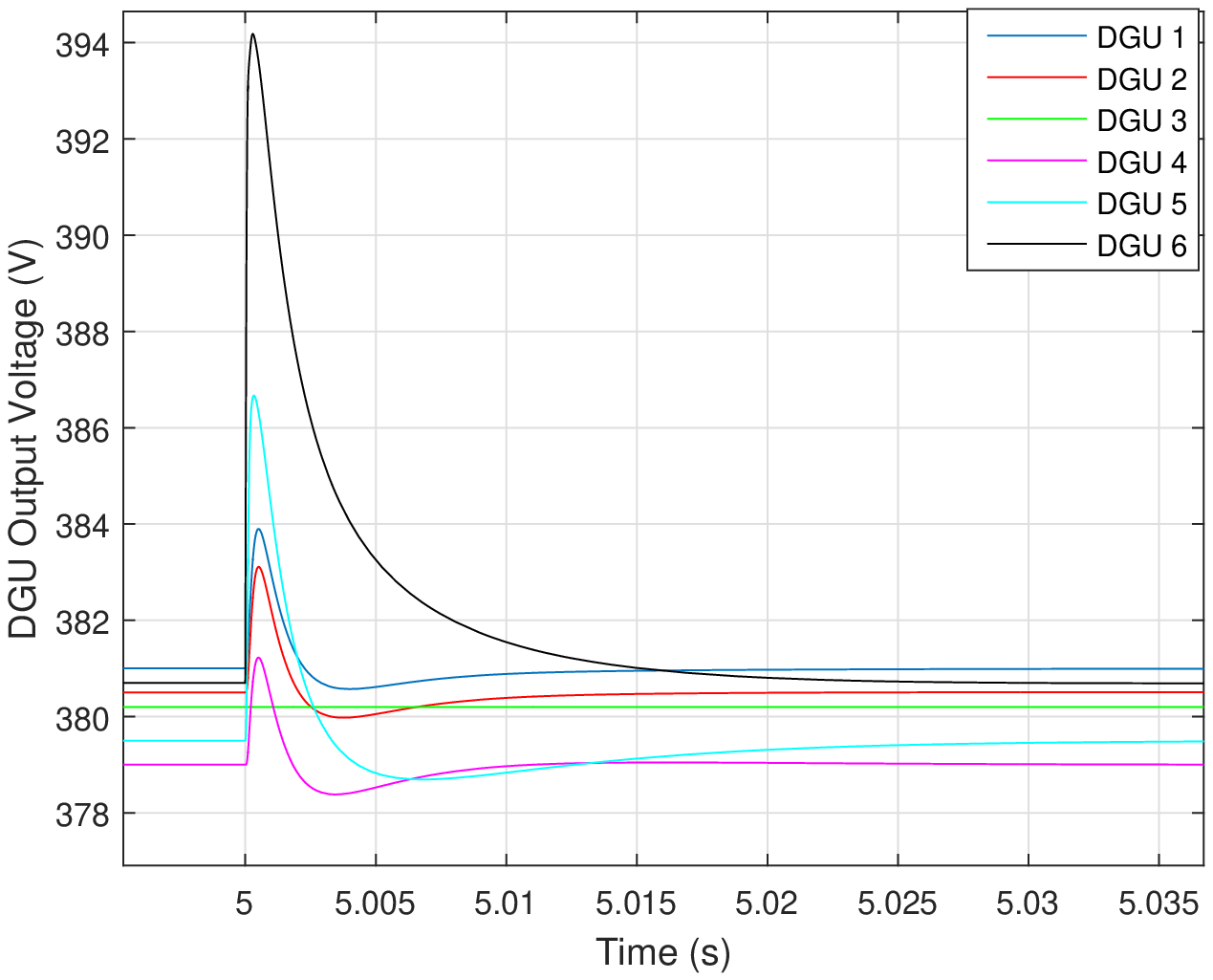}
 \caption{DGU output voltage responses to $\hat{\Sigma}_{6}^{\textrm{DGU}}$ load step of 2.5 kW to 800 W.} \label{fig:DGU6LoadChangeAv}
 \end{subfigure}
 \hspace*{\fill}
 \begin{subfigure}{0.37\textwidth}
 \includegraphics[width=\linewidth]{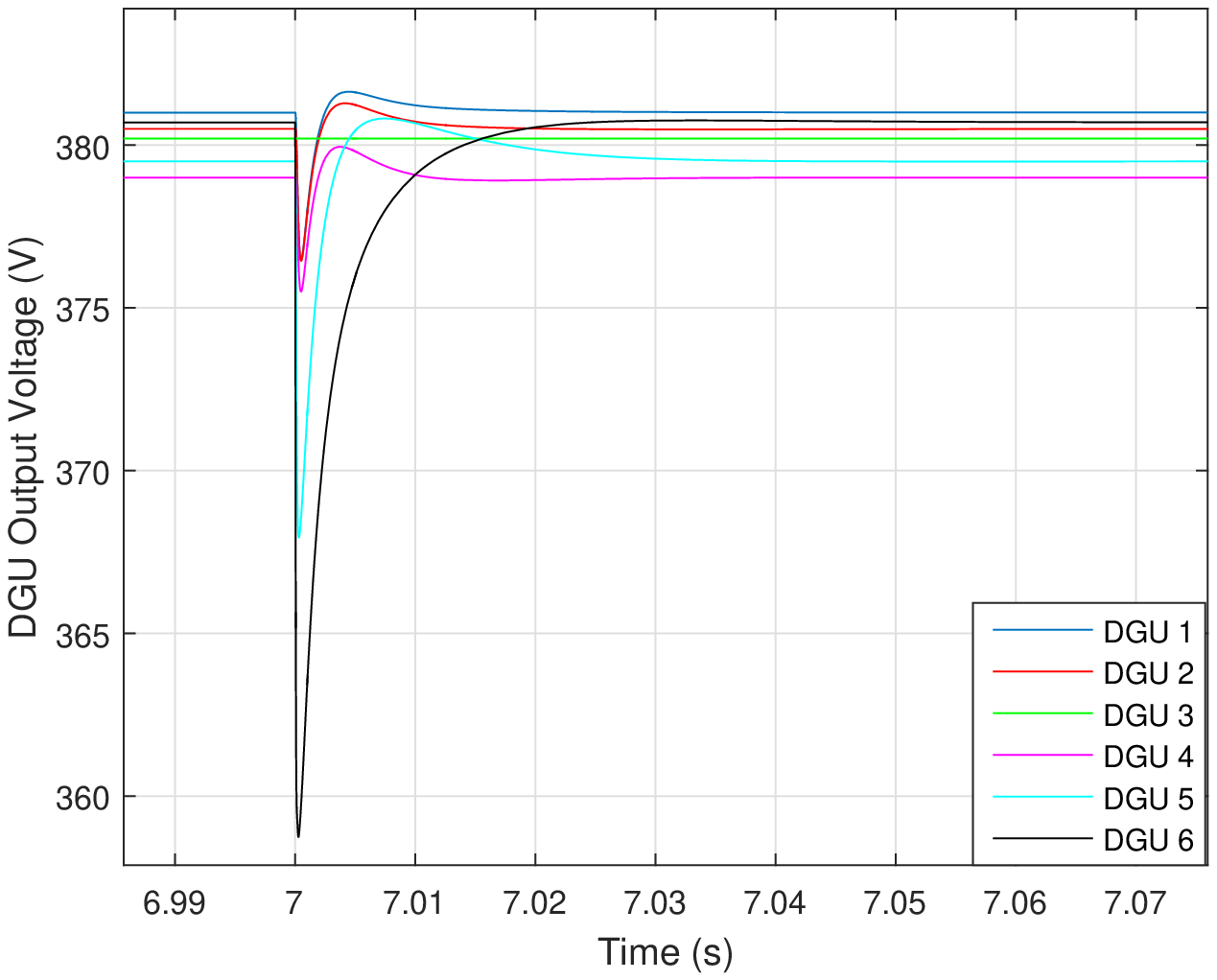}
 \caption{DGU output voltage responses to $\hat{\Sigma}_{6}^{\textrm{DGU}}$ load step of 800 W to 3.8 kW.} \label{fig:DGU6LoadChangeAv2}
 \end{subfigure}
 \begin{subfigure}{0.37\textwidth}
 \includegraphics[width=\linewidth]{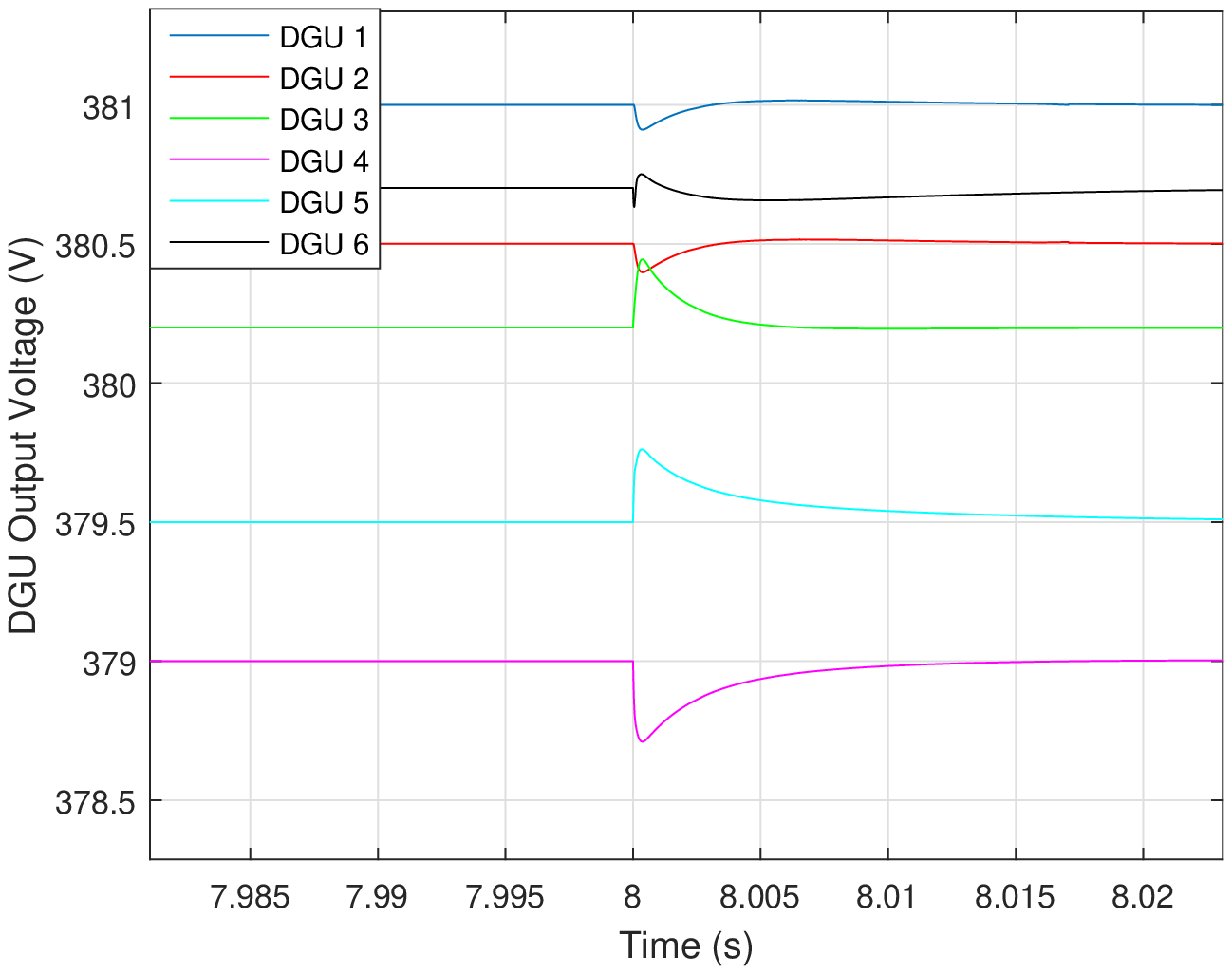}
 \caption{DGU output voltage responses to $\hat{\Sigma}_{3}^{\textrm{DGU}}$ plug-in.} \label{fig:DGUPnPDGU3_2}
 \end{subfigure}
 \hspace*{\fill}
 \begin{subfigure}{0.37\textwidth}
 \includegraphics[width=\linewidth]{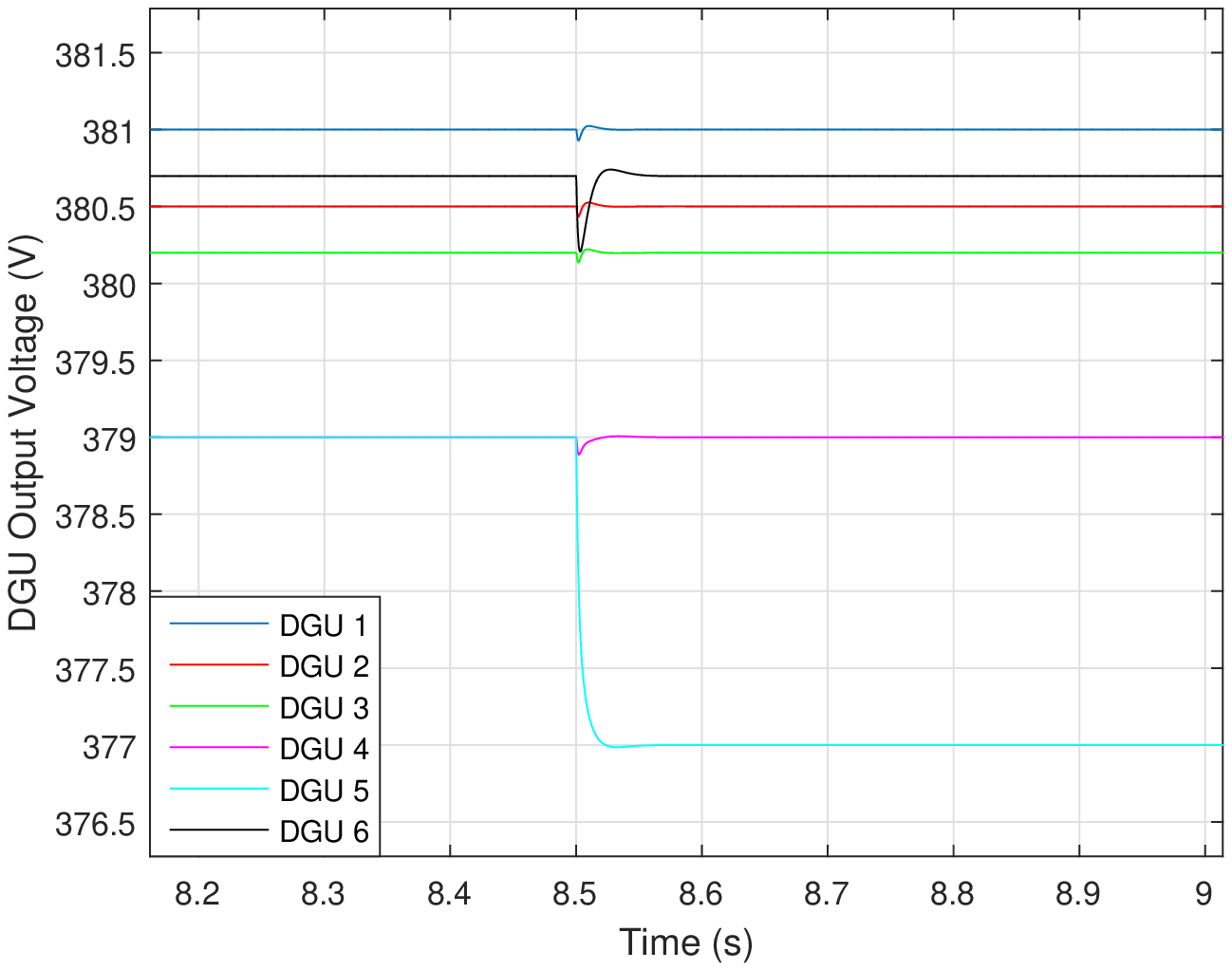}
 \caption{DGU output voltage responses to $\hat{\Sigma}_{5}^{\textrm{DGU}}$ voltage step of 379.5 V to 377 V.} \label{fig:DGUPnPDGU3_2}
 \end{subfigure}
 \caption{DGU output voltage responses using average model of (\ref{eq:AvMG})}
 \end{figure}
 
 \subsection{Robustness to Unmodelled Dynamics}\label{sec:unmod}
 
 This section demonstrates further robustness of the $\mathcal{L}_1$AC to heterogeneity, variation in parametric uncertainty and unmodelled dynamics. 
 
 The topology of each DGU is augmented by the addition of an unmodelled capacitor equivalent series resistance, $R_{c_{i}}$, in series with the output capacitor. Like the already modelled inductor equivalent series resistance, $R_{c_{i}}$ represents capacitor voltage drops associated with capacitors due to non-ideality in power converters. In fact, $R_{c_{i}}$ can be used as a design feature in order to increase output voltage damping and reduce ripple. The significance of this is that the output voltage no longer equates to the capacitor voltage and therefore controlling the capacitor voltage state does not correspond to the output voltage control. The model is derived in section 6.3.
 
 The values of $R_{c_{i}}$ are: $R_{c_{1}} = 0.02 \Omega$, $R_{c_{2}} = 0.05 \Omega$, $R_{c_{3}} = 0.15 \Omega$, $R_{c_{4}} = 0.07 \Omega$, $R_{c_{5}} = 0.09 \Omega$, $R_{c_{6}} = 0.01 \Omega$. The following results plot the response of $\hat{\Sigma}_{6}^{\textrm{DGU}}$ as it plugs in at $t$ = 0.05 s, and its neighbours $\hat{\Sigma}_{1}^{\textrm{DGU}}$ and  $\hat{\Sigma}_{5}^{\textrm{DGU}}$.
 
 \begin{figure}[!htb] % "[t!]" placement specifier just for this example
 % {Images/Small_Voltage_Ref_Diff/} }
 \begin{subfigure}{0.48\textwidth}
 \includegraphics[width=\linewidth]{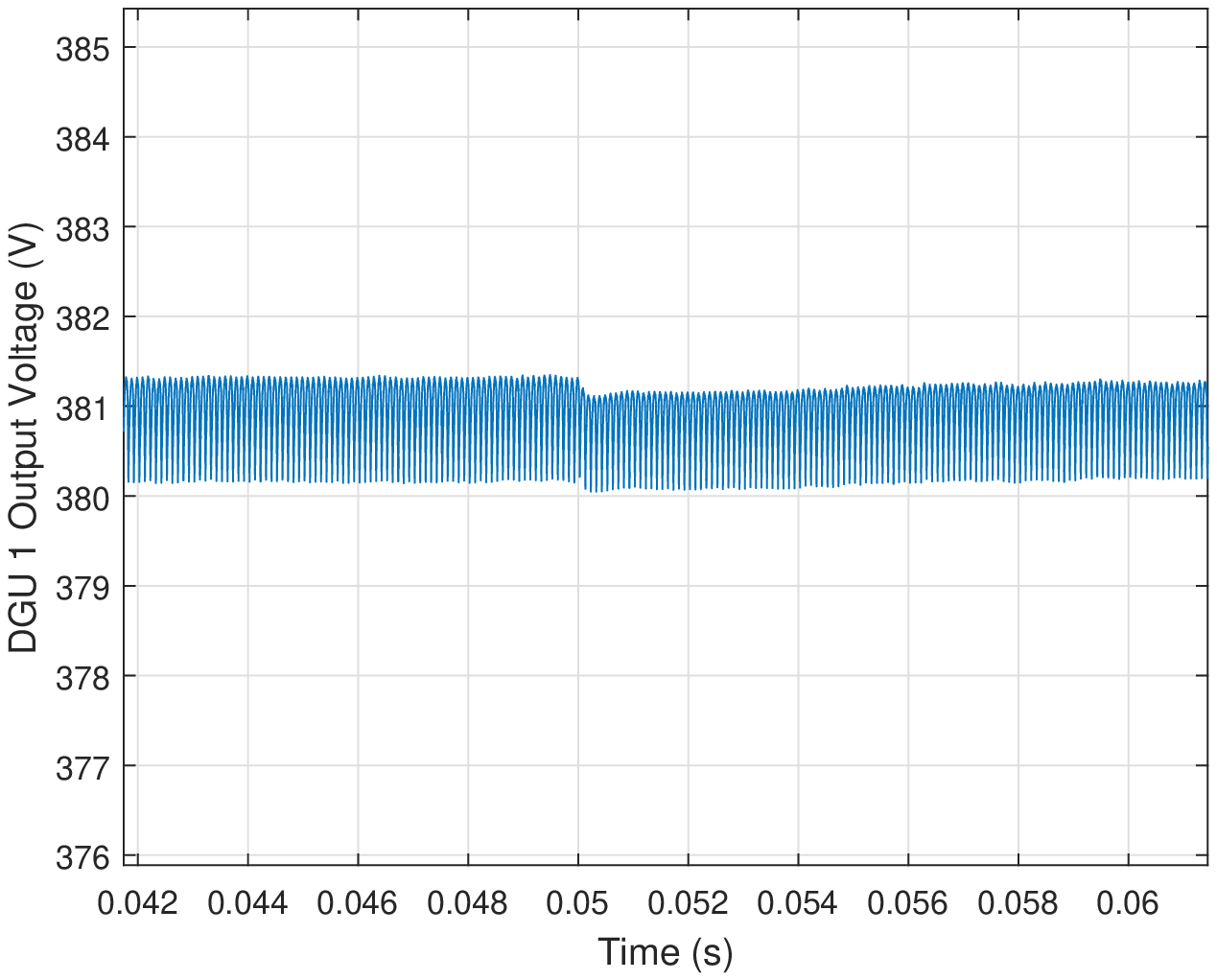}
 \caption{$\hat{\Sigma}_{1}^{\textrm{DGU}}$ output voltage} \label{fig:DGU1PnPDGU6_3}
 \end{subfigure}\hspace*{\fill}
 \begin{subfigure}{0.48\textwidth}
 \includegraphics[width=\linewidth]{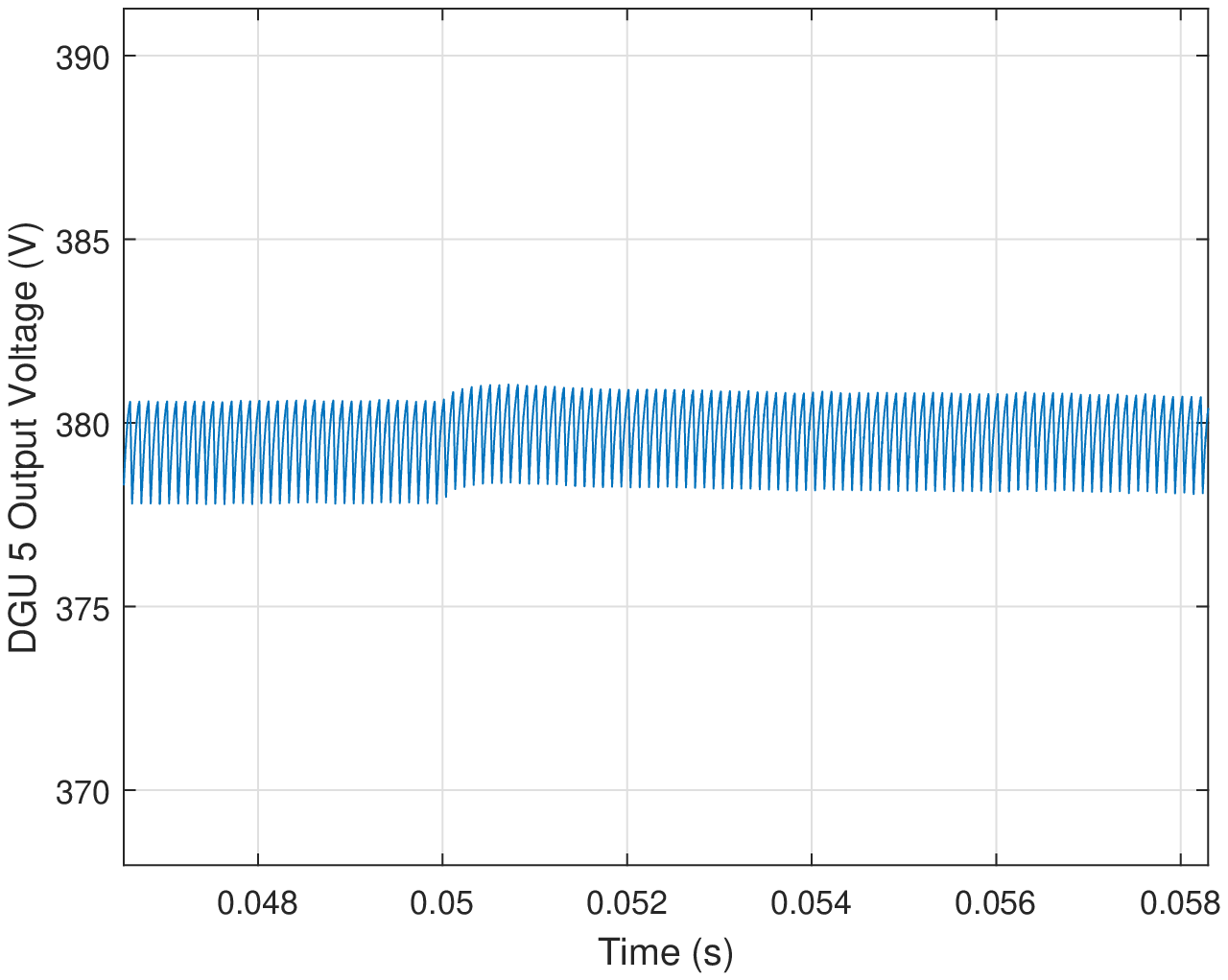}
 \caption{$\hat{\Sigma}_{5}^{\textrm{DGU}}$ output voltage} \label{fig:DGU5PnPDGU6_3}
 \end{subfigure}
 \medskip
 \centering
 \begin{subfigure}{0.48\textwidth}
 \includegraphics[width=\linewidth]{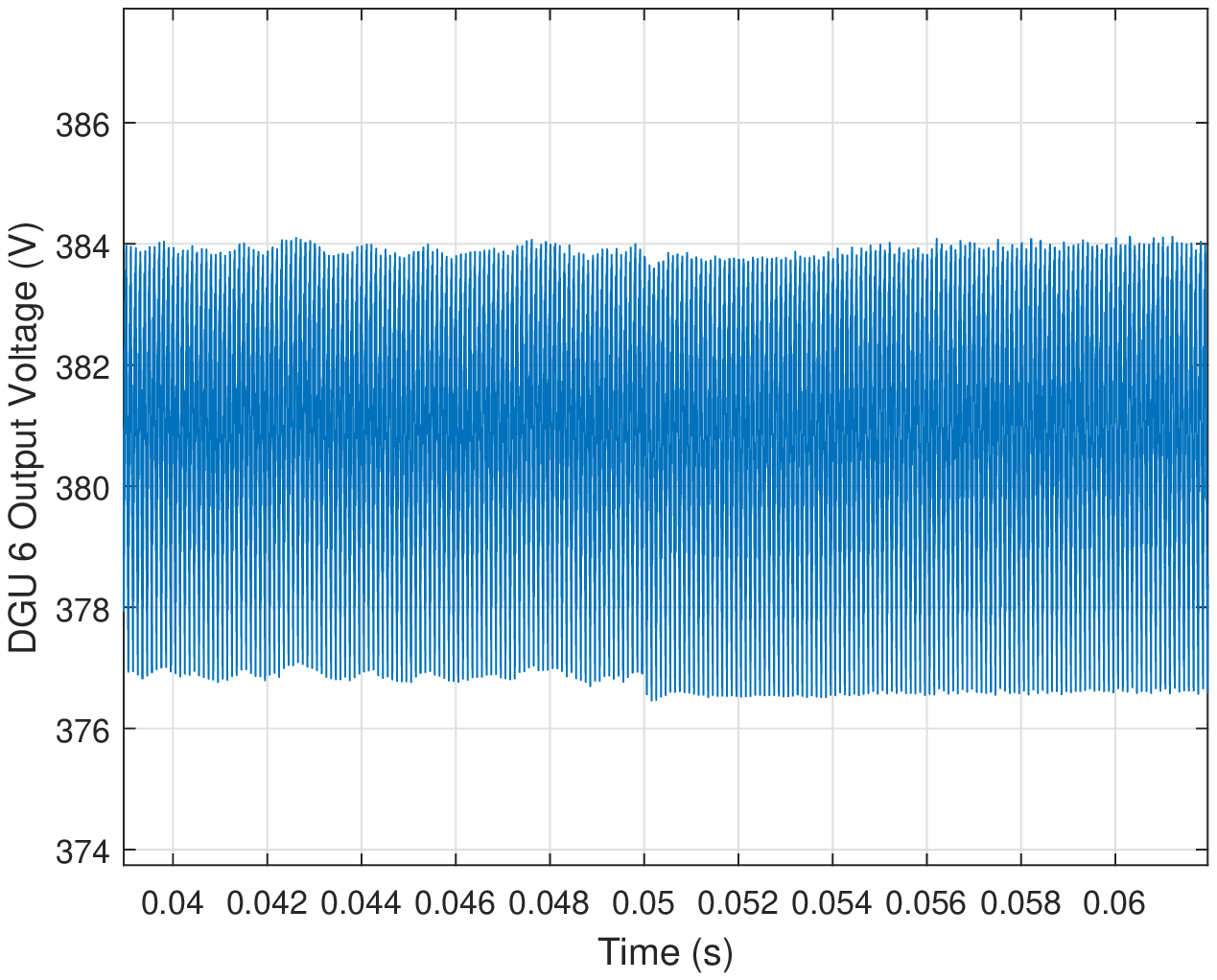}
 \caption{$\hat{\Sigma}_{6}^{\textrm{DGU}}$ output voltage} \label{fig:DGU6PnPDGU6_3}
 \end{subfigure}
 \caption{DGU output voltage responses to $\hat{\Sigma}_{6}^{\textrm{DGU}}$ plug-in with unmodelled parasitic capacitor resistance.}
 \end{figure}

 \section{Conclusion}
 This paper develops a scalable PnP decentralised $\mathcal{L}_1$ adaptive controller for augmentation of DGU baseline voltage controllers within a large-scale DC ImG. These controllers are equipped locally to each DGU at the primary control level and guarantee local asymptotic stability in the presence of parametric and topology uncertainty. Asymptotic stability of the global system can be guaranteed for the decentralised primary control level by adhering to conservative design conditions. Such conditions are determined offline by iteratively checking if the derivative of the overall Lyapunov function candidate is negative definite, or if decoupled terms are more negative definite than coupling terms.
 
 A heterogeneous DC ImG consisting of DC-DC boost converters is designed in Simulink using a radial and meshed topology to evaluate the performance of the proposed architecture. As long as appropriate bounds of uncertainty are incorporated, the $\mathcal{L}_1$AC can treat the DGU as a black-box. The control architecture demonstrates fast and robust output voltage performance when evaluated under PnP operations, unknown load changes, voltage reference step changes, and unmodelled dynamics.
 
 Future work will consider line-independent and distributed control architectures\footnote{Two separate pieces of Work on the theory and implementation of a scalable distributed control architecture that guarantees GAS in a PnP fashion and incorporates $\mathcal{L}_1$ adaptive controllers has been submitted to journals IEEE Transactions on Automatic Control and IEEE Transactions on Smart Grid. Pre-prints are found in \cite{OKeeffe2018a} and \cite{OKeeffe2018c} respectively.} in order to guarantee global asymptotic stability in a scalable, PnP fashion. Furthermore, implementing the proposed architecture in bus-connected topologies with constant-power loads is also of interest.
 
 \section{Appendix}
\subsection{Matrices in Microgrid Model \label{Model Matrices}}
\subsubsection{Two coupled boost converter DGU model}
Defining the power line dynamic equation of (\ref{eq:LineIJ}) in state space form yields,
\begin{equation}
\Sigma_{ij}^{\textrm{Line}}:
\begin{cases}
\dot{x}_{[l_{ij}]}(t) = A_{ll_{ij}}x_{l_{ij}}(t)+A_{li_{ij}}x_{[ij]}(t)
+A_{lj_{ij}}x_{[ij]}(t)
\end{cases}
\label{eq:DGULINE}
\end{equation}
where, $x_{l_{ij}} = I_{ij}$ is the line current state, $A_{li_{ij}} = \left[\begin{array}{cc}
-\frac{1}{L_{ij}} & 0
\end{array} \right]$, $A_{lj_{ij}} = \left[\begin{array}{cc}
\frac{1}{L_{ij}} & 0
\end{array} \right]$,  $A_{ll_{ji}} = -\frac{R_{ij}}{L_{ij}}$. Therefore, the overall state space model of the mG in Fig. \ref{fig:MG2} defined by (\ref{eq:AvMGSS}) prior to the QSL assumption, can be represented by,
\begin{equation}
\begin{aligned}
\underbrace{
\left[\begin{array}{cccc}
 \dot{x}_{[i]}(t)\\
 \dot{x}_{[j]}(t)\\
 \dot{x}_{[l,ij]}(t)\\
 \dot{x}_{[l,ji]}(t)\\
 \end{array}\right]}_{\dot{x}} =
 \underbrace{ \left[\begin{array}{cccc}
A_{ii} & A_{ij} & 0  & 0\\
A_{ji} & A_{jj}  & 0  & 0\\ 
A_{li_{ij}} & A_{lj_{ij}} & A_{ll_{ij}} & 0 \\
A_{li_{ji}} & A_{lj_{ji}} & 0 & A_{ll_{ji}}
 \end{array} \right]}_{A}
 \underbrace{
 \left[\begin{array}{c}
  x_{[i]}(t)\\
  x_{[j]}(t)\\
  x_{[l,ij]}(t)\\
  x_{[l,ji]}(t)\\
  \end{array} \right]}_{x} +
  \underbrace{ \left[\begin{array}{cc}
    B_i & 0\\
    0 & B_j\\
    0 & 0\\
    0 & 0 \\
    \end{array} \right
    ]}_{B}
    \underbrace{
    \left[\begin{array}{cc}
      u_{[i]}(t)\\
      u_{[j]}(t)\\
      \end{array} \right]}_{u} \\+ 
      \underbrace{
      \left[\begin{array}{cc}
          E_i & 0\\
          0 & E_j\\
          0 & 0\\
          0 & 0 \\
          \end{array} \right
          ]}_{M}\left[\begin{array}{cc}
            d_{[i]}(t)\\
            d_{[j]}(t)\\
            \end{array} \right]
      \hspace{15mm}
           \left[\begin{array}{cc}
               y_{[i]}\\
               y_{[j]}
               \end{array} \right
               ] =
               \underbrace{ \left[\begin{array}{cccc}
                 C_i & 0 & 0 & 0\\
                 0 & C_j & 0 & 0\\
                 \end{array} \right]}_{C} \left[\begin{array}{c}
                   x_{[i]}(t)\\
                   x_{[j]}(t
                   \end{array} \right]
                 \end{aligned}
 \end{equation}
where, 

\begin{equation}
\begin{aligned}
A = 
\left[\begin{array}{cccccc}
-\frac{R_{ti}}{L_{ti}} &  -\frac{(1-d_i)}{L_{ti}}& 0 & 0 & 0 & 0\\
\frac{(1-d_j)}{C_{tj}} & 0 & 0 & \frac{1}{R_{ij}C_{ti}} & 0 & 0\\ 
0 & 0 & -\frac{R_{tj}}{L_{tj}} & -\frac{(1-d_i)}{L_{ti}} & 0 & 0\\
0 & \frac{1}{R_{ji}C_{tj}} & \frac{(1-d_j)}{C_{tj}} & 0 & 0 & 0 \\ 
-\frac{1}{L_{ij}} & 0 && \frac{1}{L_{ij}} & -\frac{R_{ij}}{L_{ij}}
& 0 \\
\frac{1}{L_{ji}} & 0 &  -\frac{1}{L_{ij}} & 0 & 0 & -\frac{R_{ji}}{L_{ji}}\end{array} \right]
B = 
\left[\begin{array}{cccccc}
\frac{1}{L_{ti}} & 0 \\ 0 & 0\\
0 & \frac{1}{L_{tj}} \\
0 & 0\\
0 & 0\\
0 & 0\\
\end{array} \right] \\
C = 
\left[\begin{array}{cccccc}
0 & 1 & 0 & 0 & 0 & 0\\
0 & 0 & 0 & 1 & 0 & 0 \\
\end{array} \right]
E = 
\left[\begin{array}{cccccc}
0 & 0 \\ -\frac{1}{C_{ti}} & 0\\
0 & 0 \\
0 & -\frac{1}{C_{tj}}\\
0 & 0\\
0 & 0\\
\end{array} \right]
\end{aligned}
\end{equation} %\todoINFO{Todo: Block matrix} %\todoINFO{Todo: Ensure matrices are correct}
\vspace{3mm}
The $A$ matrix above is block triangular, meaning that stability of the mG in Fig. \ref{fig:MG2} is dependent on the union of $\left[\begin{array}{cc}A_{ii} & A_{ij} \\
A_{ij} & A_{jj}\end{array} \right]$, $A_{ll_{ij}}$ and $A_{ll_{ji}}$. As the line dynamics are asymptotically stable by virtue of positive line resistance and inductance, stability of the overall global model is exclusively dependent on the stability of local DGUs interconnected via the QSL model of (\ref{eq:DGUSS2}). Hence the QSL model is justified.

\subsubsection{Global mG model with \textit{N} DGUs}
From section,
\begin{equation}
\underbrace{\left[\begin{array}{c}
\dot{x}_{[1]} \\
\dot{x}_{[2]} \\
\dot{x}_{[3]} \\
\vdots \\
\dot{x}_{[N]} 
\end{array} \right]}_{\dot{\textbf{x}}}=
\underbrace{
\left[\begin{array}{ccccc}
A_{11} & A_{12} & A_{13}  & \cdots & A_{1N} \\
A_{21} & A_{22} & A_{23}  & \cdots & A_{2N} \\ 
A_{31} & A_{32} &  A_{33}  & \cdots & A_{3N} \\
\vdots & \vdots & \vdots & \ddots & \vdots
\\
A_{N1} & A_{N2} & A_{N3} & \cdots & A_{NN}
\end{array} \right]}_{\textbf{A}}
\underbrace{\left[\begin{array}{c}
x_{[1]} \\
x_{[2]} \\
x_{[3]} \\
\vdots \\
x_{[N]} 
\end{array} \right]}_{\textbf{x}} + \underbrace{
\left[\begin{array}{ccccc}
B_{1} & 0 & 0 & \cdots  & 0 \\
0 & B_{2} & 0 & \ddots  & \vdots  \\ 
0 & 0 &  B_{3} & \ddots & \vdots \\
\vdots & \ddots & \ddots & \cdots & 0
\\
0 & \cdots & 0 & 0 & B_{N} 
\end{array} \right]}_{\textbf{B}}
\underbrace{\left[\begin{array}{ccccc}
u_{[1]} \\
u_{[2]} \\
u_{[3]} \\
\vdots \\
u_{[N]} 
\end{array} \right]}_{\textbf{u}} 
\\
+ \underbrace{
\left[\begin{array}{ccccc}
E_{1} & 0 & 0 & \cdots  & 0 \\
0 & E_{2} & 0 & \ddots  & \vdots  \\ 
0 & 0 &  E_{3} & \ddots & \vdots \\
\vdots & \ddots & \ddots & \cdots & 0
\\
0 & \cdots & 0 & 0 & E_{N} 
\end{array} \right]}_{\textbf{E}}
\underbrace{\left[\begin{array}{ccccc}
d_{[1]} \\
d_{[2]} \\
d_{[3]} \\
\vdots \\
d_{[N]} 
\end{array} \right]}_{\textbf{d}} 
\hspace{2mm}
;
\hspace{2mm}
\underbrace{\left[\begin{array}{ccccc}
y_{[1]} \\
y_{[2]} \\
y_{[3]} \\
\vdots \\
y_{[N]} 
\end{array} \right]}_{\textbf{y}} = 
\underbrace{
\left[\begin{array}{ccccc}
C_{1} & 0 & 0 & \cdots  & 0 \\
0 & C_{2} & 0 & \ddots  & \vdots  \\ 
0 & 0 &  C_{3} & \ddots & \vdots \\
\vdots & \ddots & \ddots & \cdots & 0
\\
0 & \cdots & 0 & 0 & C_{N} 
\end{array} \right]}_{\textbf{C}} 
\underbrace{\left[\begin{array}{c}
x_{[1]} \\
x_{[2]} \\
x_{[3]} \\
\vdots \\
x_{[N]} 
\end{array} \right]}_{\textbf{x}}
\end{equation}
\subsection{A Glimpse at Instability/Stability Due to Converter Interaction Using Baseline Controllers only/Augmenting $\mathcal{L}_1$ Adaptive Controllers} \label{Instability}

The following 6 DGU DC ImG topology is used to demonstrate that decentralised baseline controllers, designed to be locally stable without accounting for interactions, can destabilise the global mG when DGUs are indeed interconnected.

Consider DGUs with the dynamics of (\ref{eq:DGUSSCL}). Electrical parameters of Table \ref{table:parameters} are used here. Decentralised baseline controllers are designed for each DGU assuming they are dynamically decoupled i.e. $\hat{\zeta}_{i}(t) = \hat{\zeta}_{j}(t) = 0$. State feedback controllers are again defined as in (\ref{eq:SFBL}).

\begin{figure}[!htb]    
\graphicspath{ {Images/} }
\centering
\includegraphics[width=8.5cm]{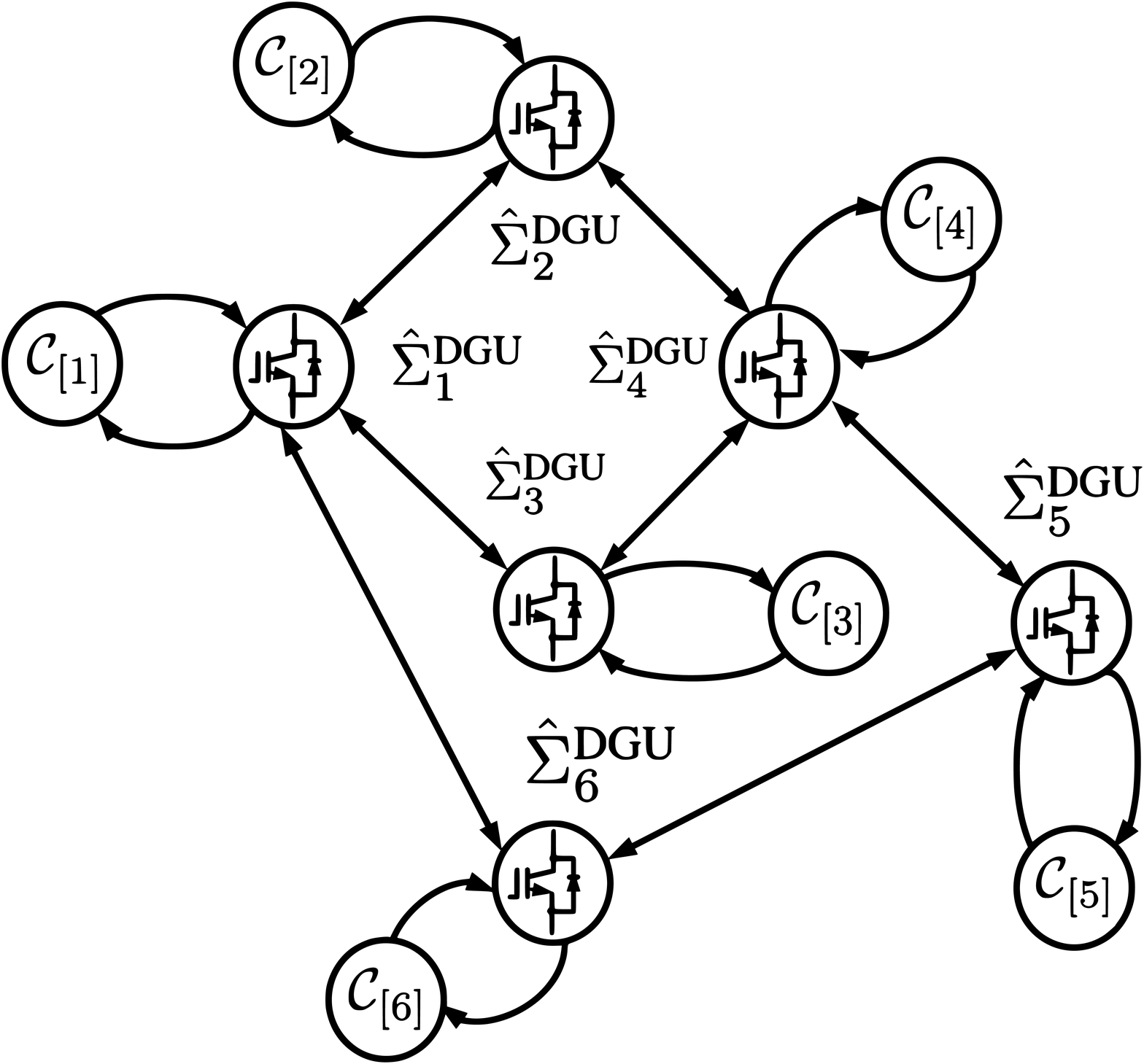}
\caption{Meshed and radial microgrid configuration with baseline controllers only - $\hat{\Sigma}_{6}^{\textrm{DGU}}$ plug-in.}
\label{fig:DGU6PnPw/oL1}
\end{figure}
At start-up, $\hat{\Sigma}_{1}^{\textrm{DGU}}, \hat{\Sigma}_{2}^{\textrm{DGU}}, \hat{\Sigma}_{3}^{\textrm{DGU}}$ and $\hat{\Sigma}_{4}^{\textrm{DGU}}$ are connected together through $RL$ power lines in a radial configuration. $ \hat{\Sigma}_{5}^{\textrm{DGU}}$ is connected to $ \hat{\Sigma}_{4}^{\textrm{DGU}}$, while $ \hat{\Sigma}_{6}^{\textrm{DGU}}$ powers a local load on its own. Controllers are designed to ensure local asymptotic stability of the closed-loop DGUs, implying that the eigenvalues of decoupled global linear representation of the decoupled DGUs is also asymptotically stable,
\begin{equation}
\underbrace{
\left[\begin{array}{cccccc}
\hat{A}_{11}^C & \underbar{0}_{3\text{x}3} & \underbar{0}_{3x3} & \underbar{0}_{3x3} & \underbar{0}_{3x3} & \underbar{0}_{3x3}\\
\underbar{0}_{3x3} & \hat{A}_{22}^C & \underbar{0}_{3x3} & \underbar{0}_{3x3} & \underbar{0}_{3x3} & \underbar{0}_{3x3}  \\ 
\underbar{0}_{3x3} & \underbar{0}_{3x3} & \hat{A}_{33}^C &\underbar{0}_{3x3} & \underbar{0}_{3x3} & \underbar{0}_{3x3} \\
\underbar{0}_{3x3} & \underbar{0}_{3x3} & \underbar{0}_{3x3} & \hat{A}_{44}^C & \underbar{0}_{3x3} & \underbar{0}_{3x3}
\\
\underbar{0}_{3x3} & \underbar{0}_{3x3} & \underbar{0}_{3x3} & \underbar{0}_{3x3} & \hat{A}_{55}^C & \underbar{0}_{3x3} \\ 
\underbar{0}_{3x3} & \underbar{0}_{3x3} & \underbar{0}_{3x3} & \underbar{0}_{3x3} & \underbar{0}_{3x3} & \hat{A}_{66}^C
\end{array} \right]}_{\textbf{A}_{CL}^{D}}
\end{equation}
where,
\begin{equation}
\hat{A}_{ii}^C = \left[\begin{array}{cc}
A_{ii} -  B_{i}K_{bl}^x & B_{i}K_{bl}^\xi \\
-C_{i} & 0
\end{array} \right]
\end{equation}
and $K_{bl}^x = [K_{bl}^i, K_{bl}^v]$. While baseline controller gains are tuned for decoupled, load-dependent DGUs, $A_{ii}$ models the line-dependent coupled DGUs as in (\ref{eq:DGUSS2}). However, the dynamic coupling at start-up of DGUs in Fig. (\ref{fig:DGU6PnPw/oL1}) means the global mG is linearly represented by the state matrix, 
\begin{equation}
\underbrace{
\left[\begin{array}{cccccc}
\hat{A}_{11} & \hat{A}_{12} & \hat{A}_{13} & \underbar{0}_{3x3} & \underbar{0}_{3x3} & \underbar{0}_{3x3}\\
\hat{A}_{21} & \hat{A}_{22} & \underbar{0}_{3x3} & \underbar{0}_{3x3} & \hat{A}_{24} & \underbar{0}_{3x3}  \\ 
\hat{A}_{31} & \underbar{0}_{3x3} & \hat{A}_{33} &\hat{A}_{34} & \underbar{0}_{3x3} & \underbar{0}_{3x3} \\
\underbar{0}_{3x3} & \hat{A}_{42} & \hat{A}_{43} & \hat{A}_{44} & \hat{A}_{45} & \underbar{0}_{3x3}
\\
\underbar{0}_{3x3} & \underbar{0}_{3x3} & \underbar{0}_{3x3} & \hat{A}_{54} & \hat{A}_{55} & \underbar{0}_{3x3} \\ 
\underbar{0}_{3x3} & \underbar{0}_{3x3} & \underbar{0}_{3x3} & \underbar{0}_{3x3} & \underbar{0}_{3x3} & \hat{A}_{66}
\end{array} \right]}_{\textbf{A}_{CL}^{C}}
\label{eq:AclCoup}
\end{equation}
Plotting the eigenvalues of both $\textbf{A}_{CL}^{D}$ and $\textbf{A}_{CL}^{C}$,
\begin{figure}[!htb]    
\graphicspath{ {Images/} }
\centering
\includegraphics[width=7.5cm]{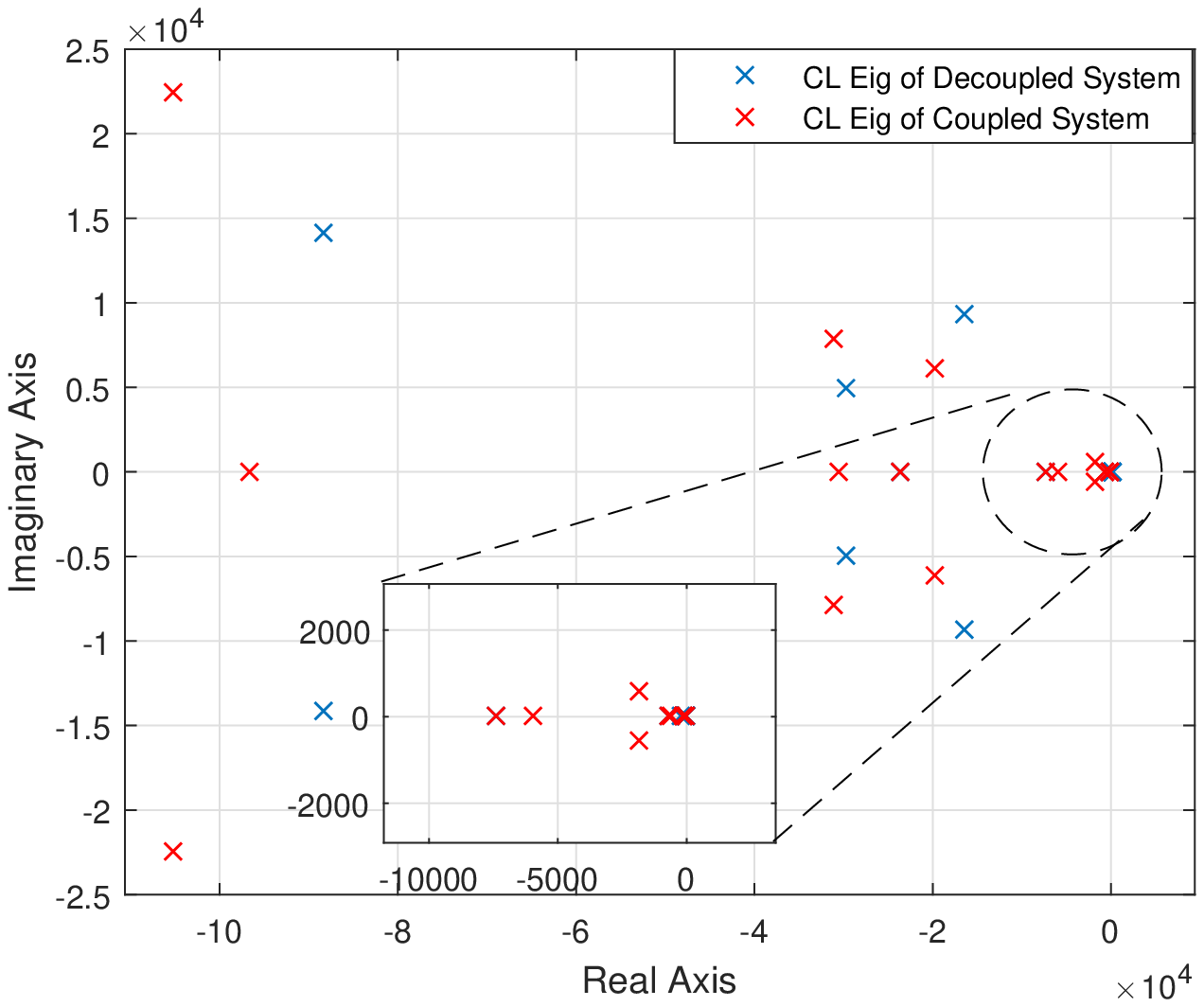}
\caption{Eigenvalues at start-up of $\textbf{A}_{CL}^{D}$ and $\textbf{A}_{CL}^{C}$.}
\label{fig:Start-upEig}
\end{figure}

Though global asymptotic stability cannot be guaranteed at start-up through the use of decentralised controllers, Fig. \ref{fig:Start-upEig} shows that the eigenvalues of the coupled linear system in (\ref{eq:AclCoup}) are in the left-half plane, resulting in a globally stable configuration. The interconnection of DGUs does however reduce the damping within the system as eigenvalues move towards the imaginary axis when compared to the eigenvalues of the decoupled system representation.   

When $ \hat{\Sigma}_{6}^{\textrm{DGU}}$ is plugged-in, connecting with $ \hat{\Sigma}_{1}^{\textrm{DGU}}$ and $ \hat{\Sigma}_{5}^{\textrm{DGU}}$, the global closed-loop system  representation changes to,
\begin{equation}
\underbrace{
\left[\begin{array}{cccccc}
\hat{A}_{11} & \hat{A}_{12} & \hat{A}_{13} & \underbar{0}_{3x3} & \underbar{0}_{3x3} & \hat{A}_{16}\\
\hat{A}_{21} & \hat{A}_{22} & \underbar{0}_{3x3} & \underbar{0}_{3x3} & \hat{A}_{24} & \underbar{0}_{3x3}  \\ 
\hat{A}_{31} & \underbar{0}_{3x3} & \hat{A}_{33} &\hat{A}_{34} & \underbar{0}_{3x3} & \underbar{0}_{3x3} \\
\underbar{0}_{3x3} & \hat{A}_{42} & \hat{A}_{43} & \hat{A}_{44} & \hat{A}_{45} & \underbar{0}_{3x3}
\\
\underbar{0}_{3x3} & \underbar{0}_{3x3} & \underbar{0}_{3x3} & \hat{A}_{54} & \hat{A}_{55} & \hat{A}_{56} \\ 
\hat{A}_{61} & \underbar{0}_{3x3} & \underbar{0}_{3x3} & \underbar{0}_{3x3} & \hat{A}_{65} & \hat{A}_{66}
\end{array} \right]}_{\textbf{A}_{CL}^{C DGU 6}}
\label{eq:AclCoup2}
\end{equation} 
Plotting the eigenvalues of $\textbf{A}_{CL}^{C DGU 6}$ against the eigenvalues plotted in Fig. \ref{fig:Start-upEig},
\begin{figure}[!htb] 
\graphicspath{ {Images/} }
\centering
\includegraphics[width=7.5cm]{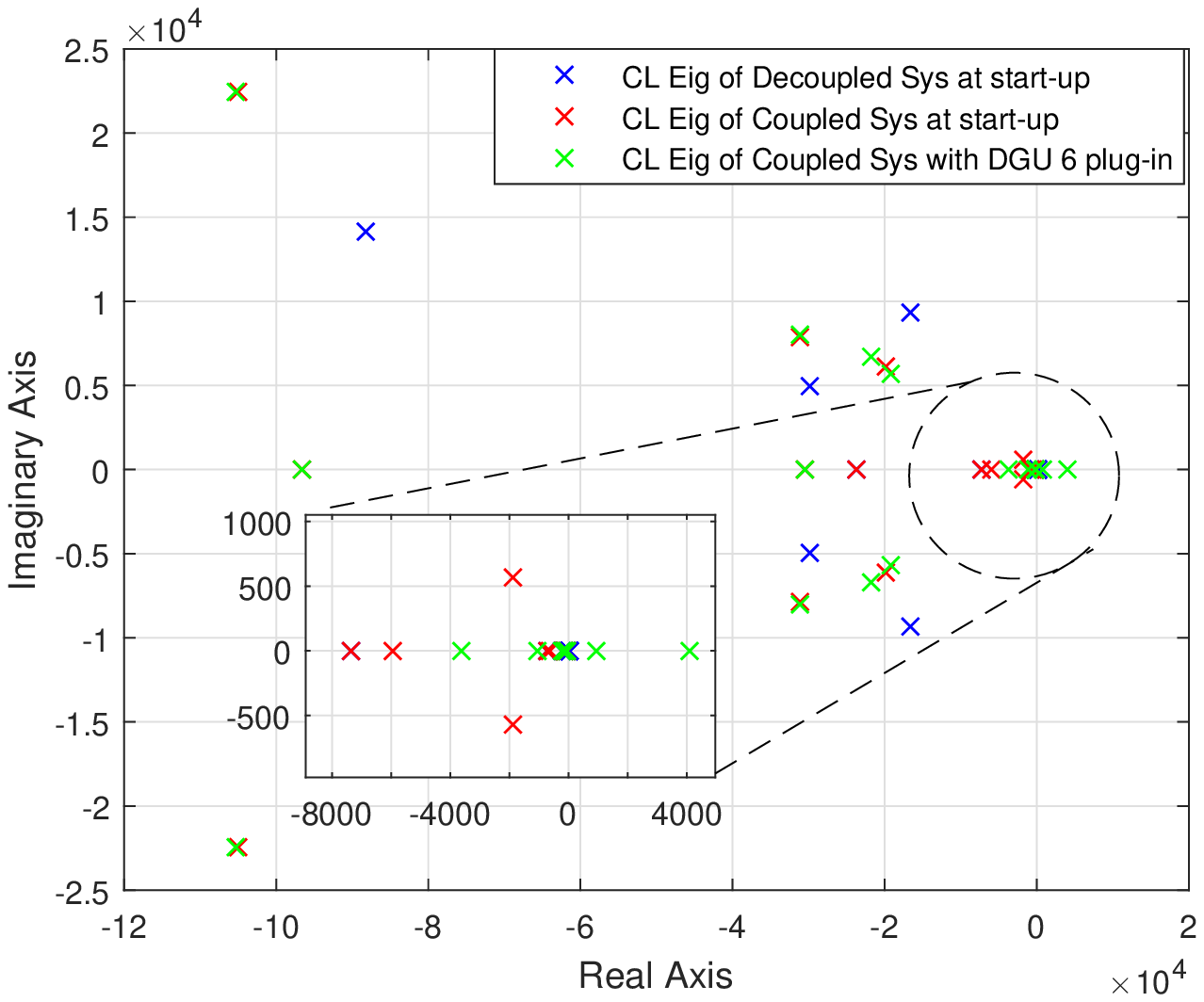}
\caption{Eigenvalues of $\textbf{A}_{CL}^{D}$ , $\textbf{A}_{CL}^{C}$ at start-up, and $\textbf{A}_{CL}^{C DGU 6}$ after $ \hat{\Sigma}_{6}^{\textrm{DGU}}$ plug-in.}
\label{fig:EigwDGU6}
\end{figure}
From Fig. (\ref{fig:EigwDGU6}), the addition of $ \hat{\Sigma}_{6}^{\textrm{DGU}}$ moves eigenvalues of the linear global system into the right-half plane, resulting in a globally unstable mG. Type III baseline controllers (each with their own tuning/bandwidths) are used to demonstrate that the non-linear switching model also becomes unstable when $ \hat{\Sigma}_{6}^{\textrm{DGU}}$ plugs into the system. The output voltage of each DGU is plotted below,

\begin{figure}[!htb] 
\graphicspath{ {Images/} }
\centering
\includegraphics[width=8.5cm]{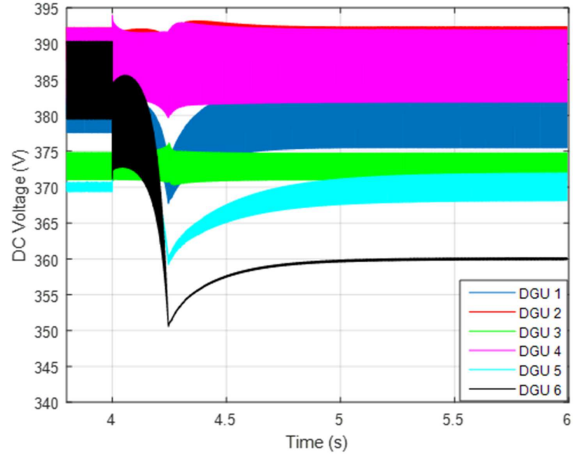}
\caption{DGU Output Voltages with $\hat{\Sigma}_{6}^{\textrm{DGU}}$ plug-in.}
\label{fig:UnstablewDGU6}
\end{figure}

At 4 seconds, $\hat{\Sigma}_{6}^{\textrm{DGU}}$ plugs-in, connecting to $\hat{\Sigma}_{1}^{\textrm{DGU}}$ and $\hat{\Sigma}_{5}^{\textrm{DGU}}$. Though $\hat{\Sigma}_{6}^{\textrm{DGU}}$ maintains a steady-state voltage at 360 V, it loses reference tracking (i.e. 385 V). Additionally, while Fig. \ref{fig:UnstablewDGU6} does
not indicate instability since other voltages in the grid maintain their voltage references, on closer inspection, the duty cycle of $\hat{\Sigma}_{6}^{\textrm{DGU}}$ clearly becomes unstable i.e. exponentially increasing to
infinity. 

\begin{figure}[!htb] 
\graphicspath{ {Images/} }
\centering
\includegraphics[width=8.5cm]{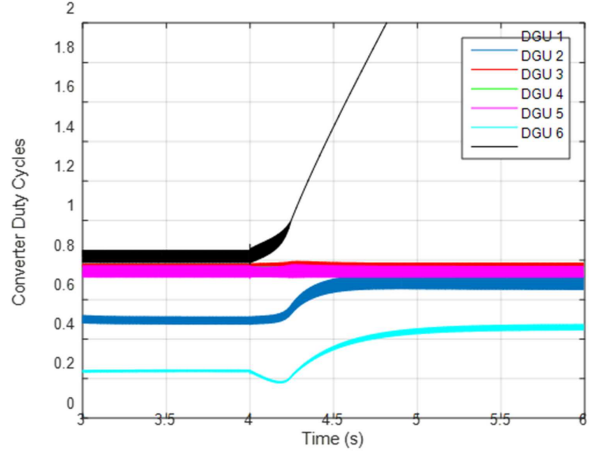}
\caption{DGU duty cycles with $\hat{\Sigma}_{6}^{\textrm{DGU}}$ plug-in.}
\label{fig:DutyCyclesUnstablewDGU6}
\end{figure}

Ultimately, as the duty cycle of $\hat{\Sigma}_{6}^{\textrm{DGU}}$ increases, its steady-state output voltage should increase as well (i.e. boost converter steady-state output voltage gain: $\frac{V_{in_{6}}}{1-D_6}$). Fig. \ref{fig:DutyCyclesUnstablewDGU6} shows $\hat{\Sigma}_{1}^{\textrm{DGU}}$ and $\hat{\Sigma}_{5}^{\textrm{DGU}}$ increasing their duty cycles in order to accommodate the fact that $\hat{\Sigma}_{6}^{\textrm{DGU}}$ is unstable by injecting more current to power the
load connected to $\hat{\Sigma}_{6}^{\textrm{DGU}}$.

Finally, the global state-space model of the system when implementing augmenting $\mathcal{L}_1$ACs and using assumption 4 is,
\begin{equation}
\underbrace{
\left[\begin{array}{cccccc}
\hat{A}_{m} & \hat{A}_{12} & \hat{A}_{13} & \underbar{0}_{3x3} & \underbar{0}_{3x3} & \hat{A}_{16}\\
\hat{A}_{21} & \hat{A}_{m} & \underbar{0}_{3x3} & \underbar{0}_{3x3} & \hat{A}_{24} & \underbar{0}_{3x3}  \\ 
\hat{A}_{31} & \underbar{0}_{3x3} & \hat{A}_{m} &\hat{A}_{34} & \underbar{0}_{3x3} & \underbar{0}_{3x3} \\
\underbar{0}_{3x3} & \hat{A}_{42} & \hat{A}_{43} & \hat{A}_{m} & \hat{A}_{45} & \underbar{0}_{3x3}
\\
\underbar{0}_{3x3} & \underbar{0}_{3x3} & \underbar{0}_{3x3} & \hat{A}_{54} & \hat{A}_{m} & \hat{A}_{56} \\ 
\hat{A}_{61} & \underbar{0}_{3x3} & \underbar{0}_{3x3} & \underbar{0}_{3x3} & \hat{A}_{65} & \hat{A}_{m}
\end{array} \right]}_{\textbf{A}_{CL}^{C DGU 6+\mathcal{L}_1}}
\label{eq:AclCoupWithL1}
\end{equation} 
Plotting the eigenvalues of (\ref{eq:AclCoupWithL1}) shows that once adaptation yields convergence to desired dynamics and local asymptotic stability then global stability can be guaranteed for the system in Fig. \ref{fig:DGU6PnPw/oL1}, i.e. global eigenvalues are in left-half plane.
\begin{figure}[!htb] 
\graphicspath{ {Images/} }
\centering
\includegraphics[width=8.5cm]{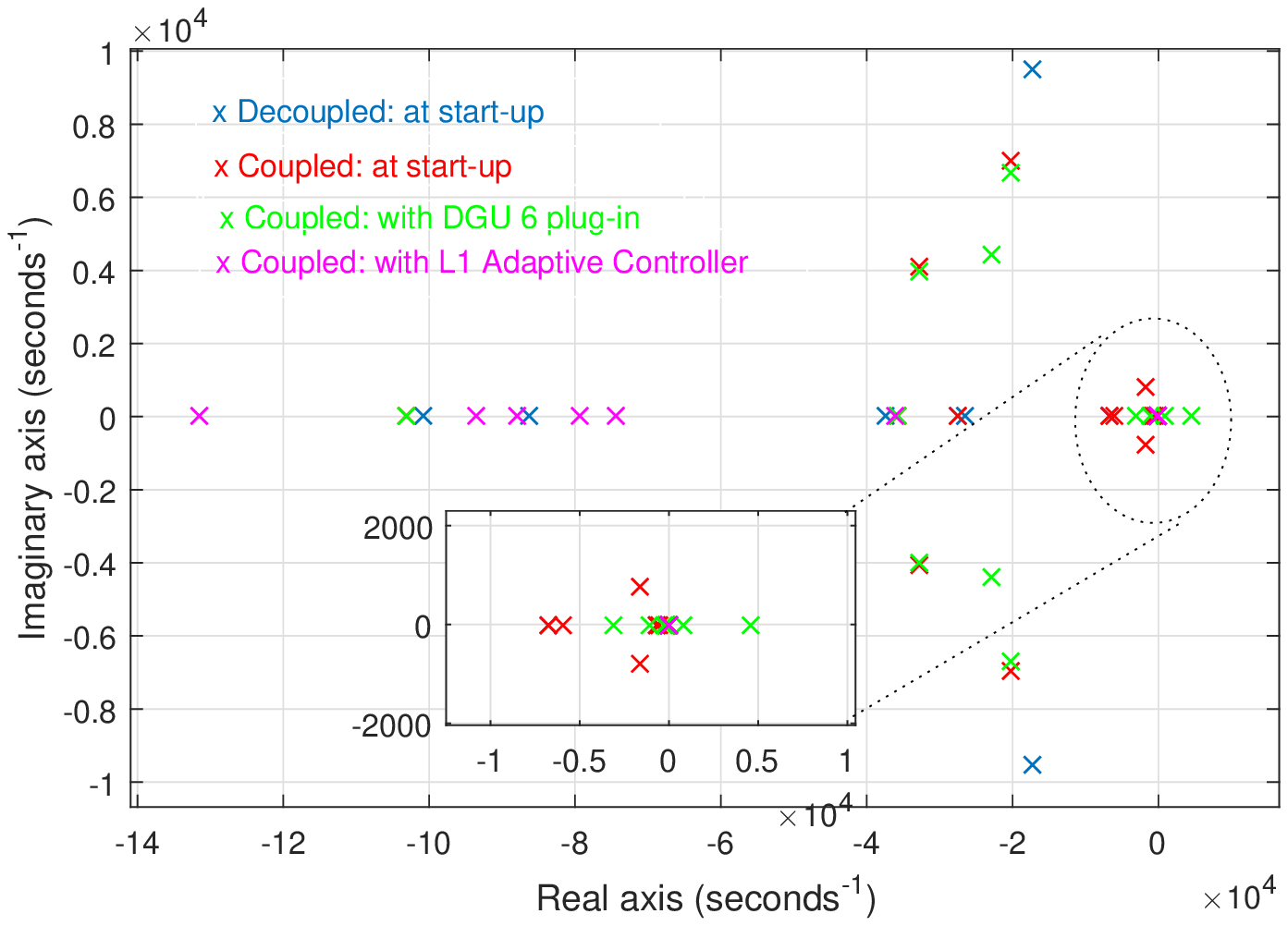}
\caption{Eigenvalues of $\textbf{A}_{CL}^{D}$ , $\textbf{A}_{CL}^{C}$, $\textbf{A}_{CL}^{C DGU 6}$ and $\textbf{A}_{CL}^{C DGU 6+\mathcal{L}_1}$.}
\label{fig:FinalEigPlot}
\end{figure}
\subsection{State-Space Model of Interconnected Boost Converters with Capacitor Equivalent Series Resistance}
In this section, the local model of a coupled boost converter is derived when the ESR of the output capacitor is included. This highlights how the model changes for investigating robustness to unmodelled dynamics in section \ref{sec:unmod}. The boost converter DGU model is shown below.
\begin{figure}[!htb] 
\graphicspath{ {Images/} }
\centering
\includegraphics[width=12.5cm]{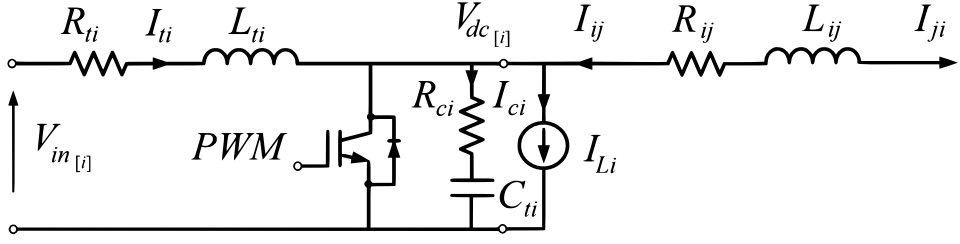}
\caption{Coupled boost converter DGU with Output Capacitor ESR.}
\label{fig:EsrCapDiag}
\end{figure}

Using QSL approximations of section \ref{sec:QSL}, the differential equations during the on-time PWM switching are,
\begin{equation}
\begin{cases}
	     	     \dfrac{d I_{ti}}{dt} = \dfrac{ 1}{L_{ti}}V_{in_{i}} - \dfrac{R_{ti}}{L_{ti}} I_{ti} \\
	     	     \\
	     \dfrac{dV_{dc_{i}}}{dt} =  \sum_{j\in\mathcal{N}_i} \left(\frac{V_{dc_{j}}-V_{dc_{i}}}{R_{ij}C_{ti}}\right) - \dfrac{1}{C_{ti}}I_{Li}
	     \end{cases}
	     \label{eq:on}
\end{equation}
which in state-space form can be written as,
\begin{equation}
\dot{x}_{{on}_{[i]}}(t) = A_{{on}_{ii}}x_{{on}_{[i]}}(t)+B_{{on}_{i}}u_{{on}_{[i]}}(t)+E_{{on}_{i}}d_{{on}_{[i]}}(t)+\Sigma_{j\in\mathcal{N}_i}A_{{on}_{ij}}x_{{on}_{[j]}}(t)
\label{eqn:onstate}
\end{equation}
where,
\begin{equation}
A_{{on}_{ii}}=
\left[ \begin{array}{cc}
-\frac{R_{ti}}{L_{ti}} & 0\\
0 & -\Sigma_{j\in\mathcal{N}_i}\frac{1}{R_{ij}C_{ti}}
\end{array}\right]
B_{{on}_{i}}=
\left[ \begin{array}{cc}
\frac{V_{in_{i}}}{L_{ti}} \\
0
\end{array}\right]
E_{{on}_{i}}=
\left[ \begin{array}{cc}
0\\
-\frac{1}{C_{ti}}
\end{array}
\right]
A_{{on}_{ii}}=
\left[ \begin{array}{cc}
0 & 0\\
0 & \Sigma_{j\in\mathcal{N}_i}\frac{1}{R_{ij}C_{ti}}
\end{array}\right]
\end{equation}
The differential equations during the off-time PWM switching are,
\begin{equation}
\begin{cases}
	     	     \dfrac{d I_{ti}}{dt} = \dfrac{ 1}{L_{ti}}V_{in_{i}} - \dfrac{R_{ti}}{L_{ti}} I_{ti} - \dfrac{ 1}{L_{ti}}V_{c_{i}}- \dfrac{R_{ci}}{L_{ti}}I_{ci}\\
	     	     \\
	     \dfrac{dV_{dc_{i}}}{dt} = \dfrac{1}{C_{ti}}I_{ti}+ \sum_{j\in\mathcal{N}_i} \left(\frac{V_{dc_{j}}-V_{dc_{i}}}{R_{ij}C_{ti}}\right) - \dfrac{1}{C_{ti}}I_{Li}
	     \end{cases}
	     \label{eq:off}
\end{equation}
From Kirchoff's current law, $I_{ci} = I_{ti}+I_{ij}-I_{Li} = I_{ti}+\left(\frac{V_{dc_{j}}-V_{dc_{i}}}{R_{ij}C_{ti}}\right)-I_{Li}$. Therefore, (\ref{eq:off}) can be written as,
\begin{equation}
\begin{cases}
	     	     \dfrac{d I_{ti}}{dt} = \dfrac{ 1}{L_{ti}}V_{in_{i}} - \dfrac{R_{ti}}{L_{ti}} I_{ti} - \dfrac{ 1}{L_{ti}}V_{c_{i}}- \dfrac{R_{ci}}{L_{ti}}\left(-I_{ti}+\sum_{j\in\mathcal{N}_i}\left(\frac{V_{dc_{i}}-V_{dc_{j}}}{R_{ij}C_{ti}}\right)+I_{Li}\right)\\
	     	     \\
	     \dfrac{dV_{dc_{i}}}{dt} = \dfrac{1}{C_{ti}}I_{ti}+ \sum_{j\in\mathcal{N}_i} \left(\frac{V_{dc_{i}}-V_{dc_{j}}}{R_{ij}C_{ti}}\right) - \dfrac{1}{C_{ti}}I_{Li}
	     \end{cases}
	     \label{eq:off2}
\end{equation}
In state-space form, (\ref{eq:off2}) can be written as,
\begin{equation}
\dot{x}_{{off}_{[i]}}(t) = A_{{off}_{ii}}x_{{off}_{[i]}}(t)+B_{{off}_{i}}u_{{off}_{[i]}}(t)+E_{{off}_{i}}d_{{on}_{[i]}}(t)+\Sigma_{j\in\mathcal{N}_i}A_{{off}_{ij}}x_{{off}_{[j]}}(t)
\label{eqn:offstate}
\end{equation}
where,
\begin{equation}
A_{{off}_{ii}}=
\left[ \begin{array}{cc}
-\frac{(R_{ti}+R_{ci})}{L_{ti}} & -\frac{(1+R_{ci})}{L_{ti}}\\
\frac{1}{C_{ti}} & -\Sigma_{j\in\mathcal{N}_i}\frac{1}{R_{ij}C_{ti}}
\end{array}\right]
B_{{off}_{i}}=
\left[ \begin{array}{cc}
\frac{V_{in_{i}}}{L_{ti}} \\
0
\end{array}\right]
E_{{off}_{i}}=
\left[ \begin{array}{cc}
\frac{R_{ci}}{L_{ti}}\\
-\frac{1}{C_{ti}}
\end{array}
\right]
A_{{off}_{ii}}=
\left[ \begin{array}{cc}
0 & -\frac{R_{ci}}{R_{ij}L_{ti}}\\
0 & \frac{1}{R_{ij}C_{ti}}
\end{array}\right]
\end{equation}
Combining (\ref{eqn:onstate}) and (\ref{eqn:offstate}) to form the average model via $\bar{\dot{x}}_{[i]}(t) = \dot{x}_{on_{[i]}}(t)d_i + \dot{x}_{on_{[i]}}(t)(1-d_i)$, where $d_i$ is the duty-cycle, yields,
\begin{equation}
\begin{aligned}
\dot{\bar{x}}_{[i]}=
\left[ \begin{array}{cc}
-\frac{(R_{ti}+(1-d_i)R_{ci})}{L_{ti}} & -\frac{(1-d_i)(1+R_{ci})}{L_{ti}}\\
\frac{(1-d_i)}{C_{ti}} & -\Sigma_{j\in\mathcal{N}_i}\frac{1}{R_{ij}C_{ti}}
\end{array}\right]\bar{x}_{[i]}+
\left[ \begin{array}{cc}
\frac{1}{L_{ti}} \\
0
\end{array}\right]V_{in_{i}}+
\left[ \begin{array}{cc}
\frac{(1-d_i)R_{ci}}{L_{ti}}\\
-\frac{1}{C_{ti}}
\end{array}
\right]\bar{d}_{[i]}\\+\sum_{j\in\mathcal{N}_i}
\left[ \begin{array}{cc}
0 & -\frac{(1-d_i)R_{ci}}{R_{ij}L_{ti}}\\
0 & \frac{1}{R_{ij}C_{ti}}
\end{array}\right]\bar{x}_{[j]}
\end{aligned}
\label{eqn:avMod}
\end{equation}
Due to the bilinear terms between states $I_{ti}$ and $V_{dc_{i}}$, and the duty-cycle control input, making the average model non-linear, the average model of (\ref{eqn:avMod}) requires linearising to form the small-signal model. Therefore, each signal is separated into its steady-state and small-signal quantities, i.e. $d_i = D_i+d_{i}^{ac}$, $\bar{x}_{[i]} = x_{[i]}+x_{[i]}^{ac} $ etc. Finally, the linear state-space model for DGU i coupled to $N$ neighbours can be written as,
\begin{equation}
\begin{aligned}
\dot{x}_{[i]}^{ac} =
\left[ \begin{array}{cc}
-\frac{(R_{ti}+(1-D_i)R_{ci})}{L_{ti}} & -\frac{(1-d_i)}{L_{ti}}(1-R_{ci}\Sigma_{j\in\mathcal{N}_i}\frac{1}{R_{ij}})\\
\frac{(1-D_i)}{C_{ti}} & -\Sigma_{j\in\mathcal{N}_i}\frac{1}{R_{ij}C_{ti}}
\end{array}\right]x_{[i]}^{ac} +
\left[\begin{array}{cc}
\frac{1}{L_{ti}}(V_{dc_{i}}+R_{ci}(I_{ti}+I_{ij}-I_{Li})) \\
-\frac{I_{ti}}{C_{ti}}
\end{array}\right]u_{[i]}\\+
\left[ \begin{array}{cc}
\frac{(1-D_i)R_{ci}}{L_{ti}}\\
-\frac{1}{C_{ti}}
\end{array}
\right]d_{[i]]}^{ac} +\sum_{j\in\mathcal{N}_i}
\left[ \begin{array}{cc}
0 & -\frac{(1-D_i)R_{ci}}{R_{ij}L_{ti}}\\
0 & \frac{1}{R_{ij}C_{ti}}
\end{array}\right]x_{[j]}^{ac}
\end{aligned}
\label{eqn:tildeMod}
\end{equation}

\subsection{Simulation results using relatively large DGU output voltages}\label{SimLargeV}
This section demonstrates similar results when DGU output voltages differ by a relatively large amount. Without droop control or coordinated secondary control, DGUs with larger output voltage (i.e. larger duty cycles) tend to provide most of the power to the mG, and 'overpower' neighbouring DGUs. Like a see-saw, DGUs with larger voltage (i.e. larger force) will push surplus power through coupling $RL$ lines to 'help' power neighbouring load. The greater the voltage difference between DGU outputs, the greater the 'overpowering' effect.

\subsubsection{Plug-and-Play Operations}
 At $t$ = 0.05 s, $\hat{\Sigma}_{6}^{\textrm{DGU}}$ is plugged-in, connecting to $\hat{\Sigma}_{1}^{\textrm{DGU}}$ and $\hat{\Sigma}_{5}^{\textrm{DGU}}$. The output voltages are given in Table II.

\begin{table}[ht]
\centering
% used for centering table
\caption{DGU voltage and load profile} 
\begin{tabular}{c c c c c}
 % centered columns (4 columns)
\hline %\hline\hline inserts double horizontal lines
 
$\hat{\Sigma}_{i}^{\textrm{DGU}}$ & Voltage reference & Input voltage & Duty cycle & Local load power \\ [0.5ex] % inserts table
%heading
\hline\hline

$\hat{\Sigma}_{1}^{\textrm{DGU}}$ & 381 V & 95 V & 0.75 & 2.5 kW\\
$\hat{\Sigma}_{2}^{\textrm{DGU}}$ & 390 V & 100 V& 0.7372 & 2 kW\\
$\hat{\Sigma}_{3}^{\textrm{DGU}}$ & 373 V & 90 V & 0.7633 & 1.8 kW\\
$\hat{\Sigma}_{4}^{\textrm{DGU}}$ & 387 V & 105 V & 0.723 & 2.5 kW\\
$\hat{\Sigma}_{5}^{\textrm{DGU}}$ & 370 V & 92 V & 0.7576 & 3 kW\\
$\hat{\Sigma}_{6}^{\textrm{DGU}}$ & 385 V & 90 V & 0.7636 & 2.5 kW\\
\hline
\end{tabular}
\label{table:DGULargeV} % is used to refer this table in the text

\end{table}

The following figures show the response of each DGU.

\begin{figure}[!htb] % "[t!]" placement specifier just for this example
\graphicspath{ {Images/} }
\begin{subfigure}{0.48\textwidth}
\includegraphics[width=\linewidth]{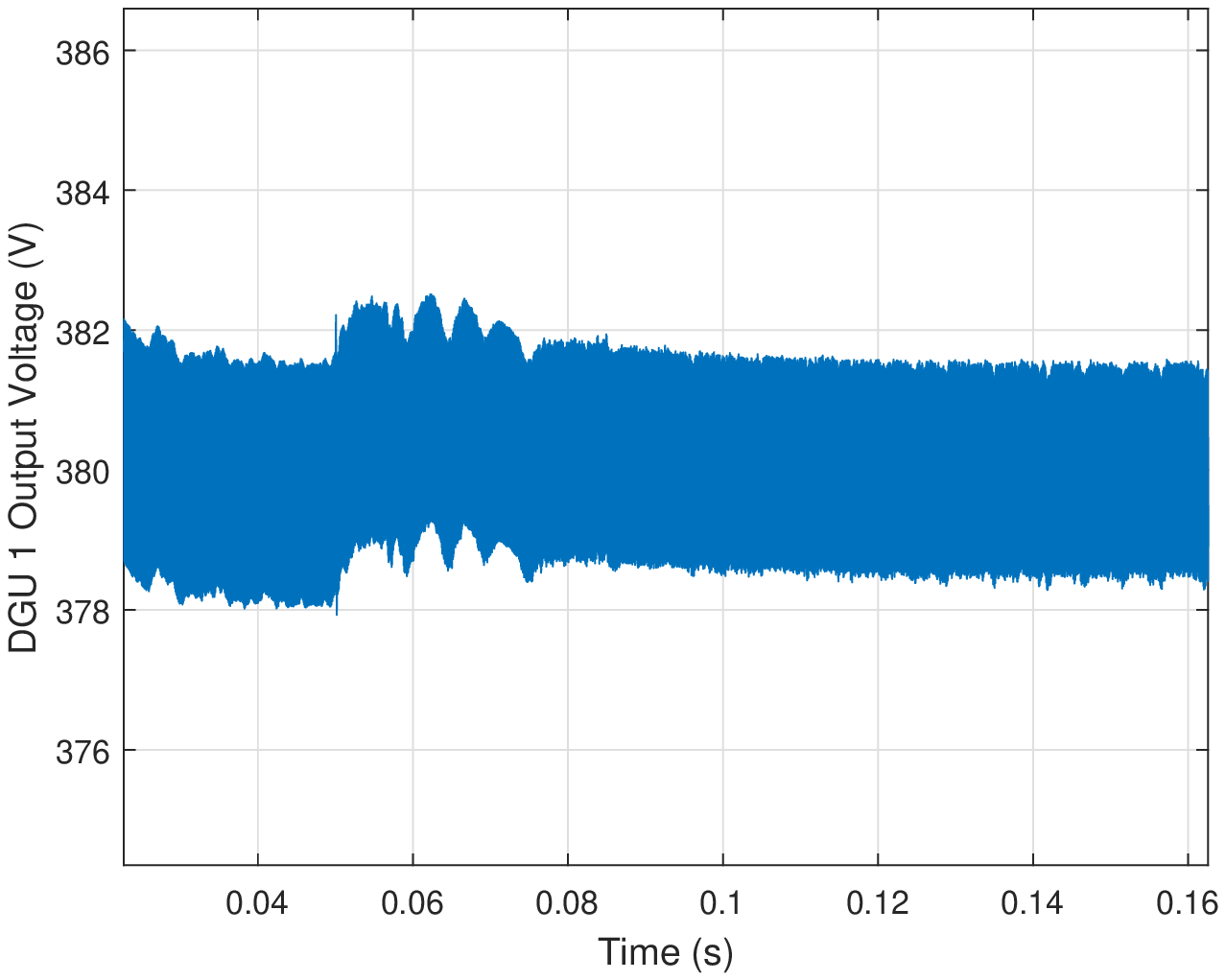}
\caption{$\hat{\Sigma}_{1}^{\textrm{DGU}}$ output voltage} \label{fig:DGU1PnPDGU63}
\end{subfigure}\hspace*{\fill}
\begin{subfigure}{0.48\textwidth}
\includegraphics[width=\linewidth]{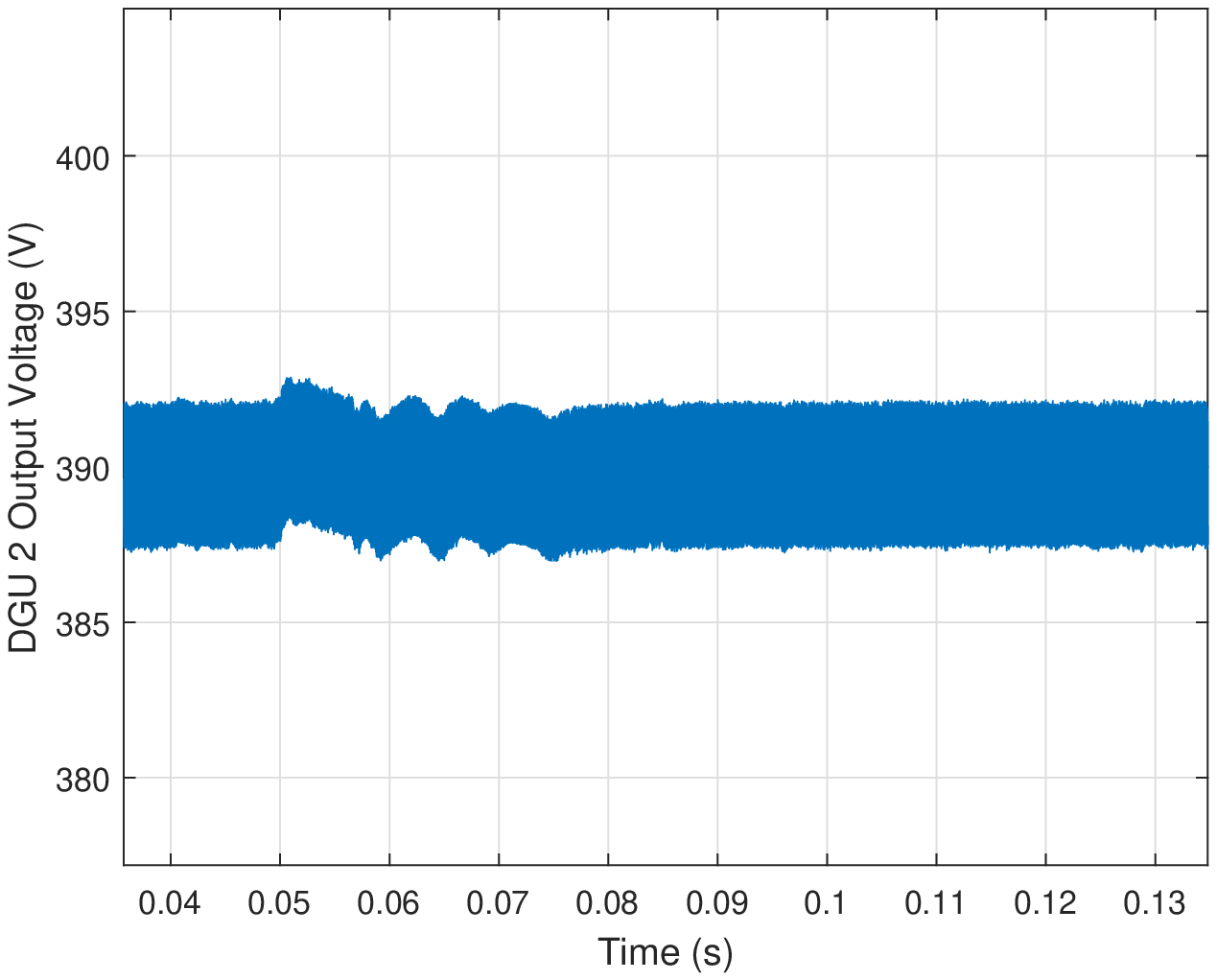}
\caption{$\hat{\Sigma}_{2}^{\textrm{DGU}}$ output voltage} \label{fig:DGU2PnPDGU63}
\end{subfigure}
\medskip
\graphicspath{ {Images/} }
\begin{subfigure}{0.45\textwidth}
\includegraphics[width=\linewidth]{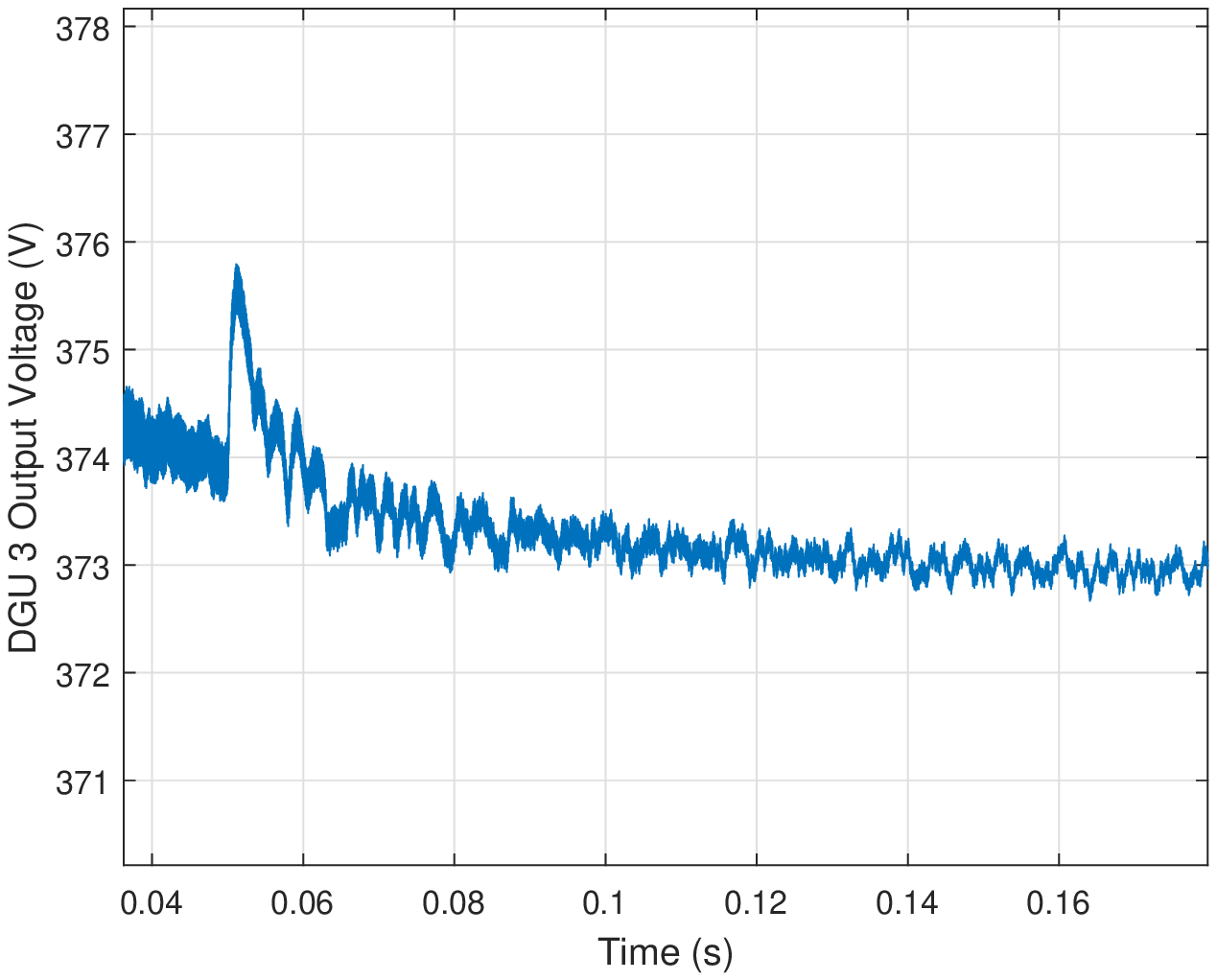}
\caption{$\hat{\Sigma}_{3}^{\textrm{DGU}}$ output voltage} \label{fig:DGU3PnPDGU63}
\end{subfigure}\hspace*{\fill}
\begin{subfigure}{0.45\textwidth}
\includegraphics[width=\linewidth]{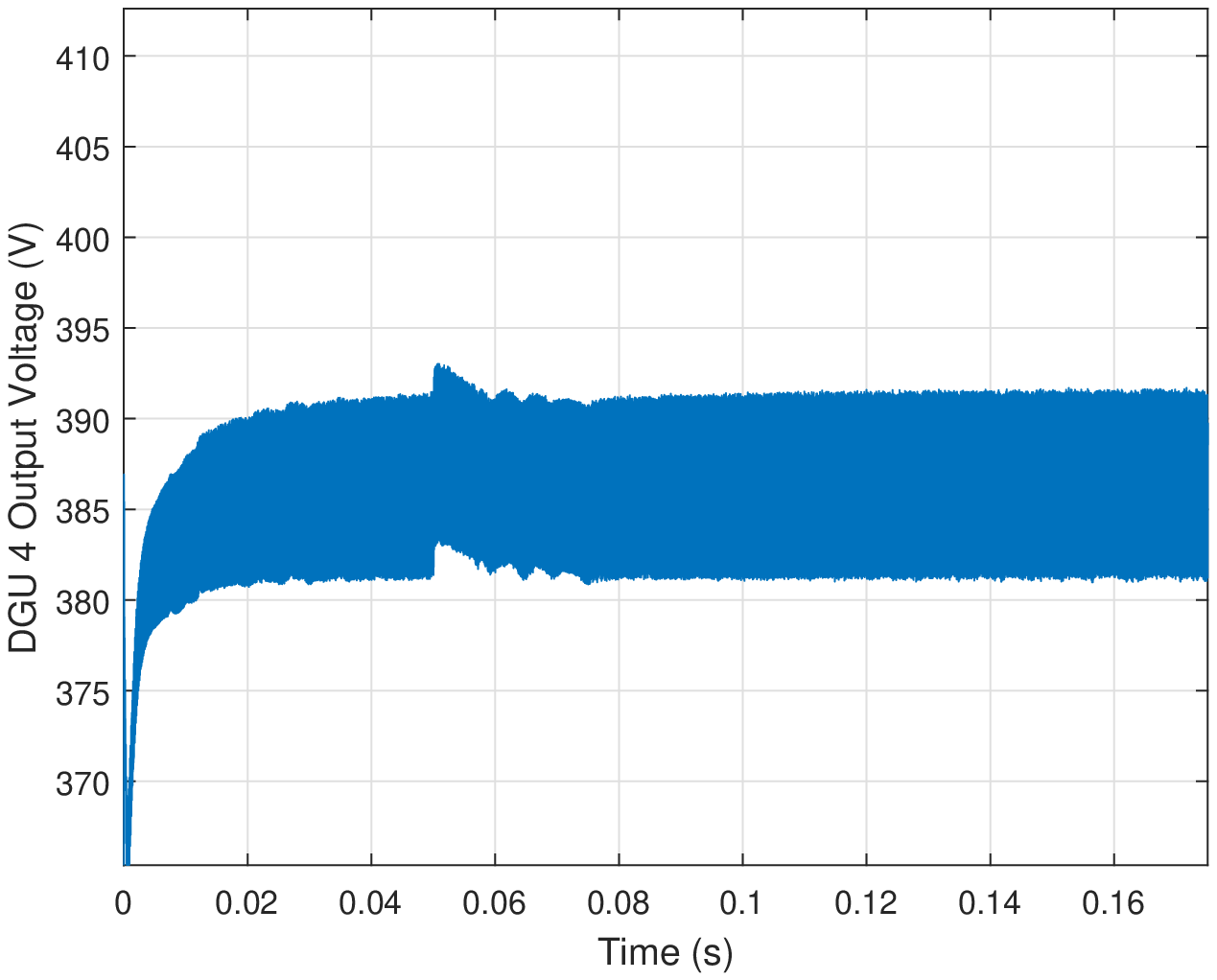}
\caption{$\hat{\Sigma}_{4}^{\textrm{DGU}}$ output voltage} \label{fig:DGU4PnPDGU63}
\end{subfigure}
\end{figure}
\medskip
\begin{figure}[!htb] % "[t!]" placement specifier just for this example
\ContinuedFloat%
\begin{subfigure}{0.45\textwidth}
\includegraphics[width=\linewidth]{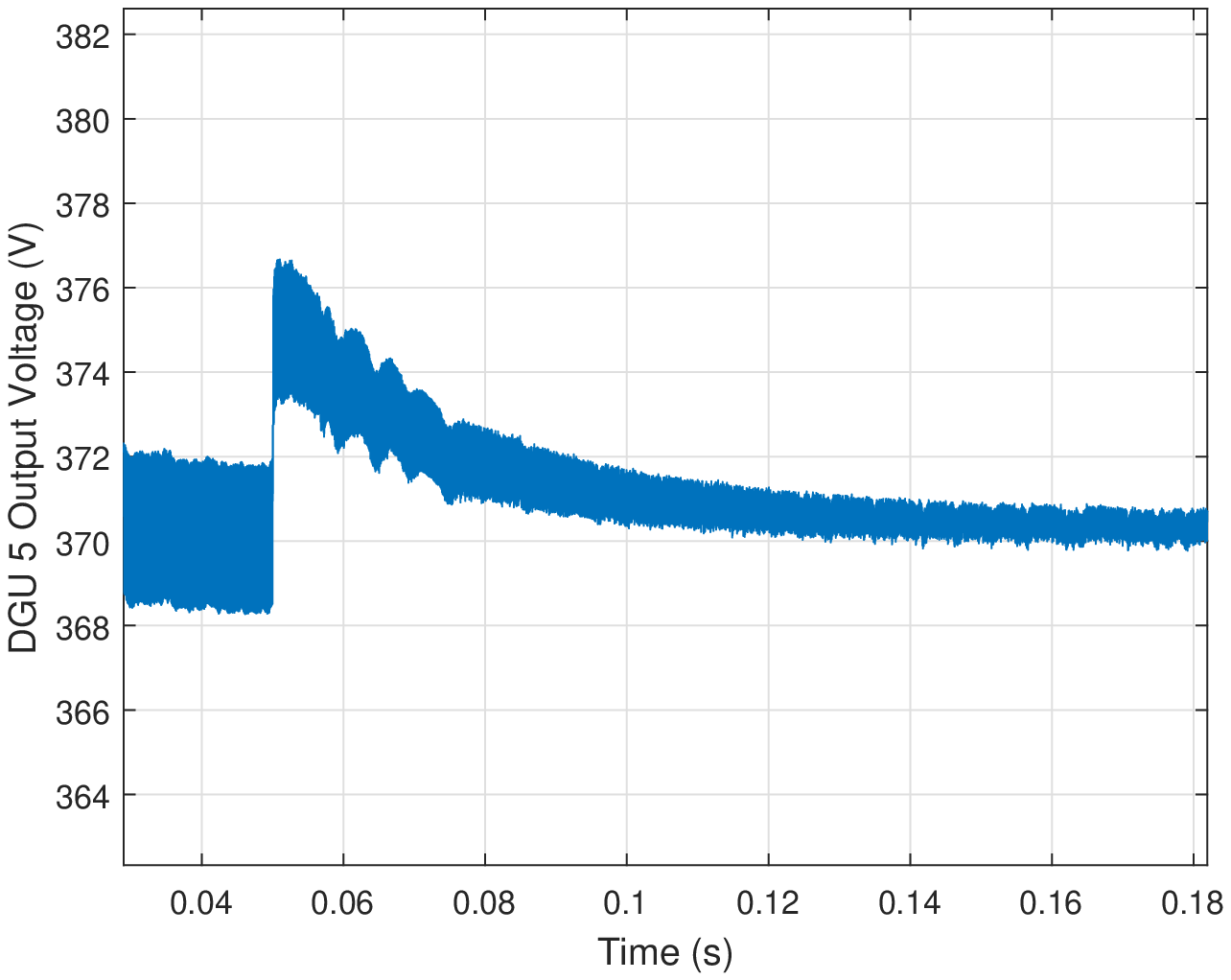}
\caption{$\hat{\Sigma}_{5}^{\textrm{DGU}}$ output voltage} \label{fig:DGU5PnPDGU63}
\end{subfigure}\hspace*{\fill}
\begin{subfigure}{0.45\textwidth}
\includegraphics[width=\linewidth]{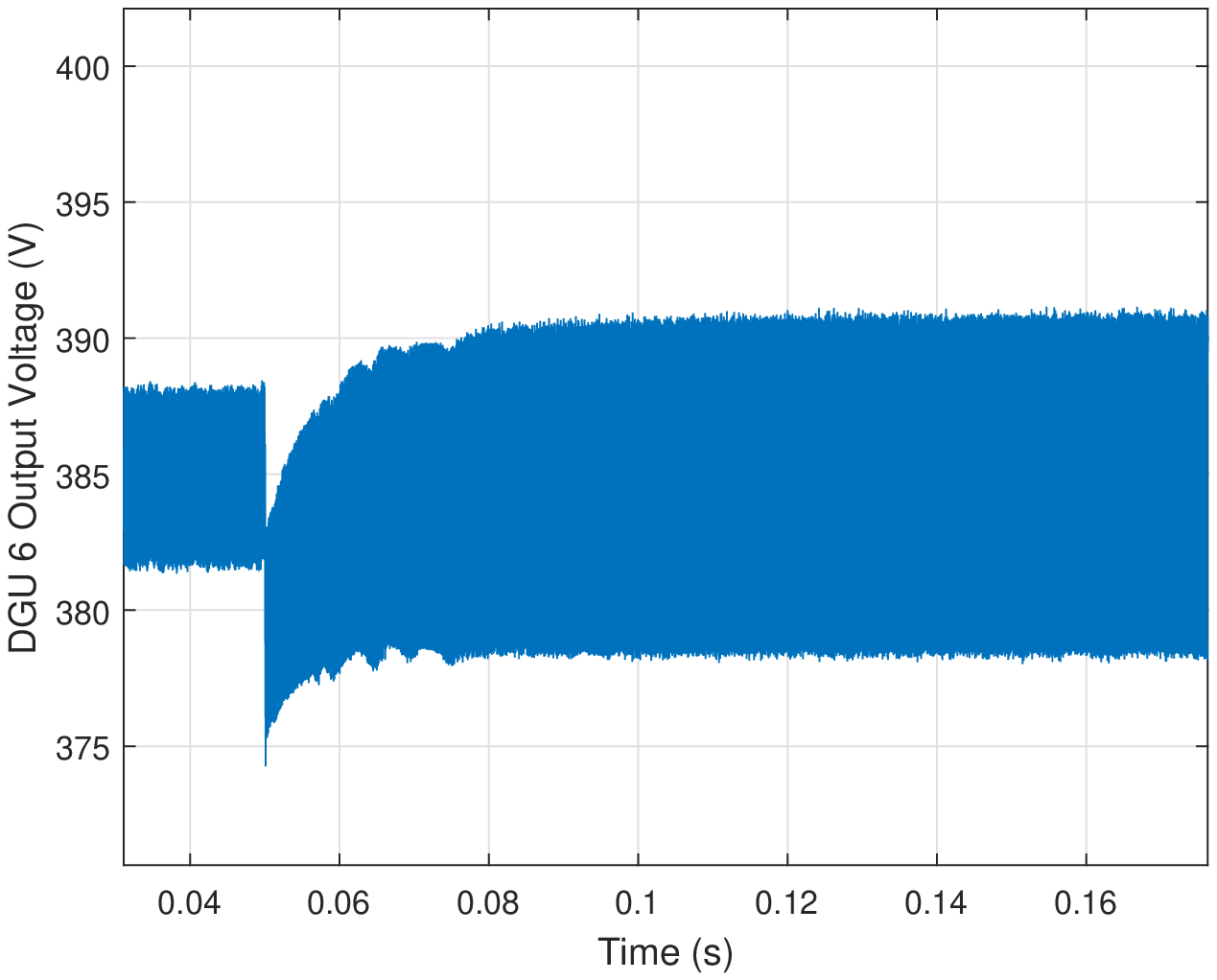}
\caption{$\hat{\Sigma}_{6}^{\textrm{DGU}}$ output voltage} \label{fig:DGU6PnPDGU63}
\end{subfigure}
\caption{DGU output voltage responses to $\hat{\Sigma}_{6}^{\textrm{DGU}}$ plugging-in with large voltage reference differences.}
\end{figure}
The responses are largely favourable. DGUs $\hat{\Sigma}_{1}^{\textrm{DGU}}$, $\hat{\Sigma}_{2}^{\textrm{DGU}}$, $\hat{\Sigma}_{4}^{\textrm{DGU}}$ and $\hat{\Sigma}_{6}^{\textrm{DGU}}$ show very fast settling times and damped responses. Though the voltage ripple of $\hat{\Sigma}_{6}^{\textrm{DGU}}$ increases from 1.3\% (or 5 V) to 3.1 \% (or 12 V), this is purely due to the effective load change as $\hat{\Sigma}_{6}^{\textrm{DGU}}$ connects to the $RL$ power lines and loads of $\hat{\Sigma}_{1}^{\textrm{DGU}}$ and $\hat{\Sigma}_{5}^{\textrm{DGU}}$. DGUs $\hat{\Sigma}_{3}^{\textrm{DGU}}$ and $\hat{\Sigma}_{5}^{\textrm{DGU}}$ have noticeably slower settling times. This can be attributed to both $\hat{\Sigma}_{3}^{\textrm{DGU}}$ and $\hat{\Sigma}_{5}^{\textrm{DGU}}$ being neighbours with DGUs that have considerably higher output voltages. As a result, $\hat{\Sigma}_{3}^{\textrm{DGU}}$ and $\hat{\Sigma}_{5}^{\textrm{DGU}}$ are in less control of their power supply capabilities, with the larger output voltages their respective neighbours,  $\hat{\Sigma}_{1}^{\textrm{DGU}}$,  $\hat{\Sigma}_{4}^{\textrm{DGU}}$ and $\hat{\Sigma}_{6}^{\textrm{DGU}}$. This is evident from the the duty cycles of each DGU  in Fig. \ref{fig:DutyCycles2}. 

\textbf{Note:} \textit{Duty cycles are plotted over the course of 1s test i.e. includes responses to subsequent tests.}

\begin{figure}[!htb] % "[t!]" placement specifier just for this example
\graphicspath{ {Images/} }
\begin{subfigure}{0.48\textwidth}
\includegraphics[width=\linewidth]{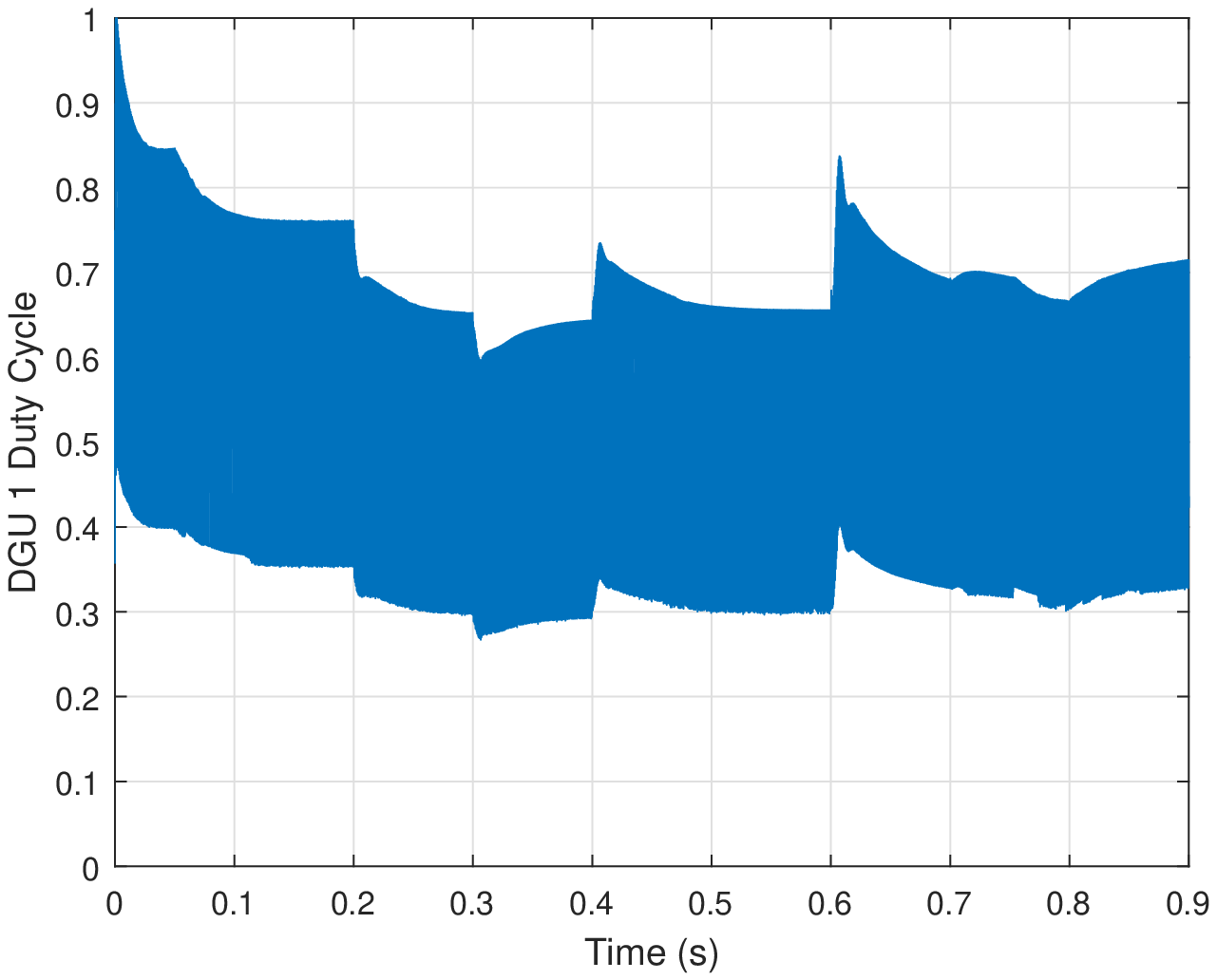}
\caption{$\hat{\Sigma}_{1}^{\textrm{DGU}}$ output voltage} \label{fig:DGU1_d}
\end{subfigure}\hspace*{\fill}
\begin{subfigure}{0.48\textwidth}
\includegraphics[width=\linewidth]{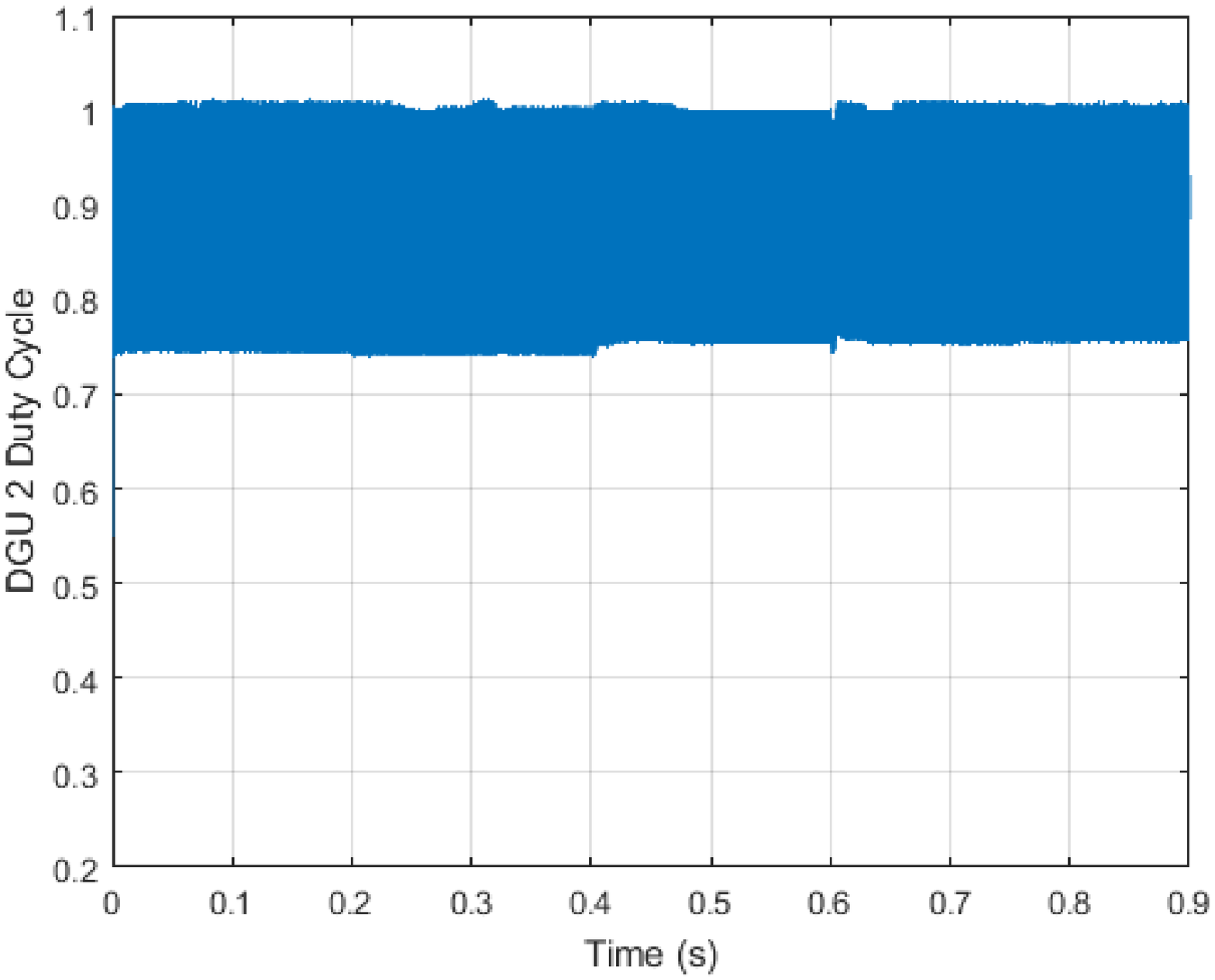}
\caption{$\hat{\Sigma}_{2}^{\textrm{DGU}}$ output voltage} \label{fig:DGU2_d}
\end{subfigure}
\medskip
\begin{subfigure}{0.45\textwidth}
\includegraphics[width=\linewidth]{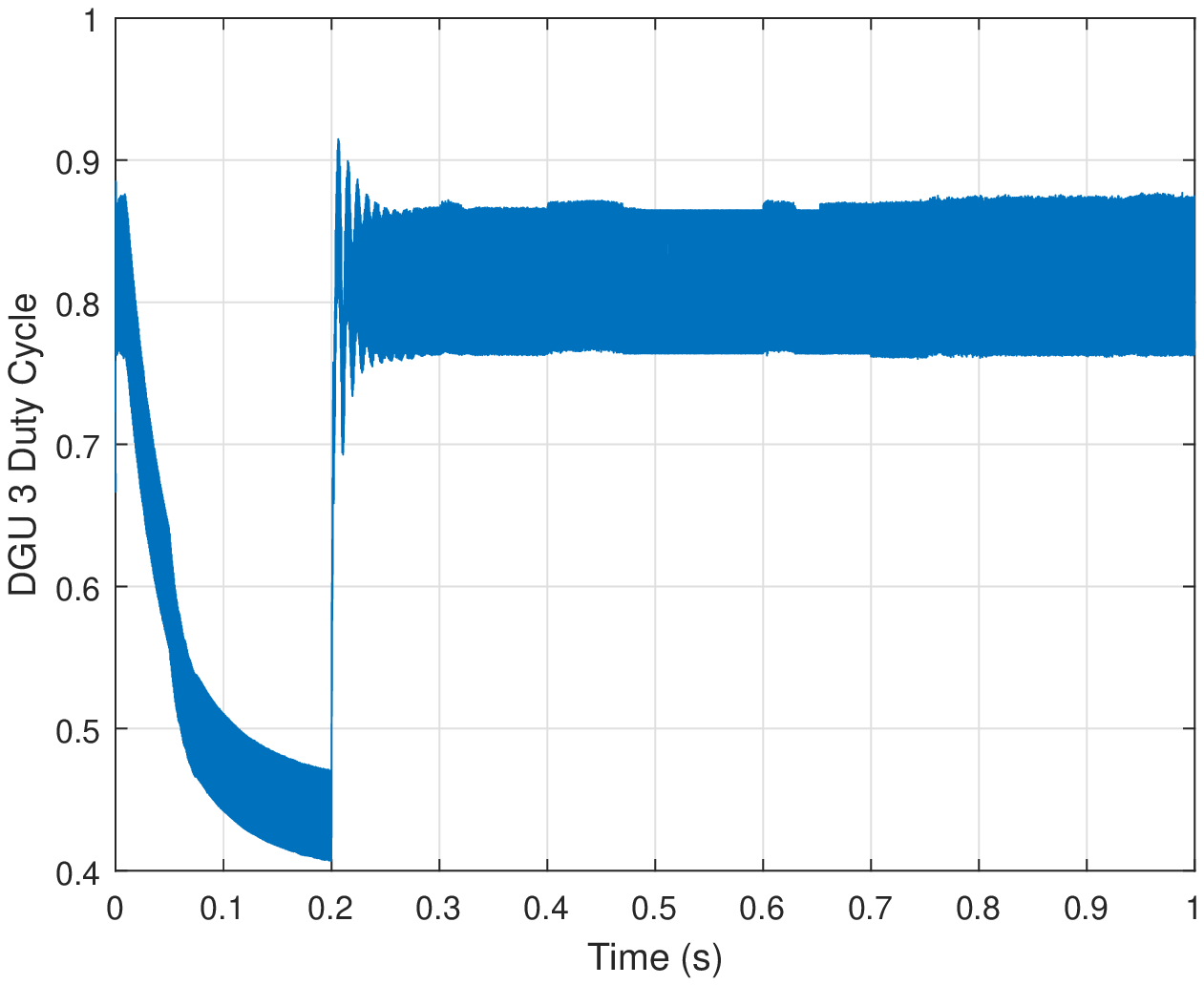}
\caption{$\hat{\Sigma}_{3}^{\textrm{DGU}}$ output voltage} \label{fig:DGU3_d}
\end{subfigure}\hspace*{\fill}
\begin{subfigure}{0.45\textwidth}
\includegraphics[width=\linewidth]{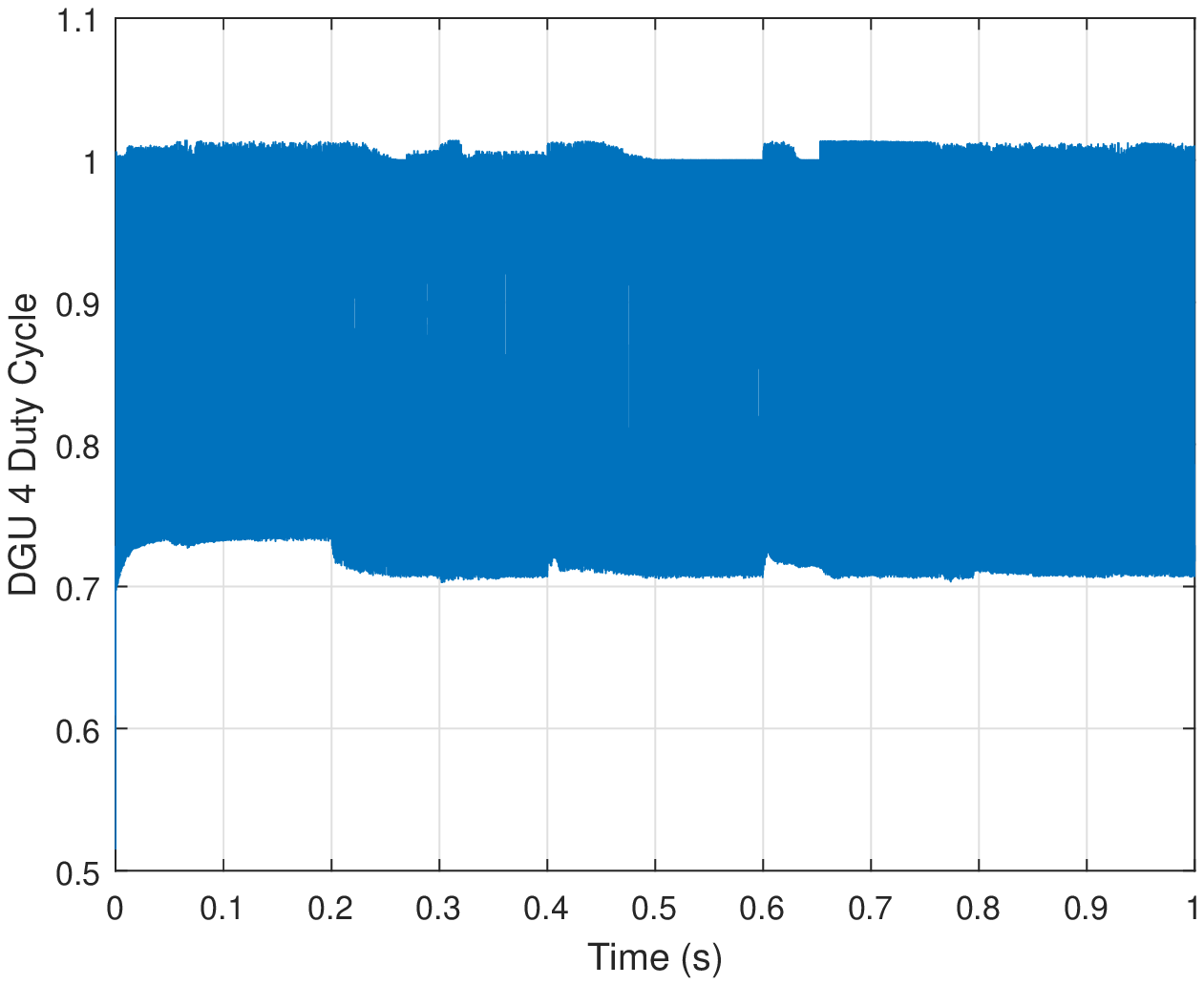}
\caption{$\hat{\Sigma}_{4}^{\textrm{DGU}}$ output voltage} \label{fig:DGU4_d}
\end{subfigure}
\end{figure}
\begin{figure}[!htb] % "[t!]" placement specifier just for this example
\ContinuedFloat%
\graphicspath{ {Images/} }
\begin{subfigure}{0.45\textwidth}
\includegraphics[width=\linewidth]{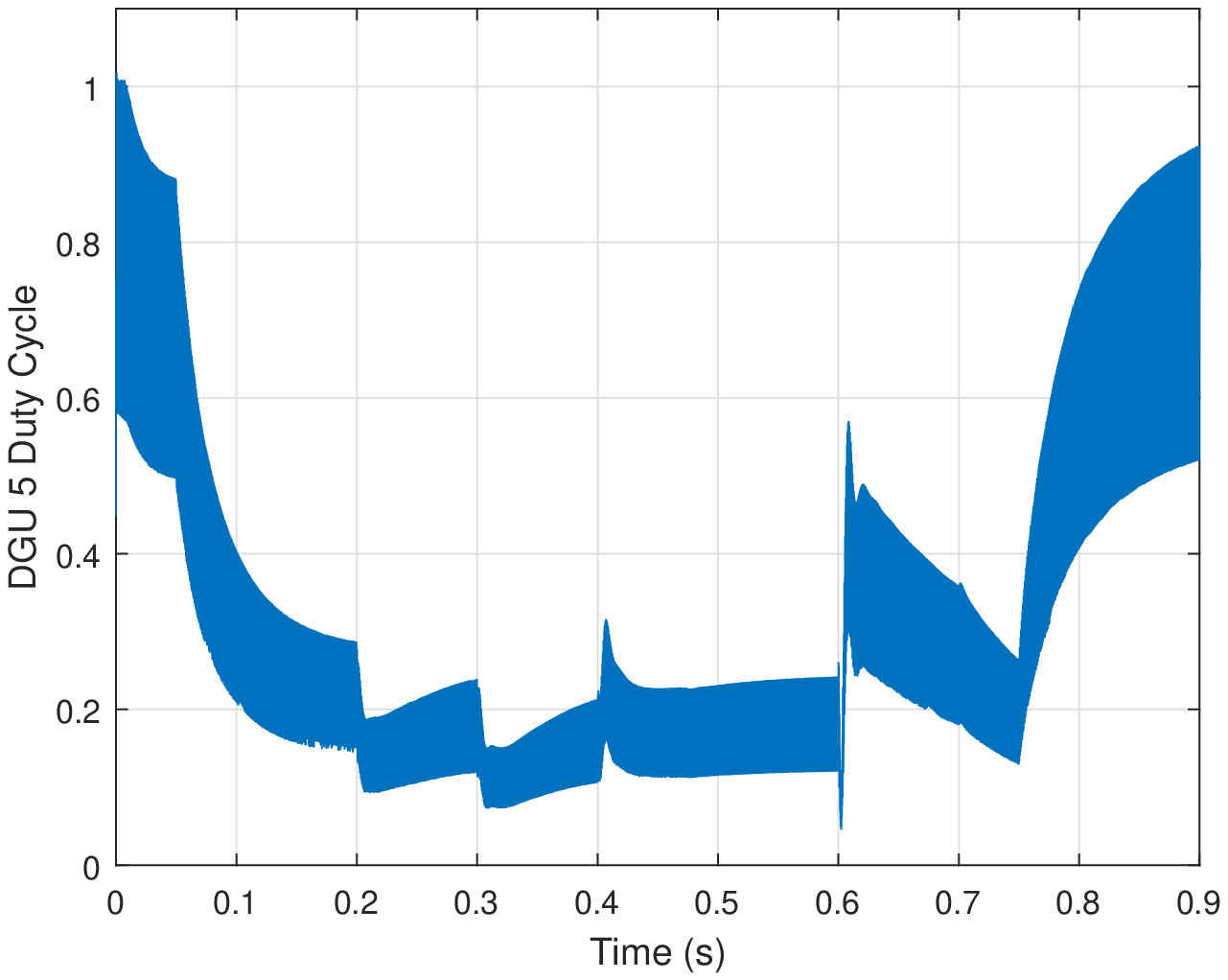}
\caption{$\hat{\Sigma}_{5}^{\textrm{DGU}}$ output voltage} \label{fig:DGU5_d}
\end{subfigure}\hspace*{\fill}
\begin{subfigure}{0.45\textwidth}
\includegraphics[width=\linewidth]{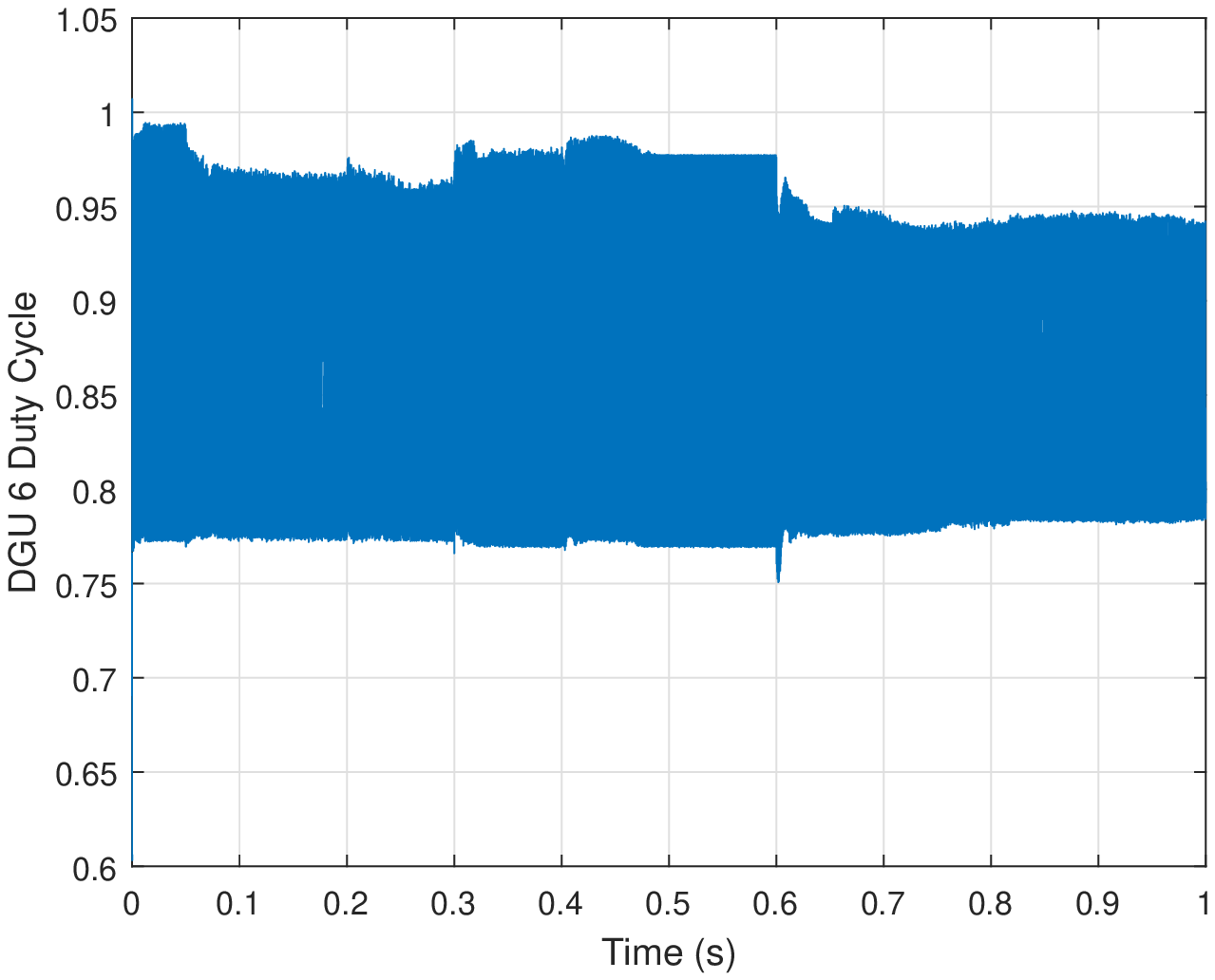}
\caption{$\hat{\Sigma}_{6}^{\textrm{DGU}}$ output voltage} \label{fig:DGU6_d}
\end{subfigure}
\caption{DGU duty cycles in response to all tests in this section} \label{fig:DutyCycles2}
\end{figure}

From Fig. \ref{fig:DutyCycles2}(c) and \ref{fig:DutyCycles2}(e), the duty cycles of $\hat{\Sigma}_{3}^{\textrm{DGU}}$ and $\hat{\Sigma}_{5}^{\textrm{DGU}}$ are reduced from their nominal steady-state values.

At $t$ = 0.2 s, $\hat{\Sigma}_{3}^{\textrm{DGU}}$ is unplugged from the rest of the mG to power a local load exclusively. The following figures show the response of each DGU.

\begin{figure}[!htb] % "[t!]" placement specifier just for this example
\graphicspath{ {Images/} }
\begin{subfigure}{0.45\textwidth}
\includegraphics[width=\linewidth]{DGU1PnPDGU3.eps}
\caption{$\hat{\Sigma}_{1}^{\textrm{DGU}}$ output voltage} \label{fig:DGU1PnPDGU3_2}
\end{subfigure}\hspace*{\fill}
\begin{subfigure}{0.45\textwidth}
\includegraphics[width=\linewidth]{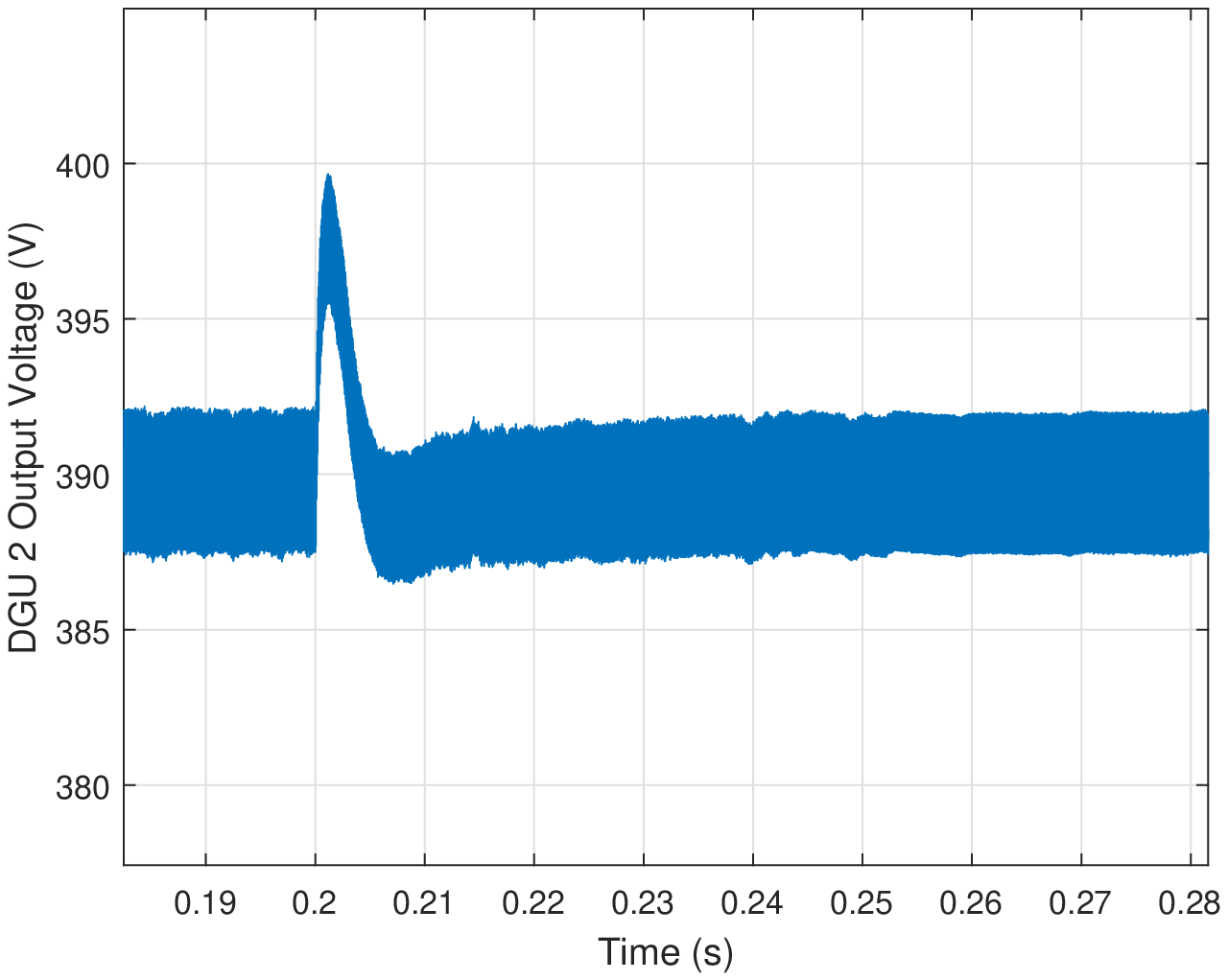}
\caption{$\hat{\Sigma}_{2}^{\textrm{DGU}}$ output voltage} \label{fig:DGU2PnPDGU3_2}
\end{subfigure}
\medskip
\begin{subfigure}{0.42\textwidth}
\includegraphics[width=\linewidth]{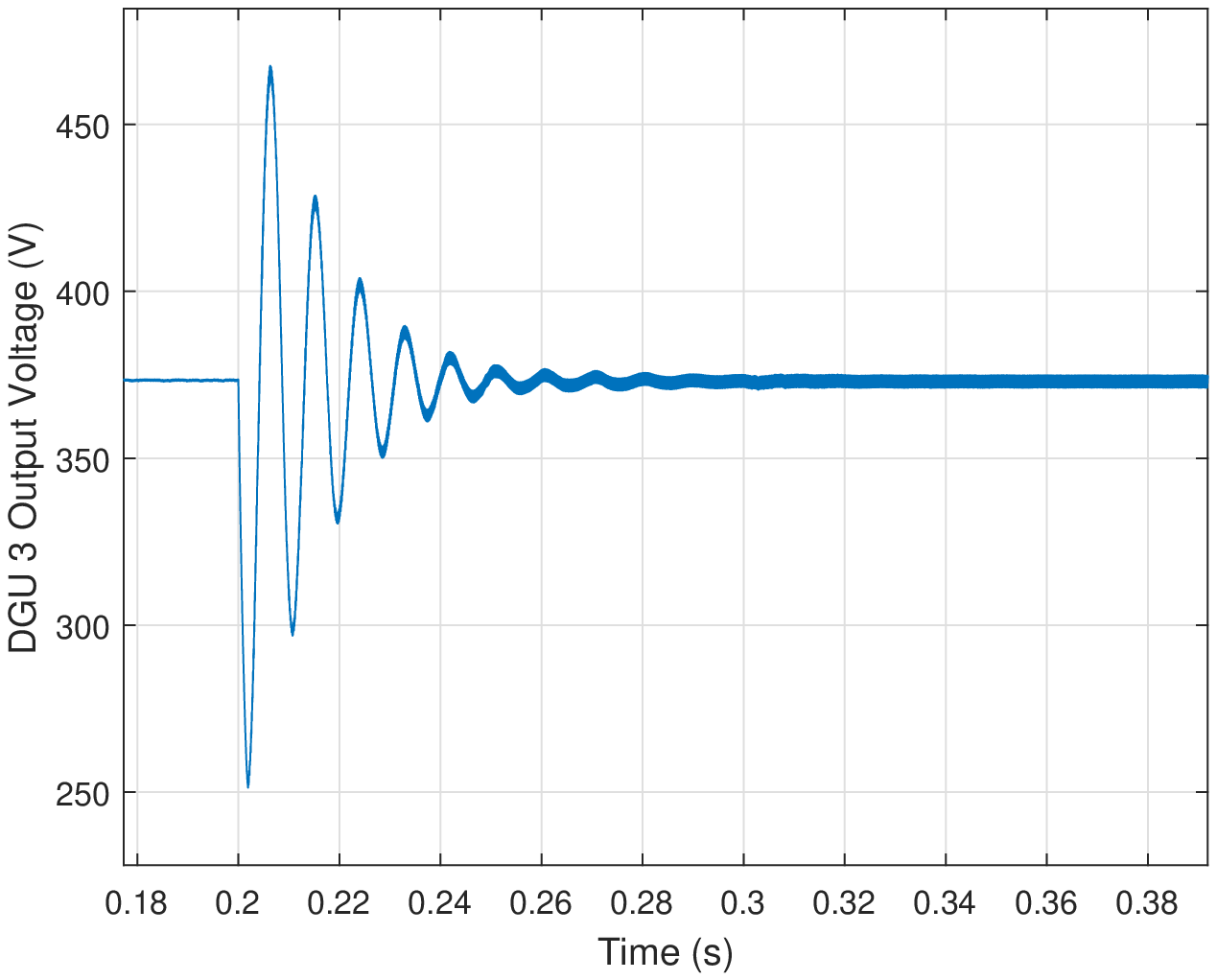}
\caption{$\hat{\Sigma}_{3}^{\textrm{DGU}}$ output voltage} \label{fig:DGU3PnPDGU3_2}
\end{subfigure}\hspace*{\fill}
\begin{subfigure}{0.42\textwidth}
\includegraphics[width=\linewidth]{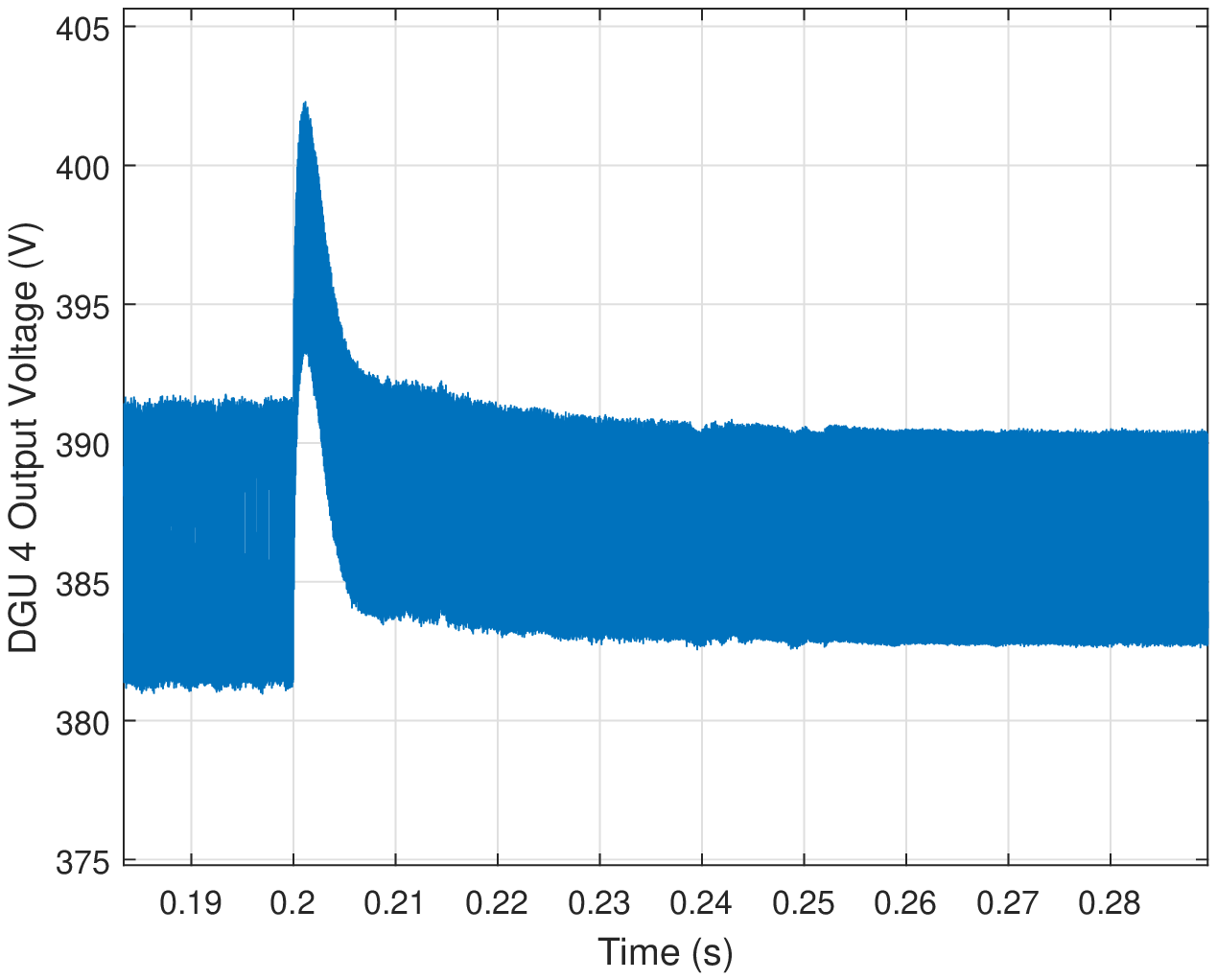}
\caption{$\hat{\Sigma}_{4}^{\textrm{DGU}}$ output voltage} \label{fig:DGU4PnPDGU3_2}
\end{subfigure}
\end{figure}
\begin{figure}[!htb] % "[t!]" placement specifier just for this example
\ContinuedFloat%
\graphicspath{ {Images/} }
\begin{subfigure}{0.42\textwidth}
\includegraphics[width=\linewidth]{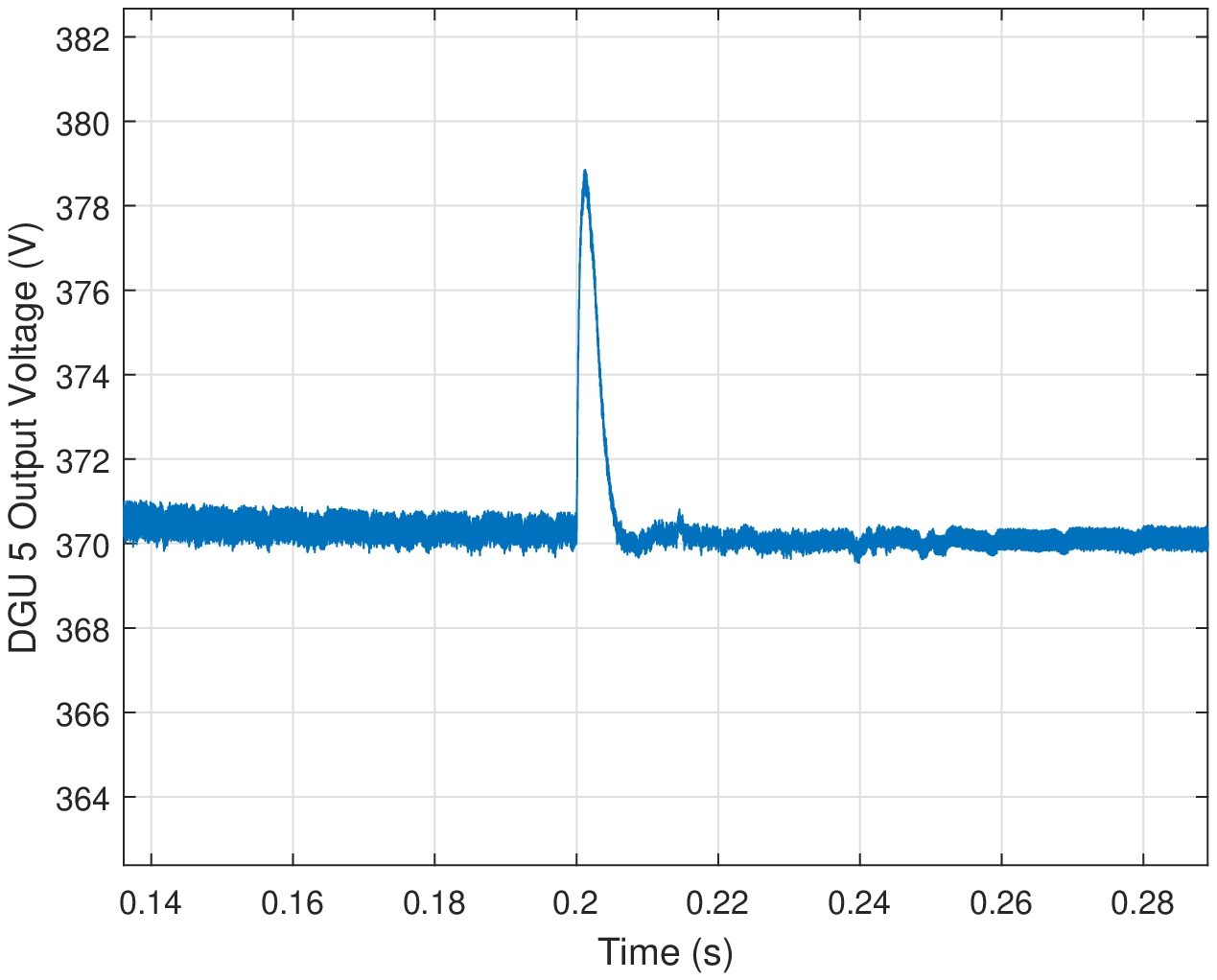}
\caption{$\hat{\Sigma}_{5}^{\textrm{DGU}}$ output voltage} \label{fig:DGU5PnPDGU3_2}
\end{subfigure}\hspace*{\fill}
\begin{subfigure}{0.42\textwidth}
\includegraphics[width=\linewidth]{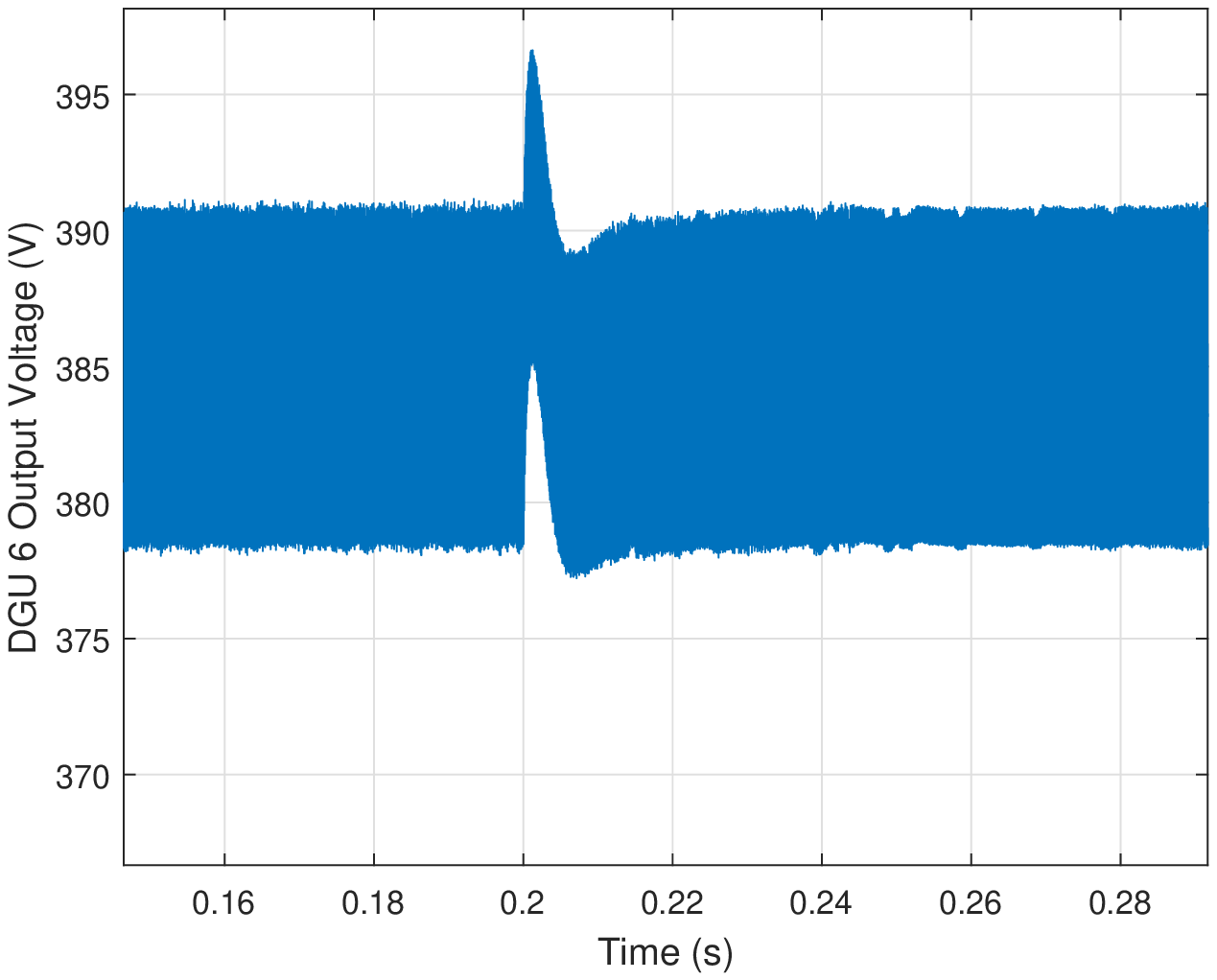}
\caption{$\hat{\Sigma}_{6}^{\textrm{DGU}}$ output voltage} \label{fig:DGU6PnPDGU3_2}
\end{subfigure}
\caption{DGU output voltage responses to $\hat{\Sigma}_{6}^{\textrm{DGU}}$ plug-out} \label{fig:DGUPnPDGU3_3}
\end{figure}

All responses, except $\hat{\Sigma}_{3}^{\textrm{DGU}}$, are favourable, with fast settling times of between 2 - 25 ms and damped overshoots of maximum 12 V ($\hat{\Sigma}_{5}^{\textrm{DGU}}$). As $\hat{\Sigma}_{3}^{\textrm{DGU}}$ is unplugged, the change in going from a coupled system requiring assistance from the $\mathcal{L}_1$AC loop to a decoupled system that only needs the baseline controller is a big enough jump to induce large oscillations, with a peak swing of 225 V. The response settles after 90 ms, which for primary voltage control is still fast. However, as $\hat{\Sigma}_{3}^{\textrm{DGU}}$ does not affect the rest of the grid, this oscillation might be tolerable as long as it is within the tolerance level of the local load. Alternatively, it is shown in Fig. \ref{fig:DGU3PnP3} that turning off the adaptation loop after $\hat{\Sigma}_{3}^{\textrm{DGU}}$ is plugged-out avoids the oscillations.

Finally, at $t$ = 0.7 s, $\hat{\Sigma}_{3}^{\textrm{DGU}}$ is plugged back into the ImG,  connecting to $\hat{\Sigma}_{1}^{\textrm{DGU}}$ and $\hat{\Sigma}_{5}^{\textrm{DGU}}$. Responses are plotted in Fig. \ref{fig:DGUPnPDGU3_2}.

\begin{figure}[!htb] % "[t!]" placement specifier just for this example
\graphicspath{ {Images/} }
\begin{subfigure}{0.42\textwidth}
\includegraphics[width=\linewidth]{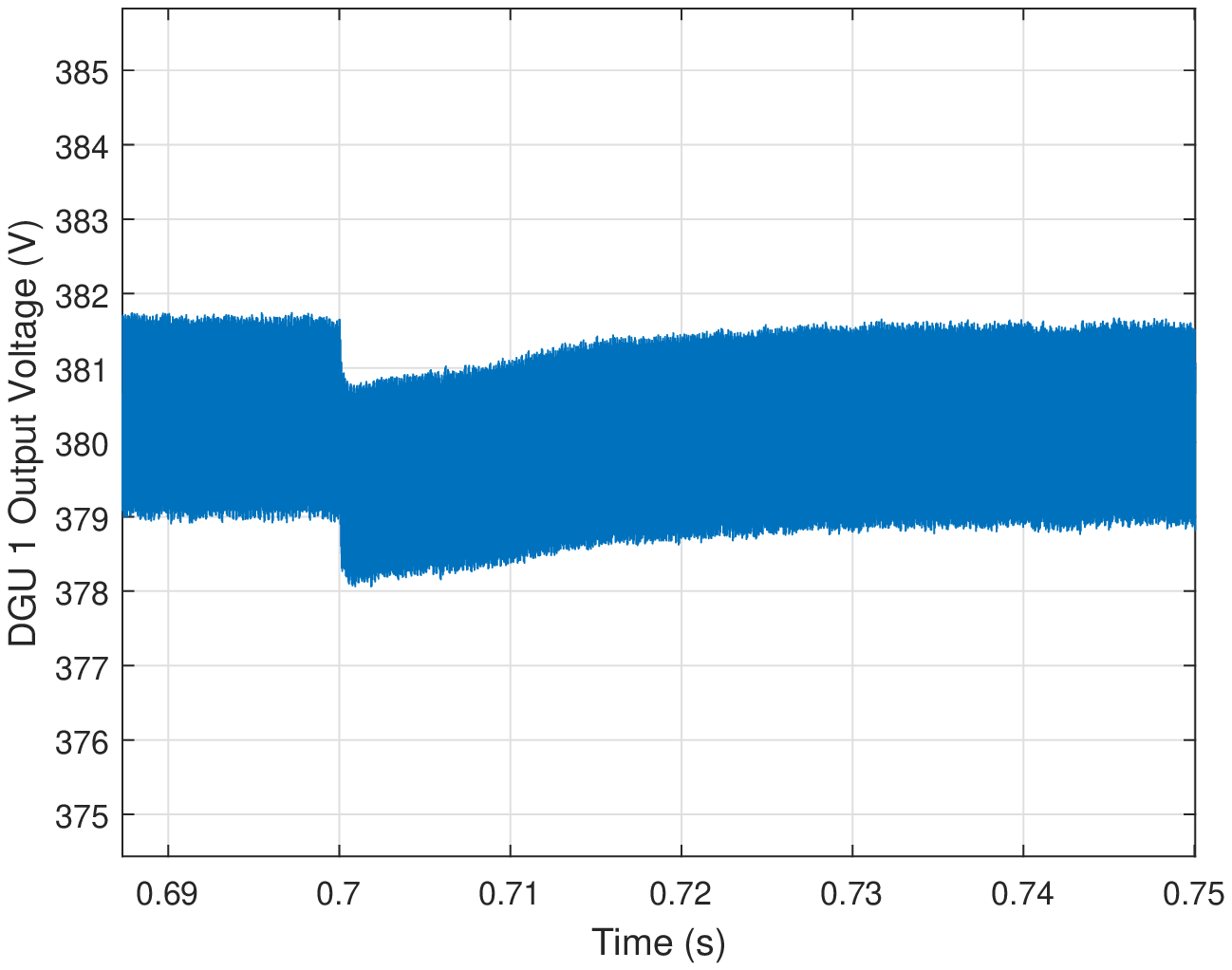}
\caption{$\hat{\Sigma}_{1}^{\textrm{DGU}}$ output voltage} \label{fig:DGU1PnPDGU3_2}
\end{subfigure}\hspace*{\fill}
\begin{subfigure}{0.42\textwidth}
\includegraphics[width=\linewidth]{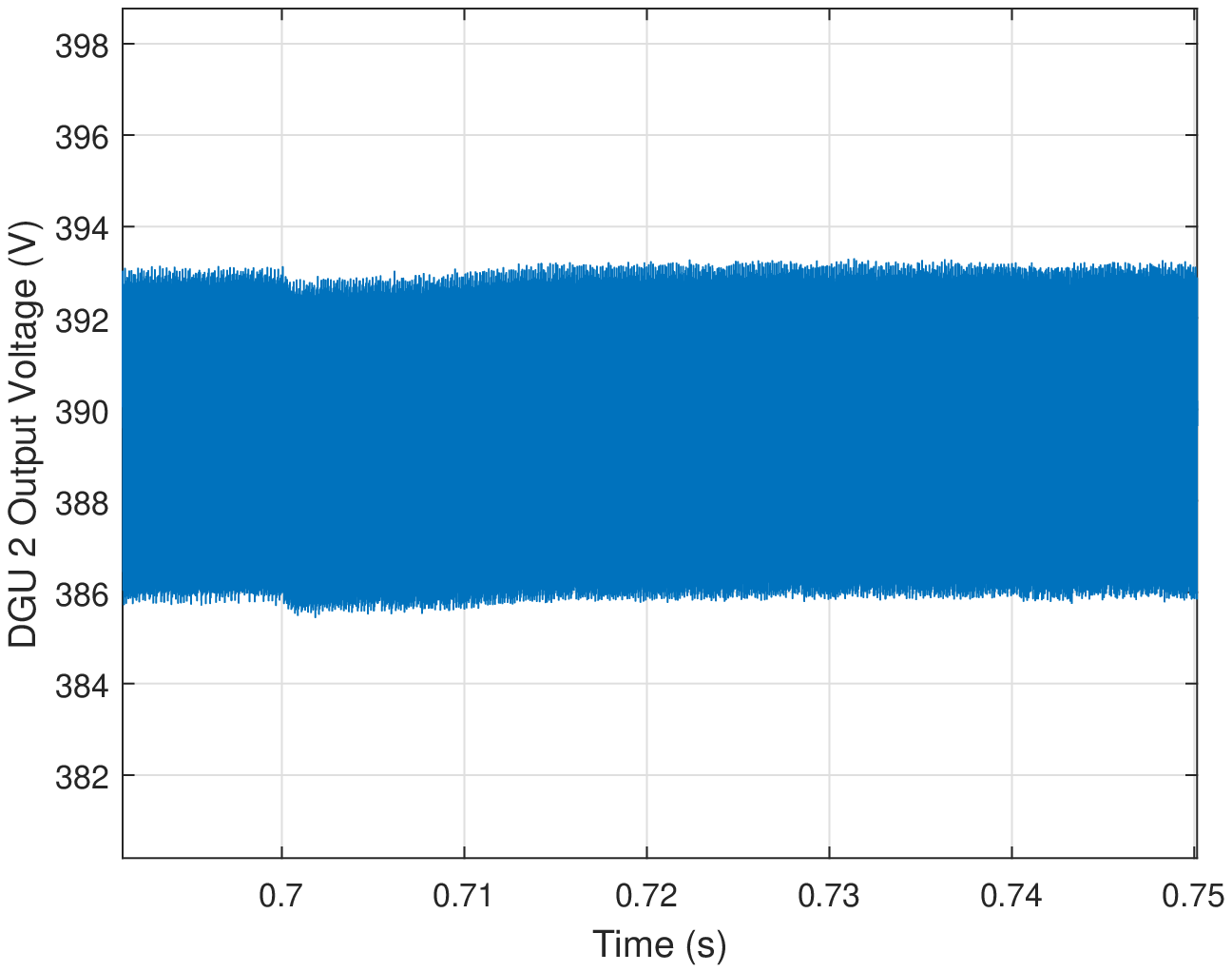}
\caption{$\hat{\Sigma}_{2}^{\textrm{DGU}}$ output voltage} \label{fig:DGU2PnPDGU3_2}
\end{subfigure}
\medskip
\begin{subfigure}{0.42\textwidth}
\includegraphics[width=\linewidth]{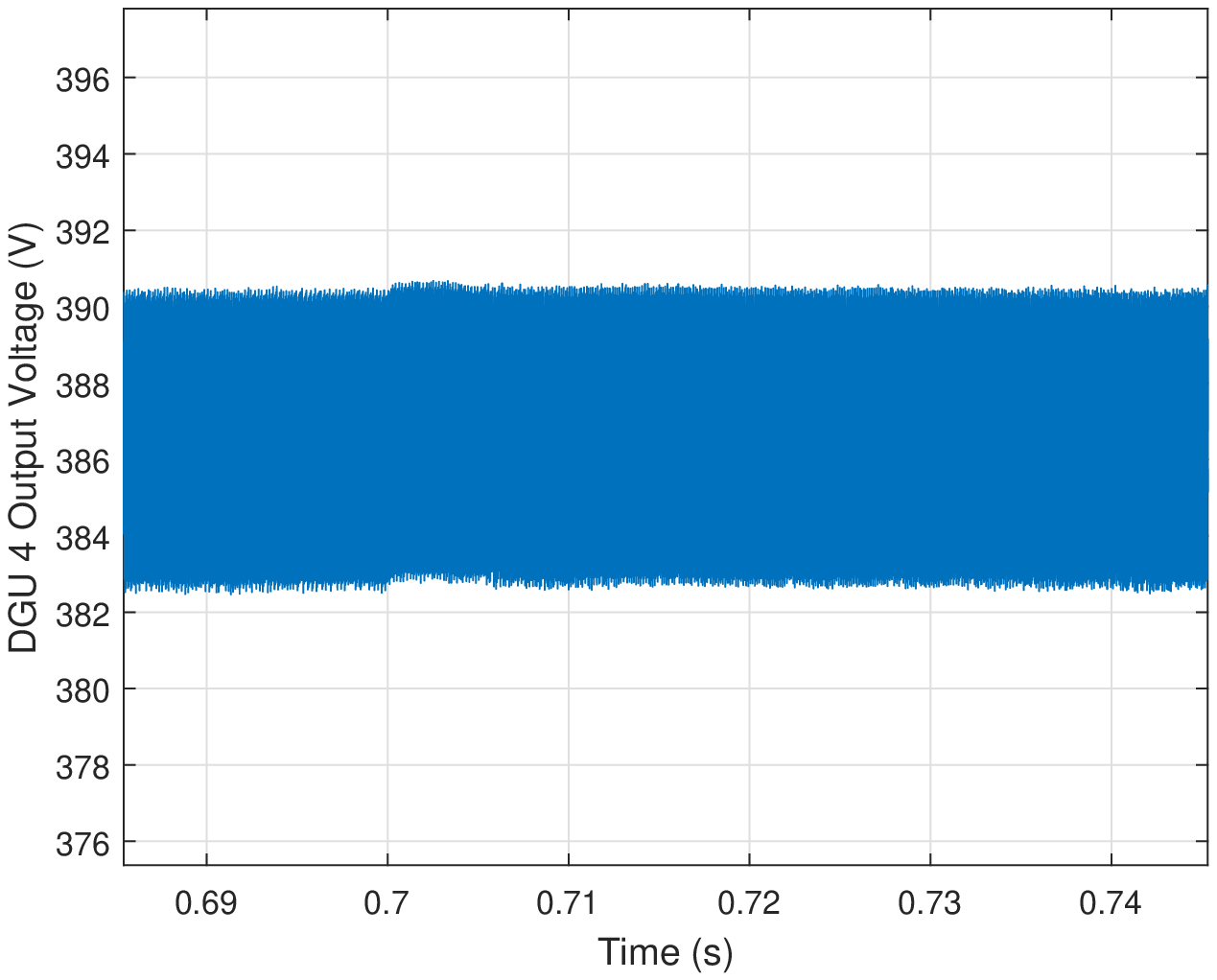}
\caption{$\hat{\Sigma}_{4}^{\textrm{DGU}}$ output voltage} \label{fig:DGU4PnPDGU3_2}
\end{subfigure}\hspace*{\fill}
\begin{subfigure}{0.42\textwidth}
\includegraphics[width=\linewidth]{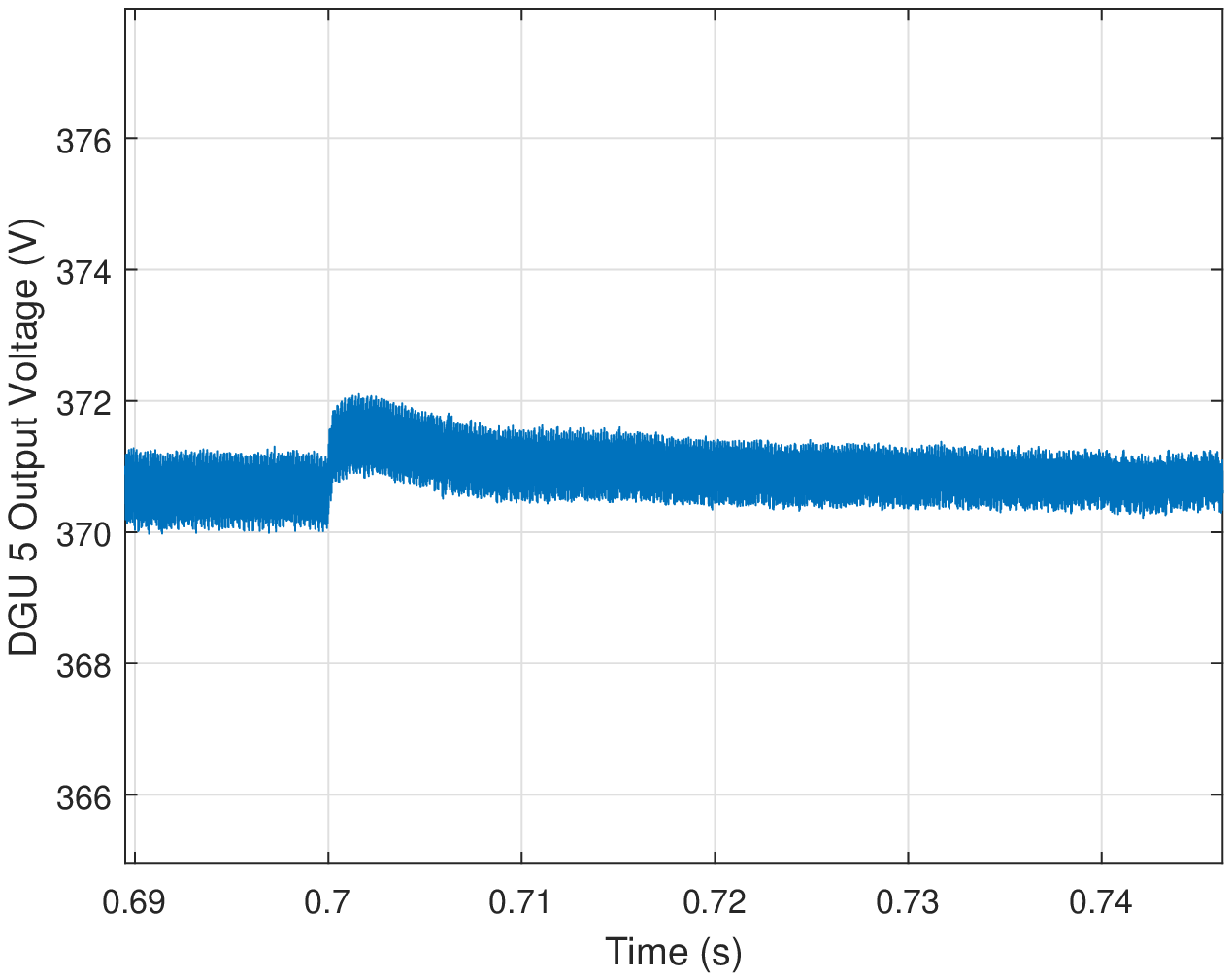}
\caption{$\hat{\Sigma}_{5}^{\textrm{DGU}}$ output voltage} \label{fig:DGU5PnPDGU3_2}
\end{subfigure}
\medskip
\centering
\begin{subfigure}{0.42\textwidth}
\includegraphics[width=\linewidth]{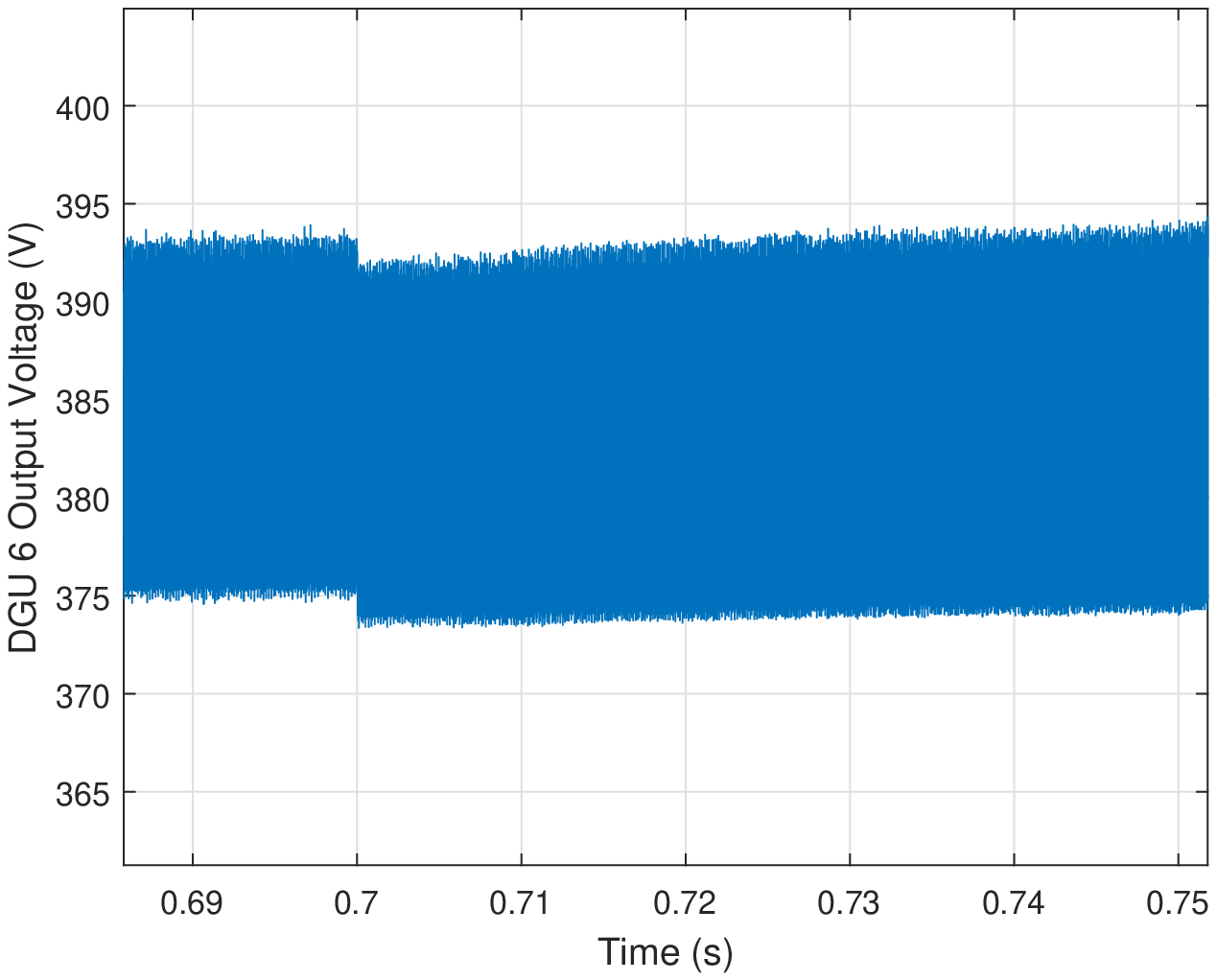}
\caption{$\hat{\Sigma}_{6}^{\textrm{DGU}}$ output voltage} \label{fig:DGU65PnPDGU3_2}
\end{subfigure}
\caption{DGU duty cycles in response to all tests in this section} \label{fig:DGUPnPDGU3_2}
\end{figure}

The responses are very good. The settling times are fast, with the longest at 50 ms associated with $\hat{\Sigma}_{1}^{\textrm{DGU}}$ and $\hat{\Sigma}_{5}^{\textrm{DGU}}$ since these are the DGUs that $\hat{\Sigma}_{3}^{\textrm{DGU}}$ connects to. 
Overall, PnP operations are satisfactory when relatively large differences between each DGU voltage reference exist. The DGU 'overpowering' effect does not adversely affect stability or performance during PnP operations - at worst, the settling times of the 'overpowered' DGUs become slower. However, it should be noted that some of the DGU duty cycles are close to saturation. Realistic duty cycles are typically limited to 80 \%.
 
\subsubsection{Robustness to Unknown Load Dynamics}

In order to examine the robustness of the global DC-ImG, the load at $\hat{\Sigma}_{6}^{\textrm{DGU}}$ is stepped from 2.5 kW to 800 W at $t$ = 0.3 s. The responses of each DGU are plotted below.

\begin{figure}[!htb] % "[t!]" placement specifier just for this example
\graphicspath{ {Images/} }
\begin{subfigure}{0.42\textwidth}
\includegraphics[width=\linewidth]{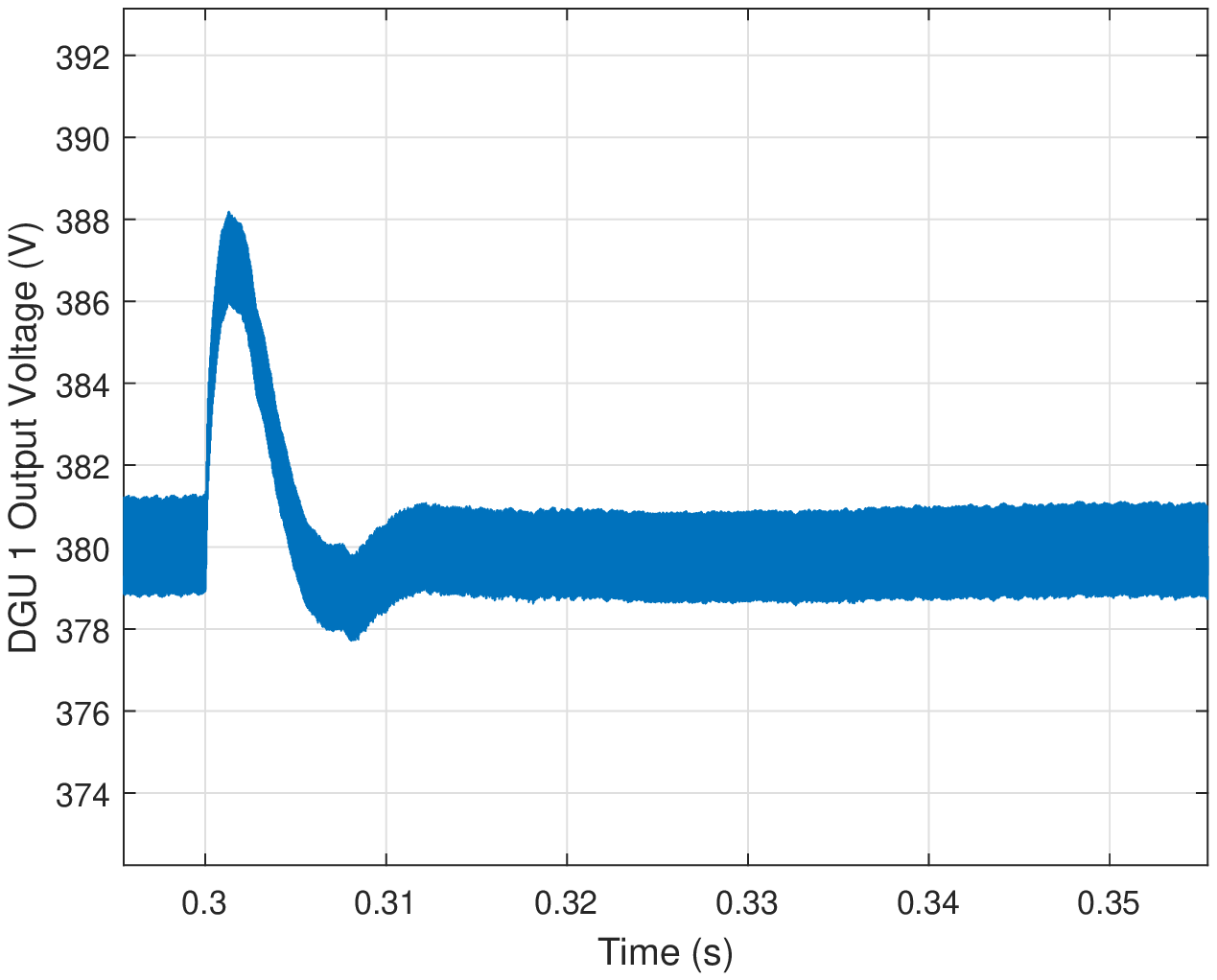}
\caption{$\hat{\Sigma}_{1}^{\textrm{DGU}}$ output voltage} \label{fig:DGU1LoadStepDGU6_2500To800W}
\end{subfigure}\hspace*{\fill}
\begin{subfigure}{0.42\textwidth}
\includegraphics[width=\linewidth]{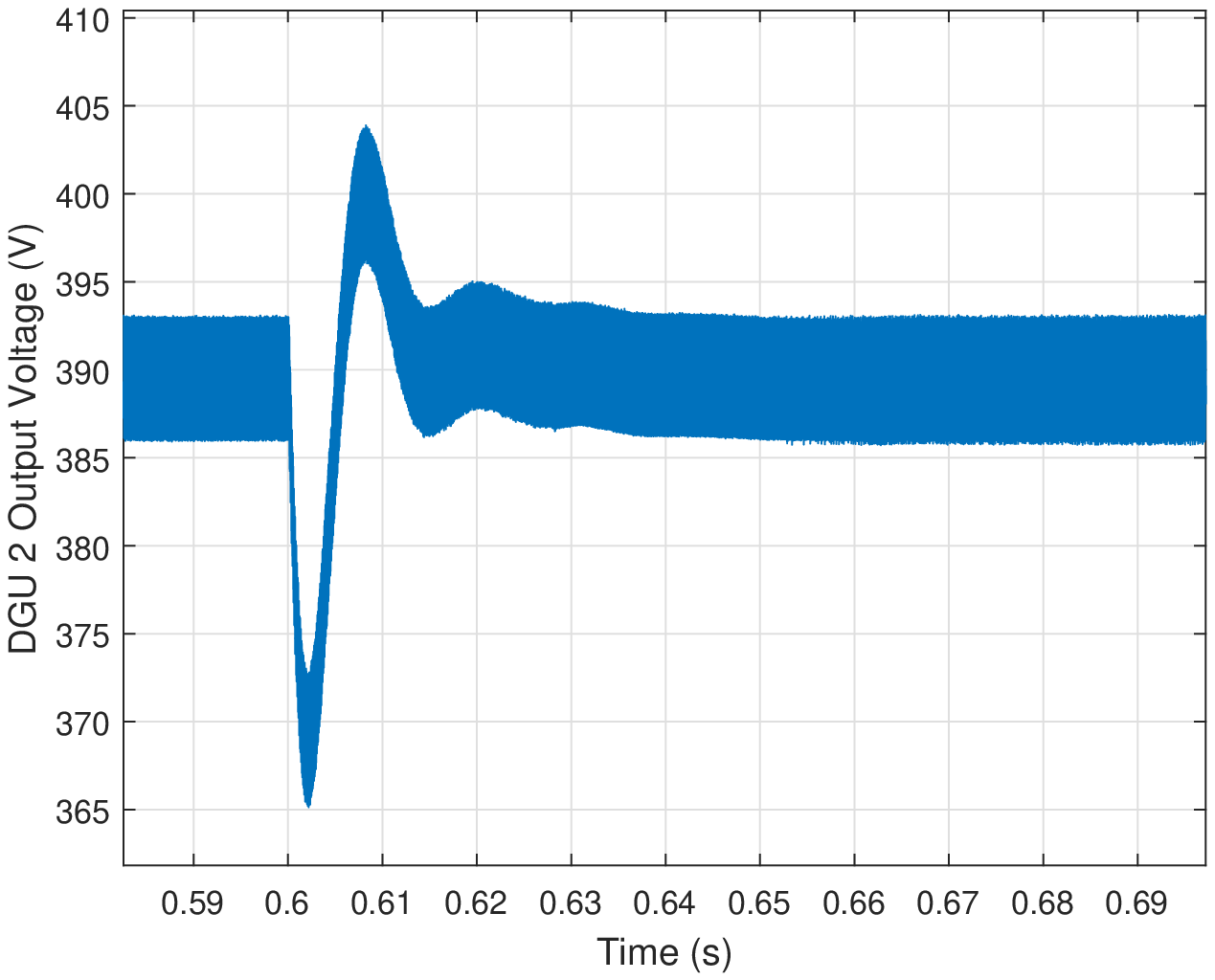}
\caption{$\hat{\Sigma}_{2}^{\textrm{DGU}}$ output voltage} \label{fig:DGU2LoadStepDGU6_2500To800W.}
\end{subfigure}
\end{figure}

\begin{figure}[!htb] % "[t!]" placement specifier just for this example
\ContinuedFloat%
\graphicspath{ {Images/} }
\begin{subfigure}{0.45\textwidth}
\includegraphics[width=\linewidth]{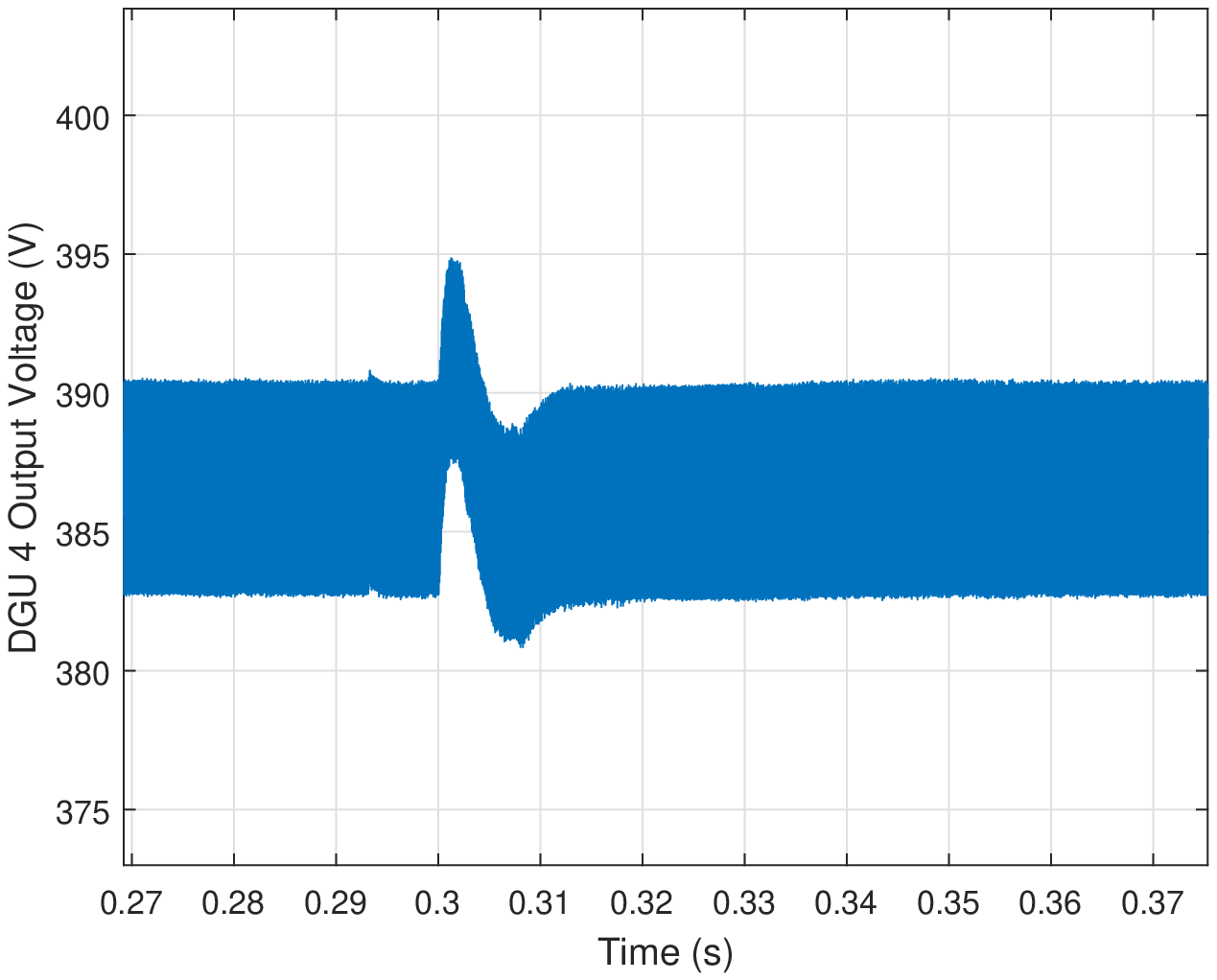}
\caption{$\hat{\Sigma}_{4}^{\textrm{DGU}}$ output voltage} \label{fig:DGU4LoadStepDGU6_2500To800W.}
\end{subfigure}\hspace*{\fill}
\medskip
\begin{subfigure}{0.45\textwidth}
\includegraphics[width=\linewidth]{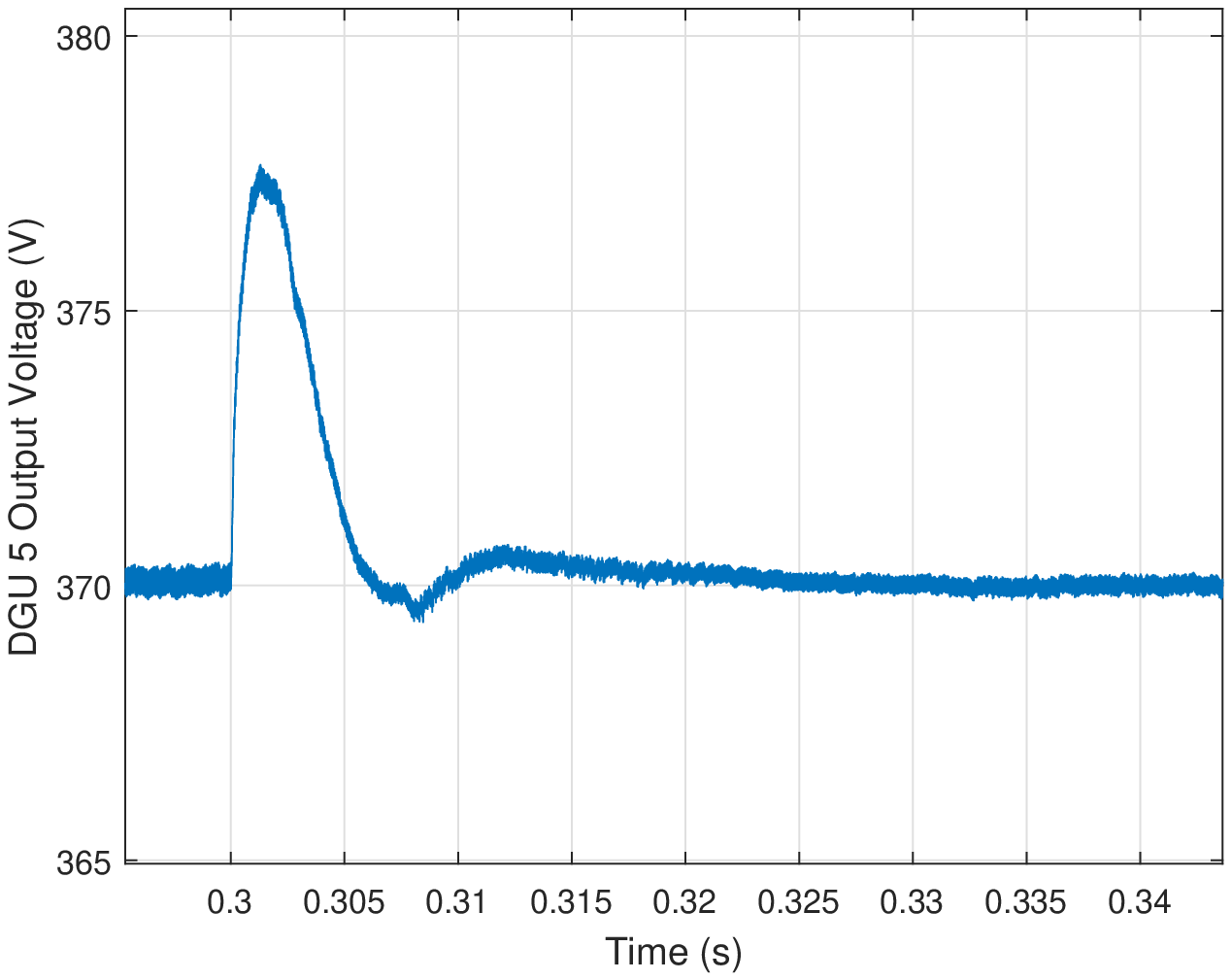}
\caption{$\hat{\Sigma}_{5}^{\textrm{DGU}}$ output voltage} \label{fig:DGU5LoadStepDGU6_2500To800W.}
\end{subfigure}
\medskip
\centering
\begin{subfigure}{0.45\textwidth}
\includegraphics[width=\linewidth]{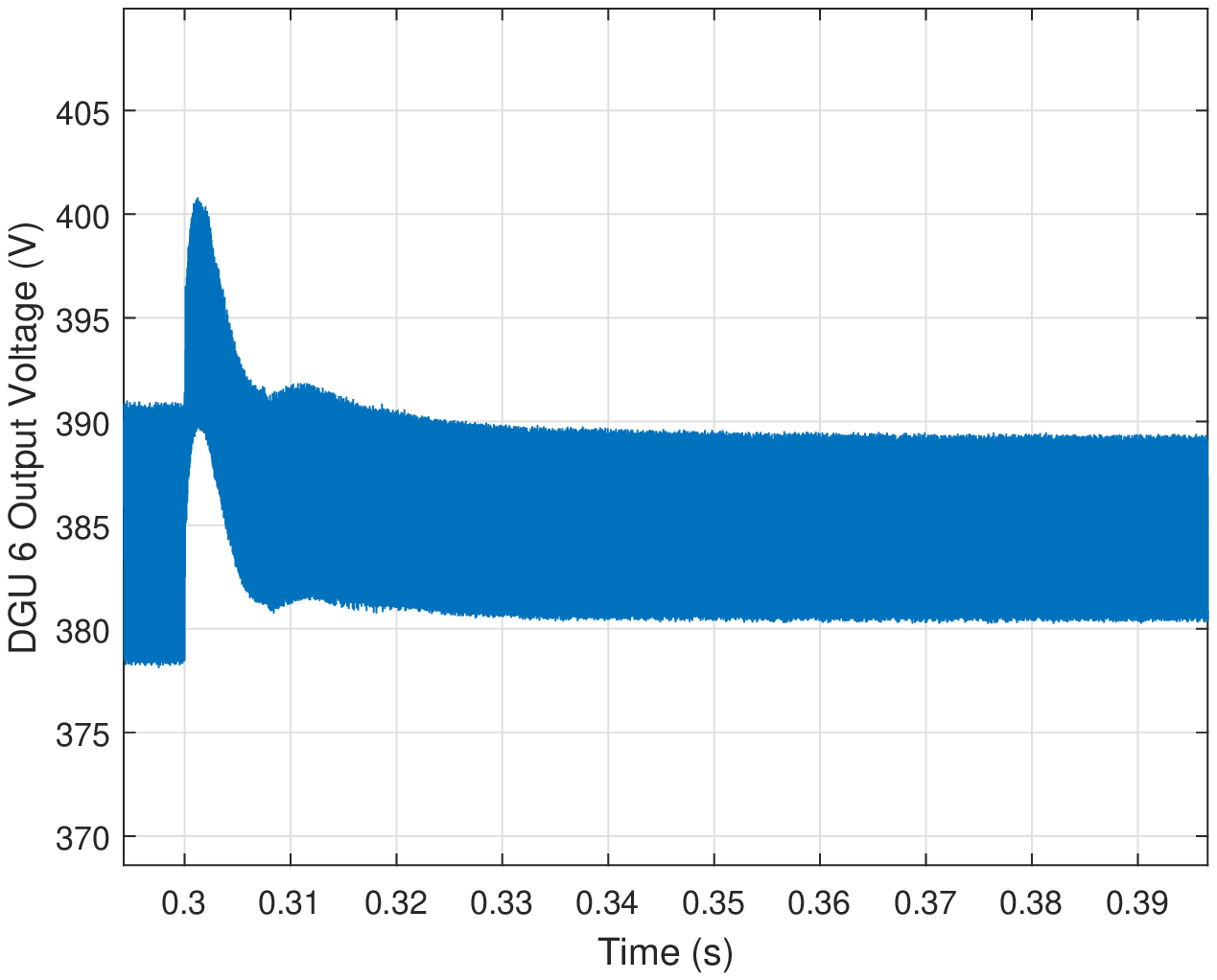}
\caption{$\hat{\Sigma}_{6}^{\textrm{DGU}}$ output voltage} \label{fig:DGU6LoadStepDGU6_2500To800W.}
\end{subfigure}
\caption{DGU output voltage responses to $\hat{\Sigma}_{6}^{\textrm{DGU}}$ load step of 2.5 kW to 800 W.} \label{fig:LoadChangeLargeV}
\end{figure}

\textbf{Note:} \textit{Output voltage response for $\hat{\Sigma}_{3}^{\textrm{DGU}}$ is not plotted at $t$ = 0.3 s, as it has been unplugged from the grid.}

As can be seen, though only one load is changed, it causes a disturbance to every DGU. All responses are favourable, with the longest settling time of 50 ms ($\hat{\Sigma}_{6}^{\textrm{DGU}}$) and largest under/overshoot 25 V ($\hat{\Sigma}_{2}^{\textrm{DGU}}$). 

Comparing the performance using the $\mathcal{L}_1$AC to the state-of-the-art PnP voltage controllers of \cite{Tucci2016c} reveals promising results. The resulting dynamics are faster than that of a 8-4 $\Omega$ load change at $\hat{\Sigma}_{6}^{\textrm{DGU}}$ in Fig. 14 of \cite{Tucci2016c}. While the dynamics of neighbouring DGUs are more oscillatory in \cite{Tucci2016c}, settling times are similar to Fig. \ref{fig:LoadChangeLargeV}(a) and \ref{fig:LoadChangeLargeV}(e), at 2 ms. However, the response of $\hat{\Sigma}_{6}^{\textrm{DGU}}$ to its load change in \cite{Tucci2016c} has a considerably slower settling time of 900 ms, while above, Fig. \ref{fig:LoadChangeLargeV}(e) shows that the settling time is 50 ms.

These tests highlight the very good performance and robustness of controllers $\mathcal{C}_{i}^{\mathcal{L}_1}, i = 1,...,6$, ensuring fast reference tracking in the presence of unknown load disturbances.

\subsubsection{Voltage Reference Tracking}

\begin{figure}[!htb] % "[t!]" placement specifier just for this example
\graphicspath{{Images/}}
\begin{subfigure}{0.48\textwidth}
\includegraphics[width=\linewidth]{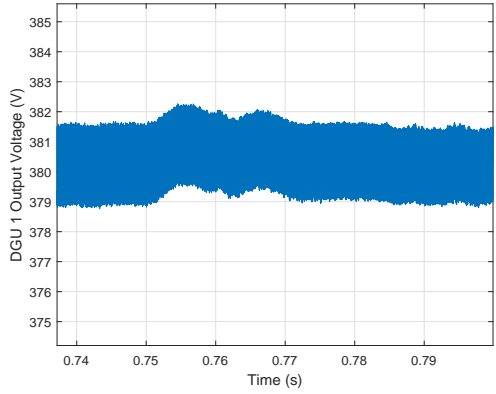}
\caption{$\hat{\Sigma}_{1}^{\textrm{DGU}}$ output voltage} \label{fig:DGU1VchangeDGU5370Vto377V}
\end{subfigure}\hspace*{\fill}
\begin{subfigure}{0.48\textwidth}
\includegraphics[width=\linewidth]{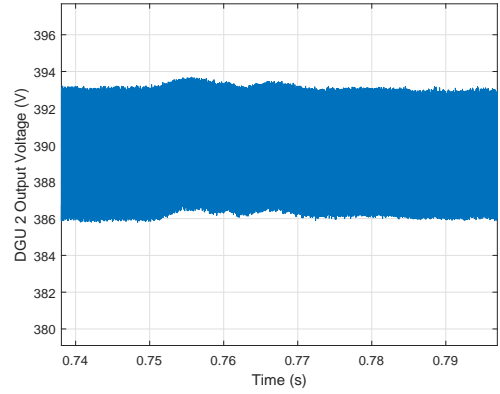}
\caption{$\hat{\Sigma}_{2}^{\textrm{DGU}}$ output voltage} \label{fig:DGU2VchangeDGU5370Vto377V}
\end{subfigure}

\medskip
\begin{subfigure}{0.48\textwidth}
\includegraphics[width=\linewidth]{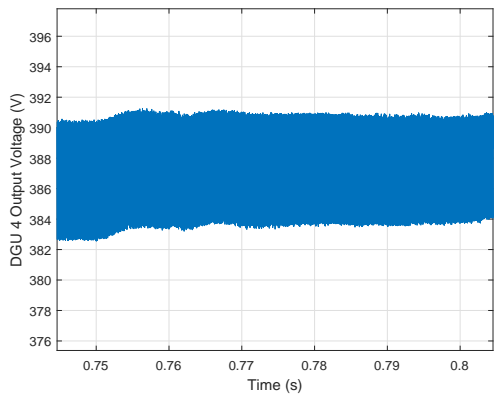}
\caption{$\hat{\Sigma}_{4}^{\textrm{DGU}}$ output voltage} \label{fig:DGU4VchangeDGU5370Vto377V}
\end{subfigure}\hspace*{\fill}
\begin{subfigure}{0.48\textwidth}
\includegraphics[width=\linewidth]{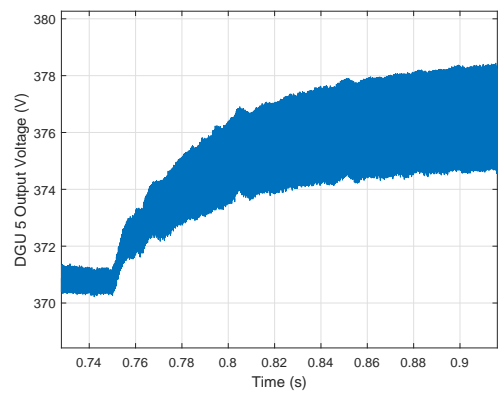}
\caption{$\hat{\Sigma}_{5}^{\textrm{DGU}}$ output voltage} \label{fig:DGU5VchangeDGU5370Vto377V}
\end{subfigure}

\medskip
\centering
\begin{subfigure}{0.48\textwidth}
\includegraphics[width=\linewidth]{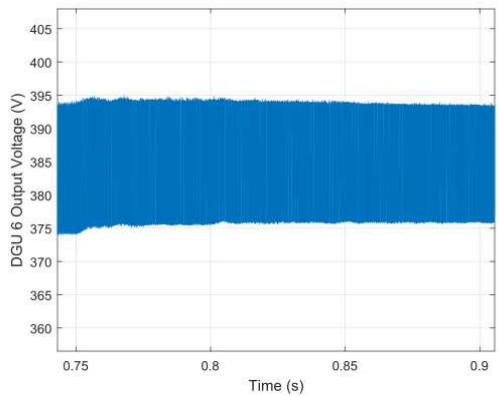}
\caption{$\hat{\Sigma}_{6}^{\textrm{DGU}}$ output voltage} \label{fig:DGU6VchangeDGU5370Vto377V}
\end{subfigure}

\caption{DGU output voltage responses to $\hat{\Sigma}_{5}^{\textrm{DGU}}$ voltage reference step of 370 V to 377 V.} \label{fig:VchangeDGU5370Vto377V}
\end{figure}

At $t$ = 0.75 s, the voltage reference of $\hat{\Sigma}_{5}^{\textrm{DGU}}$ is stepped from 370 V to 377 V. Since the impedance of the $RL$ lines is small, the voltage increase is enough to propagate an appreciable amount of power to coupled DGUs,
causing current disturbances to cascade throughout the ImG. Therefore, it is important that neighbouring DGUs are robust to this unknown disturbance also. The responses of each DGU are plotted in Fig. \ref{fig:VchangeDGU5370Vto377V}.

Fig. \ref{fig:VchangeDGU5370Vto377V}(d) shows that the controller $\mathcal{C}_{5}^{\mathcal{L}_1}$ is capable of guaranteeing fast reference tracking, with good damping and a settling time of $\approx$ 300 ms. Also, it should be noted how there is no undershoot, as expected in non-minimum phase systems. The performance of $\mathcal{C}_{5}^{\mathcal{L}_1}$ is similar to the state-of-the-art PnP voltage controller in \cite{Tucci2016c}, albeit Fig. 6 of \cite{Tucci2016c} suggests an instantaneous response. However, the settling time of a neighbouring DGU in Fig. 7 of \cite{Tucci2016c} is much
slower ($\approx 2$ s) compared to neighbouring DGUs of Fig. \ref{fig:VchangeDGU5370Vto377V} (largest being $\approx$ 20 ms).
 
    \bibliography{arXiv_ECC_Paper}

\end{document}